\documentclass[11pt,a4paper]{article}
\usepackage{subfig}
\usepackage{graphicx}
\usepackage{float}
\usepackage{afterpage}
\usepackage{epsfig,cite}
\usepackage{amssymb}
\usepackage{amsmath}
\usepackage{dsfont}
\usepackage{multirow}
\usepackage{booktabs}
\usepackage{tabularx}
\usepackage{xcolor}
\usepackage{bm}
\usepackage{textcomp}
\usepackage{caption}
\usepackage{slashed}
\usepackage{changes}
\usepackage{url}
\usepackage[colorlinks=true,linkcolor=blue,linktoc=all,citecolor=blue]{hyperref}
\usepackage{leftindex}

\usepackage[normalem]{ulem}

\reversemarginpar

\usepackage{bm}
\usepackage{slashed}
\newcommand{\ds}[1]{\slashed{{#1}}}

\newcommand{\e}{\varepsilon}
\newcommand{\ta}{\left(}
\newcommand{\qa}{\left[}

\newcommand{\tc}{\right)}
\newcommand{\qc}{\right]}

\newcommand{\tr}[1]{\text{Tr}\qa {#1} \qc}
\renewcommand{\[}{\begin{equation}}
\renewcommand{\]}{\end{equation}}
\newcommand{\nb}{{\bar{n}}}

\usepackage[most]{tcolorbox}

\definecolor{myGreen}{rgb}{0.18,0.763,0.324}
\definecolor{myWhite}{rgb}{0.98,0.98,0.98}
\definecolor{myGray}{rgb}{0.7,0.7,0.75}
\definecolor{myGold}{rgb}{0.8,0.64,0.24}
\definecolor{myPurple}{rgb}{1,0.3,0.9}

\begin{document}

\vspace{-2.0cm}
\begin{flushright}
  DESY-23-188\\IPARCOS-UCM-23-128
\end{flushright}
\vspace{.1cm}

\begin{center} 

  {\large \bf  One-loop evolution of twist-2 generalized parton distributions}
  \vspace{.7cm}

  Valerio~Bertone$^1$\footnote{\href{mailto:valerio.bertone@cea.fr}{valerio.bertone@cea.fr}},
  Rafael~F.~del~Castillo$^2$\footnote{\href{mailto:raffer06@ucm.es}{raffer06@ucm.es}},
  Miguel~G.~Echevarria$^{3,4}$\footnote{\href{mailto:miguel.garciae@ehu.eus}{miguel.garciae@ehu.eus}},
  \'{O}scar~del~R\'{i}o$^2$\footnote{\href{mailto:oscgar03@ucm.es}{oscgar03@ucm.es}},
  Simone~Rodini$^{5,6}$\footnote{\href{mailto:simone.rodini@desy.de}{simone.rodini@desy.de}}

\setcounter{footnote}{0}

\vspace{.3cm}
$^1${\it IRFU, CEA, Universit\'e Paris-Saclay, F-91191 Gif-sur-Yvette,
  France}\\
$^2${\it Departamento de F\'{i}sica Te\'{o}rica \& IPARCOS, Universidad Complutense de Madrid, E-28040 Madrid,
Spain}\\
$^3${\it
Department of Physics, University of the Basque Country UPV/EHU, 48080 Bilbao, Spain} \\
$^4${\it
EHU Quantum Center, University of the Basque Country UPV/EHU}\\
$^5${\it Deutsches Elektronen-Synchrotron DESY, Notkestr. 85, 22607 Hamburg, Germany}\\
$^6${\it CPHT, CNRS, Ecole polytechnique, Institut Polytechnique de Paris, 91120 Palaiseau, France}

\end{center}   

\begin{center}
  {\bf \large Abstract}\\
\end{center}
We revisit the evolution of generalised parton distributions (GPDs) at
the leading order in the strong coupling constant $\alpha_s$ for all
of the twist-2 quark and gluon operators. We rederive the relevant
one-loop evolution kernels, expressing them in a form suitable for
implementation, and check analytically that some basic properties,
such as DGLAP/ERBL limits and polynomiality conservation, are
fulfilled. We also present a number of numerical results obtained with
a public implementation of the evolution in the library {\tt APFEL++}
and available within the {\tt PARTONS} framework.

\newpage

\tableofcontents

\section{Introduction}\label{sec:Introduction}

Generalised parton distributions (GPDs) were introduced more than two
decades ago as a na\-tu\-ral generalisation of parton distribution
functions
(PDFs)~\cite{Muller:1994ses,Ji:1996ek,Radyushkin:1996nd,Radyushkin:1996ru,Ji:1996nm,Radyushkin:1997ki}
(see also
Refs.~\cite{Ji:1998pc,Diehl:2003ny,Belitsky:2005qn,Boffi:2007yc,Goeke:2001tz,Kumericki:2016ehc}
for comprehensive reviews). While PDFs are typically accessed through
inclusive processes, such as deep-inelastic lepton-hadron scattering,
GPDs emerge from the factorisation of deeply-virtual exclusive
processes, with deeply-virtual Compton scattering (DVCS) being the
golden channel~\cite{Ji:1996ek,Radyushkin:1996nd,Ji:1996nm}. In the
DVCS amplitudes, GPDs are encoded in Compton Form Factors (CFFs)
through their convolution over the longitudinal partonic momentum
fraction $x$ with known perturbative kernels~\cite{Braun:2020yib,
  Braun:2022bpn, Braun:2022ezk, Schoenleber:2022myb, Ji:2023xzk}.

GPDs provide a 1+2 dimensional picture of the partonic structure of
hadrons related to both the longitudinal-momentum and the
transverse-spatial distribution of
partons~\cite{Burkardt:2002hr}. Furthermore, the second Mellin moments
of unpolarised GPDs give access to the so-called Ji spin sum
rule~\cite{Ji:1996ek} and the $D$-term, which encode information on
the mechanical properties of
hadrons~\cite{Polyakov:2018zvc,Dutrieux:2021nlz,Freese:2022jlu,Lorce:2021xku,Duran:2022xag}. In
general, GPDs contain a wealth of information on hadron structure that
has led the upcoming Electron-Ion Collider
(EIC)~\cite{AbdulKhalek:2021gbh} and the JLab22
upgrade~\cite{Accardi:2023chb} to make the study of GPDs a cornerstone
of their research programs.

However, the phenomenological study of GPDs presents us with many
challenges. One of the primary hurdles in extracting meaningful
information from experimental data lies in the intricate structure of
the factorisation theorems, which renders the analysis of GPDs
considerably more difficult than that of PDFs. A pivotal aspect in the
exploration of GPDs concerns the ability to disentangle their
dependence on both the partonic longitudinal-momentum fraction $x$ and
the skewness $\xi$. This task proved exceptionally challenging in that
the convolution of GPDs with the partonic cross sections involved in
the computation of CFFs intertwines these variables, preventing a
straightforward separation. The longstanding belief that evolution
effects may help achieve this separation was finally disproven in
Ref.~\cite{Bertone:2021yyz}, where the concept of ``shadow'' GPDs,
\textit{i.e.} GPDs having an arbitrary small imprint on CCFs, was
introduced. However, more recently Ref.~\cite{Moffat:2023svr} revived
this debate. Nonetheless, it is broadly accepted that a solid
extraction of GPDs should not only rely on DVCS data but rather on a
simultaneous analysis of different processes, as routinely done for
PDFs.

This underscores the rationale behind the substantial efforts invested
in recent years to derive the perturbative structure of both the CFFs
and the evolution kernels governing the evolution of
GPDs~\cite{Mueller:2005ed,Braun:2017cih,Braun:2012bg,Braun:2012hq,Braun:2014sta,Braun:2017cih,Braun:2019qtp,Braun:2020yib,Braun:2021tzi,Braun:2022byg,Braun:2022bpn,Braun:2022ezk,Braun:2022qly,Ji:2023xzk,Schoenleber:2022myb,Moch:2021cdq,VanThurenhout:2022nmx}. This
pursuit has driven extensive investigations into the computation of
higher-order results, primarily employing conformal-space
techniques. This approach, not only enhances computational efficiency,
but also offers an independent alternative to the more traditional
Feynman-diagram-based methods.

On the phenomenological side, comparatively much less effort has been
devoted to developing GPD evolution codes. The first leading-order
(LO) momentum-space evolution code was developed by Vinnikov and
presented in Ref.~\cite{Vinnikov:2006xw}. However, the only surviving
public version of this code is available only through {\tt
  PARTONS}~\cite{Berthou:2015oaw}. Freund and
McDermott~\cite{Freund:2001bf} implemented a version of GPD evolution
specifically tailored for DVCS at next-to-LO (NLO) accuracy. However,
the code is not fully open-source and cannot be easily obtained.  On
the other hand, a public, open-source implementation of GPD evolution
in conformal space at NLO also exists~\cite{Mueller:2005ed,
  Kumericki:2007sa, gepard}.

All codes mentioned above are typically able to evolve only specific
models or class of parameterisations, and can thus hardly be used
out-of-the-box with arbitrary input GPDs. This is a significant
shortcoming in view of a possible extraction of GPDs from experimental
data based on flexible parameterisations. Another disadvantage of
these codes is that none of them includes heavy-flavour threshold
crossing. This is a limitation because much of the current
experimental data lies significantly above the charm threshold and,
all the more so, the future EIC will deliver data that will require
charm and bottom to be treated as active flavours. It is therefore
clear that the lack of open-source public codes to perform GPD
evolution in momentum space without any assumption on the
initial-scale model is now becoming a bottleneck.

With this work, we aim to provide a fully open-source implementation
for all of the twist-2 GPD evolution equations in momentum space with
no a-priori assumptions on the input models and allowing for
heavy-flavour threshold crossing. We extend the work of
Ref.~\cite{Bertone:2022frx}, devoted to the one-loop evolution of
unpolarised GPDs, re-computing and implementing the evolution kernels
for longitudinally polarised quarks and gluons, and for transversely
polarised quarks and circularly polarised gluons. These quantities had
already been computed (see \textit{e.g.} Refs.\cite{Ji:1998pc,
  Belitsky:2000yn}), but a numerical implementation only exists for
longitudinally polarised partons~\cite{Vinnikov:2006xw} (with the
limitations discussed above) and was absent so far for transversely
polarised quarks and circularly polarised gluons. The implementation
of the full set of twist-2 evolution equations is made public through
the code {\tt APFEL++}~\cite{Bertone:2013vaa, Bertone:2017gds} and
available within the numerical framework {\tt
  PARTONS}~\cite{Berthou:2015oaw}.

The paper is organised as follows. In Sect.~\ref{sec:Definitions}, we
give a brief overview of definitions and conventions on both the
structure of GPDs and their evolution equations. The explicit form of
the leading-order kernels is given in Sect.~\ref{sec:Results} and
presented in a form that is well-suited for numerical
implementation. In the same section, we also discuss some relevant
properties of the kernels. In Sect.~\ref{sec:NumResults}, we present
some numerical results of our implementation, and finally in
Sect.~\ref{sec:Conclusions} we draw some conclusions.

\section{Definitions}\label{sec:Definitions}

Let us first start with a summary of our notation. We will denote
scalar products as $a^\mu b_\mu \equiv (ab)$.  We introduce two
light-cone vectors, $n$ and $\nb$, such that $n^2 = \nb^2 = 0 $ and
$(\nb n) = 1$. We parametrise the transverse space to $n$ and $\nb$ in
terms of two vectors $R$ and $L$ that satisfy the normalisation
conditions:\footnote{An explicit parametrisation of all these vectors
  is:
\[
n^\mu=\frac{1}{\sqrt{2}} (1,0,0,-1),\quad \nb^\mu=\frac{1}{\sqrt{2}}
(1,0,0,1),\quad R^\mu = -\frac{1}{\sqrt{2}}(0,1,i,0), \quad L^\mu =
  -\frac{1}{\sqrt{2}}(0,1,-i,0)\,.
\]}
\[
(RR) = (LL) = (Rn) = (Ln) = (R\nb)= (L\nb) = 0, \quad (RL) = -1.
\]
In addition, we will use the short-hand notation: $v_+ = (vn)$,
$v_- = (v\nb)$, $v_R = (vR)$ and $v_L = (vL)$ throughout.

The bare quark and gluon GPD correlators associated with the hadron
$H$ are defined in terms of off-forward matrix elements of the
collinear operators as follows (see \textit{e.g.}
Ref.~\cite{Diehl:2003ny} for basic definitions):
\begin{equation}
\begin{split}
  \hat{F}^{ij}_{q\leftarrow H}(x,\xi,t) &= \int
  \frac{dz}{2\pi}e^{-ixP_+z }\left\langle P+\frac{\Delta}{2}
    ,\Lambda'\right| \bar{q}^j\ta \frac{zn}{2}\tc
  \mathcal{W}\qa\frac{zn}{2},-\frac{zn}{2}\qc q^i\ta -\frac{zn}{2}\tc
  \left| P-\frac{\Delta}{2},\Lambda\right\rangle\,,\\
  \\
  \hat{F}^{\mu\nu}_{g\leftarrow H}(x,\xi,t) &= \int
  \frac{dz}{2\pi}\frac{e^{-ixP_+z }}{xP_+}\left\langle
    P+\frac{\Delta}{2} ,\Lambda'\right| F_a^{\mu+}\ta \frac{zn}{2}\tc
  \mathcal{W}_{ab}\qa\frac{zn}{2},-\frac{zn}{2}\qc F_b^{\nu+}\ta
  -\frac{zn}{2}\tc \left| P-\frac{\Delta}{2},\Lambda\right\rangle\,,
\end{split}
\label{GPD_definition}
\end{equation}
where $\xi = -\Delta_+ / (2P_+)$, $t=\Delta^2$, and $\mathcal{W}$ is
the Wilson line defined as:
\begin{equation}
  \mathcal{W}\qa yn, wn\qc = \mathbb{P}\exp\left\{-ig(w-y)n^\nu T^c\int_0^1 ds\  A^c_{\nu}(swn +(1-s)yn)\right\},
\end{equation}
where $\mathbb{P}$ denotes the path ordering. The representation of
the color-group generators $T^c$ is the fundamental one in the quark
case and the adjoint one in the gluon case. In the hadronic states,
$\Lambda$ and $\Lambda'$ denote the helicity states of the incoming
and outgoing hadron $H$, respectively. The relation between the spin
vector $S$ and the helicity $\Lambda$ for a state of momentum $p$ and
mass $M$ is given by:
\[
S^\mu = \Lambda \frac{p_+ \nb^\mu -p_- n^\mu}{M} - S_R L^\mu -S_L R^\mu.
\]
At twist-2, three relevant projections are to be considered for both
quarks and gluons: unpolarised, longitudinally polarised, and
transversely (quark)/circularly (gluon) polarised. For the quark
operator, the three cases are projected out as follows:
\begin{equation}
  \hat{F}^{[\Gamma]}_{q\leftarrow H}(x,\xi,t) =
  \frac{1}{2}\tr{\hat{F}_{q\leftarrow H}(x,\xi,t)\Gamma}\,,
\end{equation}
with:
\begin{equation}
  \Gamma\in \{\ds{n},\ds{n}\gamma_5,i n_\beta\sigma^{\alpha\beta}\gamma_5\}\,.
\end{equation}
In the gluon case, the projection is instead defined as:
\begin{equation}
\hat{F}_{g\leftarrow H}^{[\Gamma]}(x,\xi,t) = \Gamma_{\mu\nu} \hat{F}_{g\leftarrow H}^{\mu\nu}(x,\xi,t)\,,
\end{equation}
where the tensor $\Gamma^{\mu\nu}$ is to be selected amongst the
following structures:
\begin{equation}
\Gamma^{\mu\nu} \in \{-g_T^{\mu\nu}\equiv -(g^{\mu\nu}-n^\mu \nb^\nu - \nb^\mu n^\nu), \
-i\e_T^{\mu\nu}\equiv -i\e^{\alpha\beta\mu\nu}\nb_\alpha n_\beta, \
-R^\mu R^\nu-L^\mu L^\nu\}\,,
\label{eq:gluonprojectors}
\end{equation}
where we use the convention $\e_T^{12} \equiv \nb_\alpha n_\beta
\e^{\alpha\beta 12} = +1$.

The evolution kernels do not depend on the specific external states
and therefore they are independent of how the correlators
$\hat{F}_{f\leftarrow H}$ are parametrised in terms of the single
scalar GPDs. In principle, one could study the evolution equations in
position space~\cite{Braun:2009mi}, where the independence of the
external states is made transparent. This observation also implies
that evolution equations for transition
GPDs~\cite{Belitsky:2005qn,Kroll:2022roq} are identical to those for
standard (flavour diagonal) GPDs.

The goal of this work is the evaluation of the one-loop (leading-order
(LO)) evolution kernels in momentum space for all of the three twist-2
GPD correlators introduced above. Although these quantities are
already known in the literature,\footnote{For example, a list of all
  relevant one-loop kernels in position space and their transformation
  in momentum space can be found in Ref.~\cite{Braun:2009mi}.} we aim
to achieve an efficient numerical implementation and to lay the
foundations for a systematic Feynman-graph approach to the
computation. For this reason, we choose to work in light-cone gauge,
which allows us to consider a significantly smaller number of
diagrams. The light-cone gauge is obtained by enforcing the following
condition on the gluon field:
\begin{equation}
  A_+^c = 0\,.
  \label{eq:LightConeGuageFix}
\end{equation}
It is well known that in light-cone gauges the condition above is not
enough to completely fix the gauge \cite{Belitsky:2002sm}. Indeed, the
transverse components of the gauge field at light-cone infinity are
left unconstrained by Eq.~(\ref{eq:LightConeGuageFix}). However, in
the context of GPDs, the specific boundary condition on these
transverse components is irrelevant. Indeed, the gauge link (Wilson
line) runs along the light-cone direction and all operators are
compact, thus shielding the GPDs from being sensitive to the boundary
conditions at light-cone infinity.

A complication of working in light-cone gauge is that the gluon
propagator has a more convoluted structure that reads:
\begin{equation}
\mathcal{D}^{\mu\nu}(k)=\frac{id^{\mu\nu}(k)}{k^2+i0}\,,
\quad d^{\mu\nu}(k)=-g^{\mu\nu} +\frac{k^\mu n^\nu + k^\nu
  n^\mu}{(nk)_{\rm reg.}}\,.
\end{equation}
The subscript ``reg.'' indicates that the linear propagator
$(nk)^{-1}$, which gives rise to the so-called rapidity divergences,
has to be regularised. These spurious divergences, which are present
in single diagrams, cancel out when summing over all diagrams, so that
the regulator can eventually be safely removed. However, at one-loop
the cancellation of the rapidity divergences is apparent and we find
it unnecessary to specify a particular regularisation procedure.

The GPD correlators defined in Eq.~\eqref{GPD_definition} require UV
renormalisation. Using dimensional regularisation in $d=4-2\e$
dimensions and the $\overline{\text{MS}}$ renormalisation scheme, GPDs
are renormalised in a multiplicative fashion by means of a set of
renormalisation constants $Z$ as follows:
\begin{equation}
F^{[\Gamma]}_{f\leftarrow H}(x,\xi,t;\mu) = \sum_{f'} \int_{-1}^1
\frac{dy}{|y|} Z_{f/f'}^{[\Gamma]}\ta \frac{x}{y},\frac{\xi}{x},\e,\alpha_s\tc
\hat{F}^{[\Gamma]}_{f'\leftarrow H}(x,\xi,t;\e),\quad \Gamma=U,L,T\,,
\label{eq:GPDrenormalisation}
\end{equation}
where $U$, $L$, and $T$ stand for unpolarised, longitudinally
polarised and transversely/circularly polarised projections,
respectively.\footnote{Here and in the following we refer to
  transversely/circularly polarised GPDs with the index $T$,
  implicitly understanding transversely polarised quark GPDs and
  circularly polarised gluon GPDs.} In addition, the sums over $f'$
and/or $f''$ run over active partons. The corresponding evolution
equations for the renormalised GPDs take the general
form~\cite{Bertone:2022frx}:
\begin{equation}
\frac{\partial F^{[\Gamma]}_{f\leftarrow H}(x,\xi,t;\mu)}{\partial \ln \mu^2} = \sum_{f'} \int_{-1}^1 \frac{dz}{|z|}\mathcal{P}_{f/f'}^{[\Gamma]}\ta \frac{x}{z},\frac{\xi}{x},\alpha_s(\mu)\tc   F^{[\Gamma]}_{f'\leftarrow H}(z,\xi,t;\mu).
\end{equation}
The evolution kernel $\mathcal{P}$ are related to the renormalisation
constants as:
\begin{equation}
\mathcal{P}_{f/f'}^{[\Gamma]}\ta
\frac{x}{z},\frac{\xi}{x},\alpha_s(\mu)\tc = \lim_{\e\to0} \sum_{f''}
\int_{-1}^1 \frac{dy}{|y|} \qa \frac{\partial}{\partial\ln \mu^2} Z_{f/f''}^{[\Gamma]}\ta \frac{x}{y},\frac{\xi}{x},\alpha_s,\e\tc \qc \left({Z_{f''/f'}^{[\Gamma]}}\right)^{-1}\ta \frac{y}{z},\frac{\xi}{y},\alpha_s,\e\tc\,.
\end{equation}
The $\overline{\text{MS}}$ renormalisation constants $Z$ can depend on
the renormalisation scale $\mu$ only through the strong coupling
$\alpha_s$. Therefore, defining $a_s = \alpha_s/(4\pi)$, we have that:
\begin{equation}
\begin{split}
\frac{\partial}{\partial\ln\mu^2} Z_{f/f''}^{[\Gamma]}\ta \frac{x}{y},\frac{\xi}{x},\alpha_s,\e\tc &= \frac{ \partial a_s}{\partial \ln\mu^2} \frac{\partial}{\partial a_s} Z_{f/f''}^{[\Gamma]}\ta \frac{x}{y},\frac{\xi}{x},\alpha_s,\e\tc \\
& = \ta-\e a_s +\beta(a_s)\tc \frac{\partial}{\partial a_s} \sum_{n=1} \sum_{p=1}^n \frac{a_s^n}{\e^p} Z^{[\Gamma],[n,p]}_{f/f''}\ta \frac{x}{y},\frac{\xi}{x}\tc \\
& = \ta-\e +\frac{\beta(a_s)}{a_s}\tc\sum_{n=1} \sum_{p=1}^n \frac{n a_s^{n}}{\e^p} Z^{[\Gamma],[n,p]}_{f/f''}\ta \frac{x}{y},\frac{\xi}{x}\tc\,.
\end{split}
\end{equation}
Expanding the evolution kernel in powers of $a_s$:
\begin{equation}
\mathcal{P}_{f/f'}^{[\Gamma]} \ta \frac{x}{y},\frac{\xi}{x},\alpha_s\tc = a_s \sum_{n=0}a_s^n \mathcal{P}^{[\Gamma],[n]}_{f/f'}\ta \frac{x}{y},\frac{\xi}{x}\tc\,,
\end{equation}
at LO we immediately obtain:
\begin{equation}
\mathcal{P}^{[\Gamma],[0]}_{f/f'} \ta\frac{x}{y},\frac{\xi}{x}\tc = -Z^{[\Gamma],[1,1]}_{f/f'}\ta\frac{x}{y},\frac{\xi}{x}\tc.
\end{equation}

Since the renormalisation constants are universal and independent of
the external states, we can compute them perturbatively using the
parton-in-parton GPDs, which are defined from the hadronic GPDs by
replacing the external hadronic states with on-shell free partonic
states~\cite{Bertone:2022frx}. Expanding the bare and renormalised
parton-in-parton GPDs in powers of $a_s$:
\begin{equation}
\begin{split}
  \hat{F}^{[\Gamma]}_{f\leftarrow f'}(x,\xi,\e) &= \sum_{n=0} a_s^n\hat{F}^{[\Gamma],[n]}_{f\leftarrow f'}(x,\xi,\e)\,, \\
  F^{[\Gamma]}_{f\leftarrow f'}(x,\xi,\mu) &= \sum_{n=0}
  a_s^nF^{[\Gamma],[n]}_{f\leftarrow f'}(x,\xi,\mu)\,,
\end{split}
\end{equation}
we can derive the renormalisation constants by plugging these
expansions into Eq.~(\ref{eq:GPDrenormalisation}) and requiring that
the renormalised GPDs be finite order by order in $a_s$. This
eventually produces the iterative set of equations:
\begin{equation}
  F^{[\Gamma],[n]}_{f\leftarrow f'}(x,\xi,\mu) = \lim_{\e\to0}\sum_{f''} \int_{-1}^1 \frac{dy}{|y|} \sum_{q=0}^n \sum_{k=1}^q  \frac{1}{\e^k} Z^{[\Gamma],[q,k]}_{f/f''} \ta\frac{x}{y},\frac{\xi}{x}\tc \hat{F}^{[\Gamma],[n-q]}_{f''\leftarrow f'}(y,\xi,\e).
\end{equation}
The coefficients $Z^{[\Gamma],[q,k]}_{f/f''}$ are obtained by matching
the UV divergences produced by the diagrammatic calculation of
$\hat{F}^{[\Gamma],[n-q]}_{f''\leftarrow f'}$ in a way that
$F^{[\Gamma],[n]}_{f\leftarrow f'}$ are all finite.

\section{Analytic results}\label{sec:Results}

As discussed above, the evolution kernels derive from the
renormalisation of the parton-in-parton GPDs and only depend on the
operator involved in their definition. As a consequence, the
unpolarised GPDs $H$ and $E$ share the same evolution kernels, and so
do the longitudinally polarised GPDs $\widetilde{H}$ and
$\widetilde{E}$, and the full set of transversely/circularly polarised
GPDs $H_T$, $E_T$, $\widetilde{H}_T$, and
$\widetilde{E}_T$.

The general form of the one-loop evolution kernels for each of these
three classes of GPDs can be presented as follows:
\[\label{KernelGPDs}
\begin{split}
\mathcal{P}_{i/k}^{[\Gamma],[0]}\ta x,\frac{\xi}{x}\tc = &\theta(1-x)\qa \theta(x+\xi) p_{i/k}^{\Gamma}\ta x,\frac{\xi}{x}\tc  + \theta(x-\xi)p_{i/k}^{\Gamma}\ta x,-\frac{\xi}{x}\tc\qc \\
& +2\delta_{ik}\delta(1-x) C_i \qa K_i + \ln\left|1-\frac{\xi^2}{x^2}\right| -2\int_0^1 \frac{dz}{1-z}\qc\,,
\end{split}
\]
where the $\theta$ function is normalised such that $\theta(0)=1$.
The constants $K_i$ and $C_i$ are the same for all polarisations
$\Gamma$ and read:
\[
\begin{split}
K_q = \frac{3}{2}, \quad K_g = \frac{11}{6} - \frac{2n_f}{3} \frac{T_R}{C_A}, \quad 
\end{split}
\]
with $C_q = C_F$ and $C_g = C_A$. Conversely, the functions $p_{i/k}^{\Gamma}$
are different for each polarisation. Those associated with the
unpolarised GPDs $H$ and $E$ have already been presented in
Ref.~\cite{Bertone:2022frx}, but we report them here for completeness:
\begin{align} 
p_{q/q}^U\left(x,\frac{\xi}{x}\right) &= C_F \frac{(x+\xi)
                       (1-x+2\xi)}{\xi (1+\xi)(1-x)}\,,\\
  p_{q/g}^U \left(x,\frac{\xi}{x}\right) &= T_R\frac{(x+\xi)
                       (1-2x+\xi)}{\xi (1+\xi) (1-\xi^2)} \,,\\
  p_{g/q}^U \left(x,\frac{\xi}{x}\right) &= C_F\frac{(x+\xi)
                       (2-x+\xi)}{x\xi (1+\xi)}\,,\\
 \label{KernelFunction_ggU} p_{g/g}^U \left(x,\frac{\xi}{x}\right) &= -C_A\frac{x^2-\xi^2}{x\xi (1-\xi^2)}\left[1-\frac{2\xi}{1-x}-\frac{2(1+x^2)}{ (x-\xi ) (1+\xi)}\right]\,.
\end{align}
In the longitudinally polarised case, instead, they read:
\begin{align}
p^L_{q/q}\ta x, \frac{\xi}{x} \tc &= C_F \frac{ (x+\xi ) ( 1-x+2 \xi)}{\xi (1+ \xi )  (1-x)}\,,\\
p^L_{q/g}\ta x,\frac{\xi}{x}\tc &= - T_R\frac{x+ \xi}{\xi(1+\xi)^2}\,,\\
p^L_{g/q}\ta x,\frac{\xi}{x}\tc &=  C_F \frac{(x+\xi)^2}{x\xi(1+\xi)}\,,\\
\label{KernelFunction_ggL}p^L_{g/g}\ta x, \frac{\xi}{x}\tc & = \frac{ C_A (\xi +x) \left(-\xi ^2 (2 \xi +1)+\xi +(\xi -3) x^2+\left(\xi ^2+3\right) x\right)}{(1-\xi^2) \xi  (1+\xi) (1-x) x }\,,
\end{align}
while in the transversely/circularly polarised case, they read:
\begin{align}
  p^T_{q/q}\ta x, \frac{\xi}{x} \tc &= 2 C_F\frac{ x+\xi}{(1+\xi)(1-x) }\,,\\
p^T_{q/g}\ta x, \frac{\xi}{x} \tc &= p^T_{g/q}\ta x, \frac{\xi}{x} \tc = 0\,,\\
\label{KernelFunction_ggT}p^T_{g/g}\ta x, \frac{\xi}{x}\tc & = 2 C_A \frac{(x+\xi )^2}{(1+\xi)^2 (1-x) x}\,.
\end{align}
It is interesting to observe that in the transversely/circularly
polarised case, due to the fact that the off-diagonal functions
$p^T_{q/g}$ and $p^T_{g/q}$ are identically zero, transversely
polarised quark GPDs and circularly polarised gluon GPD do not couple
under evolution. More details on the computation of the functions
above can be found in Appendix~\ref{sec:Appendix}.

Defining appropriate GPD combinations allows for a partial
diagonalisation of the splitting matrix that is best suited for
numerical implementations. At LO, these combinations are the
non-singlet:
\begin{equation}\label{eq:totalvalence}
  F^{[\Gamma],-} = \sum_{q=1}^{n_f}F_{q\leftarrow H}^{[\Gamma]}-F_{\overline{q}\leftarrow H}^{[\Gamma]}\,,
\end{equation}
and the singlet:
\begin{equation}
  F^{[\Gamma],+}=
\begin{pmatrix}
  \displaystyle \sum_{q=1}^{n_f}F_{q\leftarrow H}^{[\Gamma]}+F_{\overline{q}\leftarrow H}^{[\Gamma]}\\
  F_{g\leftarrow H}^{[\Gamma]}
\end{pmatrix}\,,
\end{equation}
where $n_f$ is the number of active quark flavours and
$F_{i\leftarrow H}^{[U]}=H_{i\leftarrow H}, E_{i\leftarrow H}$,
$F_{i\leftarrow H}^{[L]}=\widetilde{H}_{i\leftarrow H},
\widetilde{E}_{i\leftarrow H}$,
$F_{i\leftarrow H}^{[T]}=H_{T,i\leftarrow H}, E_{T,i\leftarrow
  H},\widetilde{H}_{T,i\leftarrow H}, \widetilde{E}_{T,i\leftarrow H}$
with $i=q,g$. In addition, we have defined anti-quark GPDs using the
charge-conjugation symmetry relations:
\[
F_{\overline{q}\leftarrow H}^{[\Gamma]}(x,\xi,t;\mu)=\mp F_{q\leftarrow H}^{[\Gamma]} (-x,\xi,t;\mu), \quad F_{g\leftarrow H}^{[\Gamma]} (x,\xi,t;\mu)=\mp F_{g\leftarrow H}^{[\Gamma]} (-x,\xi,t;\mu)\,,
\]
where the upper sign applies to the unpolarised and
transversely/circularly polarised cases ($\Gamma=U,T$), while the
lower sign applies to the longitudinally polarised case
($\Gamma=L$). Non-singlet and singlet GPD combinations obey their own
decoupled evolution equations that at one loop read:
\begin{equation}
\frac{\partial F^{[\Gamma],\pm}(x,\xi,t;\mu)}{\partial\ln\mu^2} =\frac{\alpha_s(\mu)}{4\pi}
\int_x^\infty\frac{dy}{y}\mathcal{P}^{[\Gamma],\pm,[0]}\left(y,\kappa\right)F^{[\Gamma],\pm}\left(\frac{x}{y},\xi, t;\mu\right)\,,
\label{eq:regdglap}
\end{equation}
with $\kappa=\xi/x$. The evolution kernels $\mathcal{P}$ can be
decomposed as:
\begin{equation}
\mathcal{P}^{[\Gamma],\pm,[0]}\left(y,\kappa\right) = \theta(1-y)
\mathcal{P}_1^{[\Gamma],\pm,[0]}\left(y,\kappa\right)+\theta(\kappa-1)
\mathcal{P}_2^{[\Gamma],\pm,[0]}\left(y,\kappa\right)\,,
\label{eq:KernelDecomposition}
\end{equation}
with the non-singlet combinations given by:
\begin{equation}
\begin{array}{rcl}
  \displaystyle\mathcal{P}_1^{[\Gamma],-,[0]}\left(y,\kappa\right)&=&\displaystyle
                                                                    p_{q/q}^{\Gamma}\left(y,\kappa\right)+p_{q/q}^{\Gamma}\left(y,-\kappa\right)\\
\\
&+&\displaystyle \delta(1-y) 2C_q\left[K_q-2\int_0^1 \frac{dz}{1-z}-\ln\left|1-\kappa^2\right|\right]\,,\\
\\
\mathcal{P}_2^{[\Gamma],-,[0]}\left(y,\kappa\right)&=&\displaystyle
                                                     -p_{q/q}^{\Gamma}\left(y,-\kappa\right)
                                                     \pm p_{q/q}^{\Gamma}\left(-y,-\kappa\right)\,,
\label{eq:P1andP2NS}
\end{array}
\end{equation}
and the singlet combinations given by:
\begin{equation}
\begin{array}{rcl}
\displaystyle\mathcal{P}_{1,ik}^{[\Gamma],+,[0]}\left(y,\kappa\right)&=&\displaystyle p_{i/k}^{\Gamma}\left(y,\kappa\right)+p_{i/k}^{\Gamma}\left(y,-\kappa\right) \\
\\
&+&\displaystyle \delta_{ik}\delta(1-y) 2C_i\left[K_i-2\int_0^1 \frac{dz}{1-z}-\ln\left|1-\kappa^2\right|\right]\,,\\
\\
\displaystyle\mathcal{P}_{2,ik}^{[\Gamma],+,[0]}\left(y,\kappa\right)&=&\displaystyle
                                                     -p_{i/k}^{\Gamma}\left(y,-\kappa\right)
                                                     \mp p_{i/k}^{\Gamma}\left(-y,-\kappa\right)\,,
\label{eq:P1andP2SG}
\end{array}
\end{equation}
where again the upper sign refers to unpolarised and
transversely/circularly polarised distributions ($\Gamma=U,T$), while
the lower sign refers to the longitudinally polarised ones
($\Gamma=L$). Explicit expressions for the one-loop non-polarised
splitting kernels in this notation can be found in
Ref.~\cite{Bertone:2022frx}, but we report them here for completeness:
\begin{equation}
\small
\begin{array}{l}
\left\{\begin{array}{rcl}
\mathcal{P}_{1}^{[U],-,[0]}(y,\kappa) &=& \displaystyle 2C_F\left\{\left(\frac{2}{1-y}\right)_+-\frac{1
  +y}{1-\kappa^2y^2}+\delta(1-y)\left[K_q-
  \ln\left(|1-\kappa^2|\right)\right]\right\}\,,\\
\\
\mathcal{P}_{2}^{[U],-,[0]}(y,\kappa) &=& \displaystyle 2C_F\left[\frac{1+(1+\kappa)y+(1+\kappa
      -\kappa^2)y^2}{(1+y)(1-\kappa^2y^2)}-\left(\frac{1}{1-y}\right)_{++}\right]\,,
\end{array}\right.\\
\\
\left\{\begin{array}{rcl}
\mathcal{P}_{1, qq}^{[U],+,[0]}(y,\kappa) &=& \mathcal{P}_{1}^{-,[0]}(y,\kappa)\,,\\
\\
\mathcal{P}_{2, qq}^{[U],+,[0]}(y,\kappa) &=& \displaystyle
                                           2C_F\left[\frac{1+y+\kappa y+\kappa^3y^2}{\kappa(1+y)(1-\kappa^2y^2)}-\left(\frac{1}{1-y}\right)_{++}\right]\,,
\end{array}\right.\\
\\
\left\{\begin{array}{rcl}
\mathcal{P}_{1, qg}^{[U],+,[0]}(y,\kappa) &=& \displaystyle 4n_f T_R\left[\frac{y^2+(1-y)^2-\kappa^2y^2}{(1-\kappa^2y^2)^2}\right]\,,\\
\\
\mathcal{P}_{2, qg}^{[U],+,[0]}(y,\kappa) &=& \displaystyle 4n_f T_R(1-\kappa)\left[\frac{1-\kappa(\kappa+2)y^2}{\kappa(1-\kappa^2y^2)^2}\right]\,,
\end{array}\right.\\
\\
\left\{\begin{array}{rcl}
  \mathcal{P}_{1, gq}^{[U],+,[0]}(y,\kappa) &=& \displaystyle 2C_F\left[\frac{1+(1-y)^2-\kappa^2y^2}{y(1-\kappa^2y^2)}\right]\,,\\
  \\
  \mathcal{P}_{2, gq}^{[U],+,[0]}(y,\kappa) &=& \displaystyle-2C_F\frac{(1-\kappa)^2}{\kappa(1-\kappa^2 y^2)}\,,
\end{array}\right.\\
\\
\left\{\begin{array}{rcl}
           \mathcal{P}_{1, gg}^{[U],+,[0]}(y,\kappa) &=& \displaystyle
                                                     4C_A\Bigg[\left(\frac{1}{1-y}\right)_+-\frac{1+\kappa^2y}{1-\kappa^2y^2}+\frac{1}{(1-\kappa^2y^2)^2}\left(\frac{1-y}{y}+y(1-y)\right)\\
           \\
                                                 &+&\displaystyle\delta(1-y)\frac{K_g -\ln(|1-\kappa^2|)}{2}\Bigg]\,,\\
           \\
           \mathcal{P}_{2, gg}^{[U],+,[0]}(y,\kappa) &=& \displaystyle
                                                     2C_A\left[\frac{2(1-\kappa)(1+y^2)}{(1-\kappa^2y^2)^2}+\frac{\kappa^2(1+y)}{1-\kappa^2y^2}+\frac{1-\kappa^2}{1-\kappa^2y^2}\left(2-\frac{1}{\kappa}-\frac{1}{1+y}\right)
                                                     -\left(\frac{1}{1-y}\right)_{++}\right]
                                                     \,.
\end{array}\right.
\end{array}
\label{eq:PUnp}
\end{equation}

The longitudinally polarised ones read:
\begin{equation}
  \small
\begin{array}{l}
\left\{\begin{array}{l}
\mathcal{P}_{1}^{[L],-,[0]}(y,\kappa) = \displaystyle 2C_F\left[\left(\frac{2}{1-y}\right)_+-\frac{1
  +y}{1-\kappa^2y^2}+\delta(1-y)\left[K_q-
  \ln\left(|1-\kappa^2|\right)\right]\right]\,,\\
\\
\mathcal{P}_{2}^{[L],-,[0]}(y,\kappa) = \displaystyle
                                           2C_F\left[\frac{1+y+\kappa
                                            y+\kappa^3y^2}{\kappa(1+y)(1-\kappa^2y^2)}-\left(\frac{1}{1-y}\right)_{++}\right]\,,
       \end{array}\right.\\
  \\
\left\{\begin{array}{l}
\mathcal{P}_{1, qq}^{[L],+,[0]}(y,\kappa) = \mathcal{P}_{1}^{[L],-,[0]}(y,\kappa)\,,\\
\\
\mathcal{P}_{2, qq}^{[L],+,[0]}(y,\kappa) = \displaystyle 2C_F\left[\frac{1+(1+\kappa)y+(1+\kappa
                                        -\kappa^2)y^2}{(1+y)(1-\kappa^2y^2)}-\left(\frac{1}{1-y}\right)_{++}\right]\,,
       \end{array}\right.\\
  \\
\left\{\begin{array}{l}
\mathcal{P}_{1, qg}^{[L],+,[0]}(y,\kappa) = \displaystyle 4n_f
                                            T_R\left[\frac{2y - 1 -\kappa^2y^2}{(1-\kappa^2y^2)^2}\right]\,,\\
\\
\mathcal{P}_{2, qg}^{[L],+,[0]}(y,\kappa) = \displaystyle -8n_f
                                            T_R(1-\kappa)\left[\frac{y}{(1-\kappa^2y^2)^2}\right]\,,
       \end{array}\right.\\
  \\
\left\{\begin{array}{l}
  \mathcal{P}_{1, gq}^{[L],+,[0]}(y,\kappa) = \displaystyle 2C_F\left[\frac{2-y-\kappa^2y}{1-\kappa^2y^2}\right]\,,\\
  \\
  \mathcal{P}_{2, gq}^{[L],+,[0]}(y,\kappa) =
                                              \displaystyle2C_F(1-\kappa)^2\frac{y}{1-\kappa^2 y^2}\,,
       \end{array}\right.\\
  \\
  \left\{\begin{array}{l}
           \mathcal{P}_{1, gg}^{[L],+,[0]}(y,\kappa) = \displaystyle 4 C_A\left[\left(\frac{1}{1-y}\right)_++\frac{(1-\kappa ^2 y)(1 -2y -\kappa ^2 y^2)}{(1-\kappa ^2 y^2)^2}+\delta(1-y)\frac{K_g-\ln(|1-\kappa^2|)}{2}\right]\,,\\
           \\
           \mathcal{P}_{2, gg}^{[L],+,[0]}(y,\kappa) =\displaystyle 2 C_A \left[\frac{(2 y+1)^2-4 \kappa  y (y+1) +2 \kappa ^2 y -\kappa ^4 y^3 (y+2)}{(1+y) \left(1-\kappa ^2 y^2\right)^2}-\left(\frac{1}{1-y}\right)_{++}\right]
\end{array}\right.
\end{array}
\label{eq:PLong}
\end{equation}
while the (non-zero) transversely polarised ones read:
\begin{equation}
  \small
  \begin{array}{l}
\left\{\begin{array}{l}
\mathcal{P}_{1}^{[T],-,[0]}(y,\kappa) = \displaystyle 2C_F\left[\left(\frac{2}{1-y}\right)_+-\frac{2}{1-\kappa^2y^2}+\delta(1-y)\left[K_q-
  \ln\left(|1-\kappa^2|\right)\right]\right]\,,\\
\\
\mathcal{P}_{2}^{[T],-,[0]}(y,\kappa) = \displaystyle
                                        2C_F\left[\frac{1+2\kappa y+\kappa(2-\kappa)y^2}{(1+y)(1-\kappa^2y^2)}-\left(\frac{1}{1-y}\right)_{++}\right]\,,
\end{array}\right.\\
\\
\left\{\begin{array}{l}
\mathcal{P}_{1, qq}^{[T],+,[0]}(y,\kappa) = \mathcal{P}_{1}^{[T],-,[0]}(y,\kappa)\,,\\
\\
\mathcal{P}_{2, qq}^{[T],+,[0]}(y,\kappa) = \displaystyle
                                           2C_F\left[\frac{1+2y+\kappa^2y^2}{(1+y)(1-\kappa^2y^2)}-\left(\frac{1}{1-y}\right)_{++}\right]\,,
\end{array}\right.\\
    \\
    \left\{\begin{array}{l}
           \mathcal{P}_{1, gg}^{[T],+,[0]}(y,\kappa) =\displaystyle 4 C_A\left[\left(\frac{1}{1-y}\right)_+-\frac{ \left(1-\kappa ^2 y\right) \left(1+\kappa^2 y^2\right)}{\left(1-\kappa ^2 y^2\right)^2} +\delta(1-y)\frac{K_g-\ln(|1-\kappa^2|)}{2}\right]\,,\\
           \\
           \mathcal{P}_{2, gg}^{[T],+,[0]}(y,\kappa) =\displaystyle 2 C_A \left[\frac{1+2 y- 4 \kappa ^2 y^2 (\kappa -1) -
    \kappa^3 y^3 (\kappa  y+4)+2 \kappa ^2 y^3}{(1+y) \left(1-\kappa ^2 y^2\right)^2}-\left(\frac{1}{1-y}\right)_{++}\right]\,.
           \end{array}\right.
    \end{array}
\label{eq:PTrans}
\end{equation}
The $+$-distribution (with round brackets) in the expressions above is
defined as:
\begin{equation}
\int_x^1dy\left(\frac{1}{1-y}\right)_+f(y) =
\int_x^1dy\frac{f(y)-f(1)}{1-y} + f(1)\ln(1-x)\,,
\label{eq:plusprescr1}
\end{equation}
while the $++$-distribution instead is defined as:
\begin{equation}
\int_x^\infty dy\left(\frac{1}{1-y}\right)_{++}f(y)=\int_x^\infty\frac{dy}{1-y}\left[f(y)-f(1)\left(1+\theta(y-1)\frac{1-y}{y}\right)\right]+f(1)\ln(1-x)\,.
\label{eq:plusplusdist}
\end{equation}

\subsection{DGLAP limit}

One of the fundamental requirements of the GPD evolution kernels is
that, in the limit of vanishing $\xi$, the well-known DGLAP splitting
functions have to be recovered. This limit amounts to taking
$\kappa\rightarrow 0$ and, given the decomposition in
Eq.~(\ref{eq:KernelDecomposition}), it is such that
$\mathcal{P}_2^{[\Gamma],\pm,[0]}$ drops, leaving only the term
proportional to $\mathcal{P}_1^{[\Gamma],\pm,[0]}$. The presence of
$\theta(1-y)$ in this term reduces the integral in the r.h.s. of
Eq.~(\ref{eq:regdglap}) to a ``standard'' Mellin convolution which is
precisely what enters the DGLAP evolution equations. What is left to
verify is that $\mathcal{P}_1^{[\Gamma],\pm,[0]}$ in the limit
$\kappa\rightarrow 0$ tend to the known one-loop DGLAP splitting
functions. This has already been verified in
Ref.~\cite{Bertone:2022frx} in the unpolarised case. Using
Eq.~(\ref{eq:PLong}), for the longitudinally polarised evolution
kernels we find:
\begin{equation}
\begin{array}{l}
\displaystyle \lim_{\kappa\rightarrow 0}
  \mathcal{P}_{1}^{[L],-,[0]}(y,\kappa) = \lim_{\kappa\rightarrow
  0}\mathcal{P}_{1, qq}^{[L],+,[0]}(y,\kappa) = \displaystyle 2C_F\left[\left(\frac{2}{1-y}\right)_+-(1
  +y)+\delta(1-y)K_q\right]\,,\\
  \\
\displaystyle \lim_{\kappa\rightarrow 0}\mathcal{P}_{1, qg}^{[L],+,[0]}(y,\kappa) = \displaystyle 4n_fT_R\left(2y - 1\right)\,,\\
\\
\displaystyle \lim_{\kappa\rightarrow 0}\mathcal{P}_{1, gq}^{[L],+,[0]}(y,\kappa) = \displaystyle 2C_F\left(2-y\right)\,,\\
  \\
\displaystyle \lim_{\kappa\rightarrow 0} \mathcal{P}_{1, gg}^{[L],+,[0]}(y,\kappa) = \displaystyle 4 C_A\left[\left(\frac{1}{1-y}\right)_++1 -2y+\delta(1-y)\frac{K_g}{2}\right]\,,
 \end{array}
\end{equation}
that indeed coincide with the corresponding DGLAP splitting
functions~\cite{Altarelli:1977zs}. For the transversely polarised
evolution kernels we take the limit for $\kappa\rightarrow 0$ of
Eq.~(\ref{eq:PTrans}) and find:
\begin{equation}
  \begin{array}{l}
\displaystyle \lim_{\kappa\rightarrow 0}\mathcal{P}_{1}^{[T],-,[0]}(y,\kappa) = \lim_{\kappa\rightarrow 0}\mathcal{P}_{1, qq}^{[T],+,[0]}(y,\kappa) = \displaystyle 4C_F\left[\left(\frac{1}{1-y}\right)_+-1+\delta(1-y)\frac{K_q}2\right]\,,\\
\\
 \displaystyle \lim_{\kappa\rightarrow 0}\mathcal{P}_{1, gg}^{[T],+,[0]}(y,\kappa) =\displaystyle 4 C_A\left[\left(\frac{1}{1-y}\right)_+-1 +\delta(1-y)\frac{K_g}{2}\right]\,.
    \end{array}
\end{equation}
These expressions coincide with those from Refs.~\cite{Delduc:1980ef,
  Artru:1989zv,Vogelsang:1998yd}.\footnote{Note that in Eq.~(13) of
  Ref.~\cite{Vogelsang:1998yd} the term proportional to the
  $\delta$-function is correct for the $\Delta_TP_{qq}^{(0)}$, while
  it should be equal to $K_g$ in the case of $\Delta_LP_{gg}^{(0)}$
  (see Refs.~\cite{Delduc:1980ef, Artru:1989zv}).}

\subsection{ERBL limit}

The evolution equations in Eq.~(\ref{eq:regdglap}) can be
alternatively written in a form that resembles the ERBL
equation~\cite{Lepage:1980fj,Bertone:2022frx}:
\begin{equation}
\frac{\partial F^{[\Gamma],\pm}(x,\xi,t;\mu)}{\partial\ln\mu^2} =
\frac{\alpha_s(\mu)}{4\pi}\int_{-1}^1\frac{dy}{|\xi|} \mathbb{V}^{[\Gamma],\pm,[0]}\left(\frac{x}{\xi},\frac{y}{\xi}\right) F^{[\Gamma],\pm}(y,\xi, t;\mu)\,,
\label{eq:ERBLlikeeveq}
\end{equation}
with:
\begin{equation}
\begin{array}{rcl}
\displaystyle \frac{1}{|\xi|}\mathbb{V}_{ik}^{[\Gamma],+,[0]}\left(\frac{x}{\xi}, \frac{y}{\xi}\right) &=& \displaystyle
\frac{1}{y}\bigg\{\left[\theta(x-\xi) \theta(y-x)-\theta(-x-\xi)
                                              \theta(x-y)\right] \left[p_{ik}^{\Gamma}\left(\frac{x}{y}, \frac{\xi}{x}\right)+p_{ik}^{\Gamma}\left(\frac{x}{y}, -\frac{\xi}{x}\right)\right]\\
\\
&+&\displaystyle \theta(\xi-x)\theta(x+\xi) \left[\theta(y-x) p_{ik}^{\Gamma}\left(\frac{x}{y}, \frac{\xi}{x}\right)-\theta(x-y) p_{ik}^{\Gamma}\left(\frac{x}{y}, -\frac{\xi}{x}\right)\right]\bigg\}\\
\\
&+&\displaystyle \delta\left(1-\frac{x}{y}\right)
    \delta_{ik}2C_i\left[K_i+\int_\xi^{x}\frac{dz}{z-x}+\int_{-\xi}^{x}\frac{dz}{z-x}\right]\,,\\
\\
\displaystyle \frac{1}{|\xi|}\mathbb{V}^{[\Gamma],-,[0]}\left(\frac{x}{\xi}, \frac{y}{\xi}\right)&=&\displaystyle \frac{1}{|\xi|} \mathbb{V}_{qq}^{[\Gamma],+,[0]}\left(\frac{x}{\xi}, \frac{y}{\xi}\right)\,.
\end{array}
\label{eq:evkernelERBLgenxi}
\end{equation}
Taking the $\xi\rightarrow 1$ limit and performing the changes of
variable $x=2v-1$ and $y=2u-1$, the GPD evolution equations turn into
ERBL evolution equations~\cite{Lepage:1980fj} for distribution
amplitudes (DAs):
\begin{equation}
  \frac{\partial \Phi^{[\Gamma],\pm}(v,t;\mu)}{\partial\ln\mu^2} =
  \frac{\alpha_s(\mu)}{4\pi}\int_{0}^1du\, V^{[\Gamma],\pm,[0]}(v,u) \Phi^{[\Gamma],\pm}(u,t;\mu)\,,
\end{equation}
where the DAs $\Phi^{[\Gamma],\pm}$ are related to the GPDs through
the following identity:
\begin{equation}
\Phi^{[\Gamma],\pm}(v,t;\mu) = \lim_{\xi\rightarrow 1}F^{[\Gamma],\pm}(2v-1,\xi,t;\mu)\,,
\end{equation}
and the corresponding evolution kernels are defined as:
\begin{equation}
V^{[\Gamma],\pm,[0]}(v,u) = \lim_{\xi\rightarrow1}\frac{1}{|\xi|}\mathbb{V}^{[\Gamma],\pm,[0]}\left(\frac{2v-1}{\xi}, \frac{2u-1}{\xi}\right)\,.
\end{equation}
Their explicit expressions in the unpolarised case are given
by:
\begin{equation}
  \begin{array}{rcl}
    V^{[U],-,[0]}(v,u) &=&\displaystyle   V_{qq}^{+,[0]}(v,u)=
                       -C_F\left[\theta(u-v)\left(\frac{1-v}{u}-\frac{1}{u-v}\right)+\theta(v-u)\left(\frac{v}{1-u}-\frac{1}{v-u}\right)\right]_+\,,\\
    \\
    V_{qg}^{[U],+,[0]}(v,u)&=&\displaystyle -2n_f T_R\frac{2
                           u-1}{2u(1-u)}\left[\theta (u-v)\frac{v (u
                           -2 v+1) }{u}-\theta (v-u)\frac{(1-v) ( 2 v-u) }{1-u}\right]\,,\\
    \\
    V_{gq}^{[U],+,[0]}(v,u)&=&\displaystyle \frac{2C_F}{2 v-1}\left[\theta (u-v)\frac{v (2 u-v) }{u}-\theta (v-u)\frac{(1-v) (v-2 u+1) }{ 1-u}\right]\,,\\
    \\
    V_{gg}^{[U],+,[0]}(v,u)&=& \displaystyle C_A\Bigg[\theta (u-v)\frac{ \left(t^3 (4 u-2)-v^2 (u+2) (2
                           u-1)+v (2 u ((2 u-3) u+2)-1)+(u-1)
                           u\right)}{(2 v-1) (u-1)
                           u^2}\\
    \\
                   &-&\displaystyle \theta (v-u)\frac{\left(v^3 (2-4 u)+v^2 (2 u-1)
                       (u+3)-v (2 u-1) \left(2 u^2+1\right)+((4 u-5) u+2)
                       u\right)}{(2 v-1) (u-1)^2 u}\\
    \\
                   &+&\displaystyle \left[\frac{\theta (u-v)}{u-v}\right]_++\left[\frac{\theta (v-u)}{v-u}\right]_+\Bigg]+\delta(v-u)C_A K_g\,.
\end{array}
\end{equation}
In Ref.~\cite{Bertone:2022frx}, it was verified that, at least in the
non-singlet case, these expressions coincide with those present in the
literature. The longitudinally polarised expressions instead read:
\begin{equation}
  \begin{array}{rcl}
      V^{[L],-,[0]}(v,u) &=&\displaystyle   V_{qq}^{[L],+,[0]}(v,u)=
                         -C_F\left[\theta(u-v)\left(\frac{1-v}{u}-\frac{1}{u-v}\right)+\theta(v-u)\left(\frac{v}{1-u}-\frac{1}{v-u}\right)\right]_+\,,\\
    \\
    V_{qg}^{[L],+,[0]}(v,u)&=&\displaystyle -2n_f T_R\frac{2 u-1}{2}\left[\theta (u-v)\frac{v }{u^2}-\theta (v-u)\frac{1-v}{(1-u)^2}\right]\,,\\
\\
  V_{gq}^{[L],+,[0]}(v,u)&=&\displaystyle \frac{2C_F}{2v-1}\left[\theta (u-v)\frac{v^2 }{u}-\theta (v-u)\frac{(1-v)^2 }{1-u}\right]\,,\\
    \\
    V_{gg}^{[L],+,[0]}(v,u)&=& \displaystyle C_A \Bigg[
                             \theta (u-v)\frac{\left(v^2 ((6 u-7) u+2)-v
                             (1-2 u)^2+(u-1) u\right)}{(2 v-1) (u-1)
                             u^2}\\
    \\
    &-&\displaystyle \theta (v-u)\frac{\left(v^2 ((5-6 u) u-1)+v
        \left(8 u^2-6 u+1\right)+u (2-3 u)\right)}{(2 v-1) (u-1)^2
        u}\\
    \\
    &+&\displaystyle \left[\frac{\theta (u-v)}{u-v}\right]_++\left[\frac{\theta (v-u)}{v-u}\right]_+\Bigg]+\delta(v-u)C_A K_g\,,
\end{array}
\end{equation}
while the transversely/circularly polarised ones are:
\begin{equation}
  \begin{array}{rcl}
      V^{[T],-,[0]}(v,u) &=&\displaystyle V_{qq}^{[T],+,[0]}(v,u)=
                           C_F\Bigg[-\frac{\theta (u-v)
                           }{u}-\frac{\theta (v-u)
                           }{1-u}\\
    \\
    &+&\displaystyle \left[\frac{\theta (u-v)
                           }{u-v}\right]_++\left[\frac{\theta (v-u)
                           }{v-u}\right]_+\Bigg]+ \delta(v-u)C_F K_q\,,\\
    \\
    V_{gg}^{[T],+,[0]}(v,u)&=& \displaystyle C_A\Bigg[\frac{-2 v u+v+u}{2 v-1}\left(
                             \frac{\theta (u-v)}{u^2}- \frac{\theta (v-u)}{ (1-u)^2}\right)\\
    \\
    &+&\displaystyle \left[\frac{\theta (u-v) }{u-v}\right]_++\left[\frac{\theta (v-u) }{v-u}\right]_+\Bigg]+\delta(v-u)C_A K_g\,.
\end{array}
\end{equation}
The $+$-prescription in the expressions above, this time with square
brackets, is defined differently from Eq.~(\ref{eq:plusprescr1}) and reads:
\begin{equation}
  [f(v,u)]_+=f(v,u) - \delta(u-v)\int_0^1dv\, f(v,u)\,.
  \label{eq:plusprescr2}
\end{equation}

To the best of our knowledge, the one-loop ERBL kernels for the full
set of twist-2 distributions for both singlet and non-singlet
combinations have not been presented anywhere.

\subsection{Continuity at $x=\xi$ and spurious divergences}\label{subsec:continuity}

As it can be verified explicitly, all of the $\mathcal{P}_2$ functions
in Eqs.~(\ref{eq:PUnp})-(\ref{eq:PTrans}) are such that:
\begin{equation}
  \mathcal{P}_2^{[\Gamma],\pm,[0]}(y,k)\propto (1-\kappa)\,.
\label{eq:continuity}
\end{equation}
This property ensures that the r.h.s. of Eq.~(\ref{eq:regdglap}) is
continuous at $\kappa=1$, \textit{i.e.} at the crossover point
$x=\xi$. This is essential to ensure the continuity of GPDs at the
crossover point.

Of course, GPD continuity also requires that the integral in the
r.h.s. of Eq.~(\ref{eq:regdglap}) converges for all values of
$\kappa$. However, as it can be seen from
Eqs.~(\ref{eq:PUnp})-(\ref{eq:PTrans}), all of the single expressions
for $\mathcal{P}_1$ and $\mathcal{P}_2$ are affected by a spurious
pole at $y = 1/\kappa$. For $\kappa\leq 1$, $\mathcal{P}_2$ does not
contribute to the evolution, while the pole of $\mathcal{P}_1$, that
is to be integrated only up to $y=1$, falls outside the integration
region. As a consequence, the integral in this region converges. For
$\kappa> 1$, both $\mathcal{P}_1$ and $\mathcal{P}_2$ contribute. As
shown in Ref.~\cite{Bertone:2022frx} in the unpolarised case, it so
happens that the coefficients of the poles at $y = 1/\kappa$ of
$\mathcal{P}_1$ and $\mathcal{P}_2$ are equal in absolute value but
opposite in sign, such that the singularity cancels out leaving a
finite result also for $\kappa > 1$.

In the following, we will prove that the same cancellation takes place
also for the longitudinally and transversely/circularly polarised
evolution kernels. Having ascertained that the cancellation needs to
happen only for $\kappa > 1$, we concentrate on this region. For each
evolution kernel we compute the following quantities:
\begin{equation}
\mathcal{L}^{[L]/[T],\pm}=\lim_{y\rightarrow \kappa^{-1}} \mathcal{P}_1^{[L]/[T],\pm,[0]}(y,\kappa)+\mathcal{P}_2^{[L]/[T],\pm,[0]}(y,\kappa)\,,
\end{equation}
and verify that they are finite. In the longitudinally polarised case
we find:
\begin{equation}
  \begin{array}{l}
    \displaystyle \mathcal{L}^{[L],-}= C_F \frac{1-5 \kappa ^2}{\kappa(1-\kappa^2)}\,,\\
    \\
    \displaystyle \mathcal{L}_{qq}^{[L],+}= C_F \frac{3 \kappa ^2 +1}{\kappa^2-1}\,, \\
    \\
    \displaystyle \mathcal{L}_{qg}^{[L],+}= -n_fT_R\,,\\
    \\
    \displaystyle \mathcal{L}_{gq}^{[L],+}= 2C_F\,,\\
    \\
    \displaystyle \mathcal{L}_{gg}^{[L],+}= C_A \frac{1-5 \kappa ^2}{1-\kappa ^2}\,,
  \end{array}
\end{equation}
while in the transversely/circularly polarised case, we have:
\begin{equation}
  \begin{array}{l}
    \displaystyle \mathcal{L}^{[T],-}= 2 C_F\frac{ \kappa ^2+1 }{\kappa ^2-1}\,,\\
    \\
    \displaystyle \mathcal{L}_{qq}^{[T],+}= 4 C_F\frac{\kappa  }{\kappa ^2-1}\,,\\
    \\
    \displaystyle \mathcal{L}_{gg}^{[T],+}= C_A\frac{\kappa (\kappa -1)  }{\kappa +1}\,,
  \end{array}
\end{equation}
that are indeed all finite for $\kappa>1$, thus guaranteeing that the
evolution at one loop leaves GPDs continuous at the crossover point
$x=\xi$.

\subsection{Sum rules}

In this section, we discuss the sum rules. Specifically, it can be
shown that polynomiality of GPDs implies some integral constraints of
the evolution kernels. Ref.~\cite{Bertone:2022frx} discusses these
constraints in the unpolarised case. In the longitudinally polarised
case the conservation of the first moment leads to:\footnote{Note that
  these constraints apply to all orders in perturbation theory and not
  only to the one-loop contribution.}
\begin{equation}
  \int_0^1dz\left[\mathcal{P}_{1,ij}^{[L],+,[0]}\left(z,\frac{\xi}{yz}\right)+
 \frac{\xi}{y} \,
  \mathcal{P}_{2,ij}^{[L],+,[0]}\left(\frac{z\xi}{y},\frac{1}{z}\right)\right]=\mbox{constant
in }\xi\,,
\end{equation}
where the independence of $\xi$ also implies the independence of
$y$. Indeed, we find:
\begin{equation}
\label{L_sum_rules}
\begin{array}{l}
\displaystyle  \int_0^1dz\left[\mathcal{P}_{1,qq}^{[L],+,[0]}\left(z,\frac{\xi}{yz}\right)+
 \frac{\xi}{y} \,
  \mathcal{P}_{2,qq}^{[L],+,[0]}\left(\frac{z\xi}{y},\frac{1}{z}\right)\right]=0\,,\\
\\
\displaystyle \int_0^1dz\left[\mathcal{P}_{1,qg}^{[L],+,[0]}\left(z,\frac{\xi}{yz}\right)+
 \frac{\xi}{y} \,
  \mathcal{P}_{2,qg}^{[L],+,[0]}\left(\frac{z\xi}{y},\frac{1}{z}\right)\right] =0\,,\\
\\
\displaystyle\int_0^1dz\left[\mathcal{P}_{1,gq}^{[L],+,[0]}\left(z,\frac{\xi}{yz}\right)+
 \frac{\xi}{y} \,
  \mathcal{P}_{2,gq}^{[L],+,[0]}\left(\frac{z\xi}{y},\frac{1}{z}\right)\right]
  = 3 C_F\,,\\
\\
\displaystyle\int_0^1dz\left[\mathcal{P}_{1,gg}^{[L],+,[0]}\left(z,\frac{\xi}{yz}\right)+
 \frac{\xi}{y} \,
  \mathcal{P}_{2,gg}^{[L],+,[0]}\left(\frac{z\xi}{y},\frac{1}{z}\right)\right]=\frac{11
  C_A- 4 n_f T_R}{3}\,.
\end{array}
\end{equation}
It is interesting to observe that, in the cases $ij = qq, qg$, the
integrals above evaluate to zero. This implies that, at least at
one-loop accuracy, the first moment the longitudinally polarised quark
GPDs is independent of the scale $\mu$ and can thus be identified with
a physical observable. This is indeed related to the anti-symmetric
part of the energy-momentum tensor. It is known that the
anti-symmetric form factor is related to the axial form factor that at
one loop does not need renormalisation (see \textit{e.g.}
Refs.~\cite{Leader:2013jra,Lorce:2017wkb,Freese:2022jlu}). The same
does not hold for the gluon part because at the level of the
energy-momentum tensor no anti-symmetric and gauge invariant operator
exists. Therefore, the third and fourth integrals in
Eq.~\eqref{L_sum_rules} are not forced to vanish.

The conservation of the longitudinally polarised second moment instead
implies:
\begin{equation}
  \int_0^1dz\,z\left[\mathcal{P}_{1}^{[L],-,[0]}\left(z,\frac{\xi}{yz}\right)+
 \frac{\xi^2}{y^2} \,
  \mathcal{P}_{2}^{[L],-,[0]}\left(\frac{z\xi}{y},\frac{1}{z}\right)\right]=\mbox{constant
in }\xi\,,
\end{equation}
and indeed we find:
\begin{equation}
  \int_0^1dz\,z\left[\mathcal{P}_{1}^{[L],-,[0]}\left(z,\frac{\xi}{yz}\right)+
 \frac{\xi^2}{y^2} \,
  \mathcal{P}_{2}^{[L],-,[0]}\left(\frac{z\xi}{y},\frac{1}{z}\right)\right]=-\frac{8}{3}C_F\,.
\end{equation}

In the transversely/circularly polarised case, and accounting for the
fact that the $qg$ and $gq$ splitting kernels are identically zero,
the sum rules imply that:
\begin{equation}
  \begin{array}{l}
\displaystyle  \int_0^1dz\left[\mathcal{P}_{1}^{[T],-,[0]}\left(z,\frac{\xi}{yz}\right)+
 \frac{\xi}{y} \,
  \mathcal{P}_{2}^{[T],-,[0]}\left(\frac{z\xi}{y},\frac{1}{z}\right)\right]=\mbox{constant
    in }\xi\,,\\
    \\
    \displaystyle  \int_0^1dz\,z\left[\mathcal{P}_{1,qq}^{[T],+,[0]}\left(z,\frac{\xi}{yz}\right)+
    \frac{\xi^2}{y^2} \,
    \mathcal{P}_{2,qq}^{[T],+,[0]}\left(\frac{z\xi}{y},\frac{1}{z}\right)\right]=\mbox{constant
    in }\xi\,,\\
    \\
    \displaystyle  \int_0^1dz\,z\left[\mathcal{P}_{1,gg}^{[T],+,[0]}\left(z,\frac{\xi}{yz}\right)+
    \frac{\xi^2}{y^2} \,
    \mathcal{P}_{2,gg}^{[T],+,[0]}\left(\frac{z\xi}{y},\frac{1}{z}\right)\right]=\mbox{constant
    in }\xi\,,
  \end{array}
\end{equation}
and indeed we find:
\begin{equation}
  \begin{array}{l}
    \displaystyle  \int_0^1dz\left[\mathcal{P}_{1}^{[T],-,[0]}\left(z,\frac{\xi}{yz}\right)+
    \frac{\xi}{y} \,
    \mathcal{P}_{2}^{[T],-,[0]}\left(\frac{z\xi}{y},\frac{1}{z}\right)\right]=-C_F\,,\\
    \\
    \displaystyle  \int_0^1dz\,z\left[\mathcal{P}_{1,qq}^{[T],+,[0]}\left(z,\frac{\xi}{yz}\right)+
    \frac{\xi^2}{y^2} \,
    \mathcal{P}_{2,qq}^{[T],+,[0]}\left(\frac{z\xi}{y},\frac{1}{z}\right)\right]=-3C_F\,,\\
    \\
    \displaystyle  \int_0^1dz\,z\left[\mathcal{P}_{1,gg}^{[T],+,[0]}\left(z,\frac{\xi}{yz}\right)+
    \frac{\xi^2}{y^2} \,
    \mathcal{P}_{2,gg}^{[T],+,[0]}\left(\frac{z\xi}{y},\frac{1}{z}\right)\right]=-\frac{7C_A+4
    n_f T_R}{3}\,.
  \end{array}
\end{equation}

The fulfilment of the sum rules provides a strong check of the
correctness of the evolution kernels derived here.

\subsection{Conservation of polynomiality}\label{subsec:PolynomCons}

In this section, we prove analytically that GPD polynomiality is
conserved by the evolution. The proof presented below is limited to
the unpolarised non-singlet quark GPD, but a similar demonstration can
be given also for the other cases. The polynomiality property reads:
\begin{equation}
\int_{-1}^1dx\,x^{2n}F^{[U],-}(x,\xi,t;\mu) = \sum_{k=0}^{n}A_{k}^{(n)}(t,\mu)\xi^{2k}\,.
\end{equation}
Using Eq.~(\ref{eq:ERBLlikeeveq}), one finds:
\begin{equation}
\sum_{k=0}^{n}\frac{\partial A_{k}^{(n)}(t,\mu)}{\partial\ln\mu^2} \xi^{2k}=
\frac{\alpha_s(\mu)}{4\pi}\int_{-1}^1dy\left[\int_{-1}^1 \frac{dx}{|\xi|} \,x^{2n}\mathbb{V}^{[U],-,[0]}\left(\frac{x}{\xi},\frac{y}{\xi}\right)\right]F^{[U],-}(y,\xi,t;\mu)\,.
\end{equation}
In order for this equality to be fulfilled, the following identity has
to hold:
\begin{equation}
\int_{-1}^1 \frac{dx}{|\xi|}\,x^{2n}\mathbb{V}^{[U],-,[0]}\left(\frac{x}{\xi},\frac{y}{\xi}\right)
= \mathcal{V}_n^{[U],-,[0]}y^{2n}\,,
\label{eq:PolynomialityCondition}
\end{equation}
where $\mathcal{V}_n^{[U],-,[0]}$ is a constant to be evaluated. The
integral in the l.h.s. of the equation above can be computed using the
results in Appendix~C of Ref.~\cite{Bertone:2022frx}, thus proving
that the equality in Eq.~(\ref{eq:PolynomialityCondition}) is indeed
true with:
\begin{equation}
\mathcal{V}_n^{[U],-,[0]} = 2C_F\left[\frac32+\frac{1}{(n+1)(n+2)}-2\sum_{k=1}^{n+1}\frac{1}{k}\right]\,.
\end{equation}
Interestingly, this allows us to derive evolution equations for the
coefficients $A_k ^{(n)}$ that read:
\begin{equation}
\frac{\partial A_{k}^{(n)} (t,\mu)}{\partial\ln\mu^2} = \frac{\alpha_s(\mu)}{4\pi}\mathcal{V}_n^{[U],-,[0]}A_{k}^{(n)}(t,\mu)\,,
\end{equation}
and that admit the solution:
\begin{equation}
A_{k}^{(n)} (t,\mu) =
\exp\left[\frac{\mathcal{V}_n^{[U],-,[0]}}{4\pi}\int_{\mu_0}^{\mu}d\ln\mu'^2\alpha_s(\mu')\right]A_{k}^{(n)} (t,\mu_0)\,.
\end{equation}

\section{Numerical results}\label{sec:NumResults}

Having presented the expression for the evolution kernels in
Sect.~\ref{sec:Results}, we are now in a position to implement them in
the numerical code {\tt APFEL++}~\cite{Bertone:2013vaa,
  Bertone:2017gds}.

To showcase the effect of the evolution, we have used as a set of
initial-scale GPDs the realistic model of
Refs.~\cite{Goloskokov:2005sd,Goloskokov:2007nt,Goloskokov:2009ia},
referred to as Goloskokov-Kroll (GK) model, as implemented in {\tt
  PARTONS}~\cite{Berthou:2015oaw}. For the unpolarised evolution we
selected the GPD $H$, in the longitudinally polarised case we instead
used $\widetilde{H}$, while in the transversely/circularly polarised
case we used $H_T$. Since the GK model does not provide a circularly
polarised gluon GPD, we used the unpolarised $H_g$ as a proxy to test
the evolution. Since the evolution of the circularly polarised gluon
GPD is completely decoupled, no spurious effects are introduced in the
evolution of the transversely polarised quark distributions that also
evolve independently. GPDs are evolved from $\mu_0=2$~GeV to
$\mu=10$~GeV in the variable-flavour-number scheme (VFNS),
\textit{i.e.} allowing for heavy-flavour threshold crossing, with
charm and bottom thresholds set to $m_c=2.1$~GeV and $m_b=4.75$~GeV,
respectively.\footnote{The unusually large value of the charm
  threshold is due to the fact that the lowest available scale
  accessible to the GK model is
  $\mu_0=2$~GeV~\cite{Goloskokov:2005sd,Goloskokov:2007nt,Goloskokov:2009ia}. However,
  at this scale, no distribution associated with the charm quark is
  provided. Therefore, we assumed $n_f=3$ active flavours at $\mu_0$
  which required setting $m_c>\mu_0$.} The strong coupling is
consistently evolved at LO in the VFNS using $\alpha_s(M_Z)=0.118$ as
a boundary condition. We set the value of the momentum transfer
squared, that does not directly participate in the evolution, to
$t=-0.1$~GeV$^2$ throughout.

\begin{figure}[h]
\centering
\subfloat[a][]{\includegraphics[width = 0.33\textwidth]{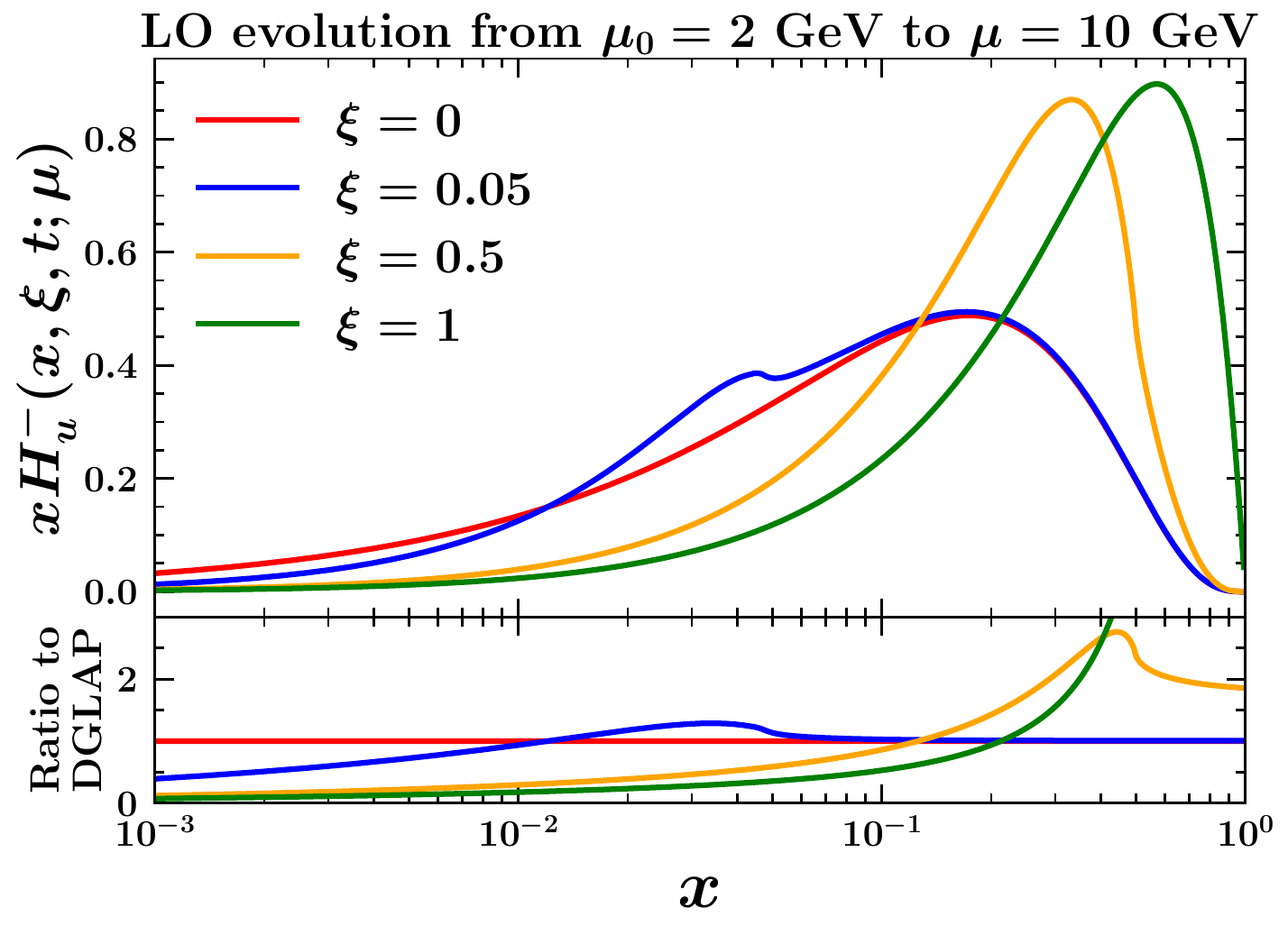}\label{GPDEvolutionUnpNonSinglet}}
\subfloat[b][]{\includegraphics[width = 0.33\textwidth]{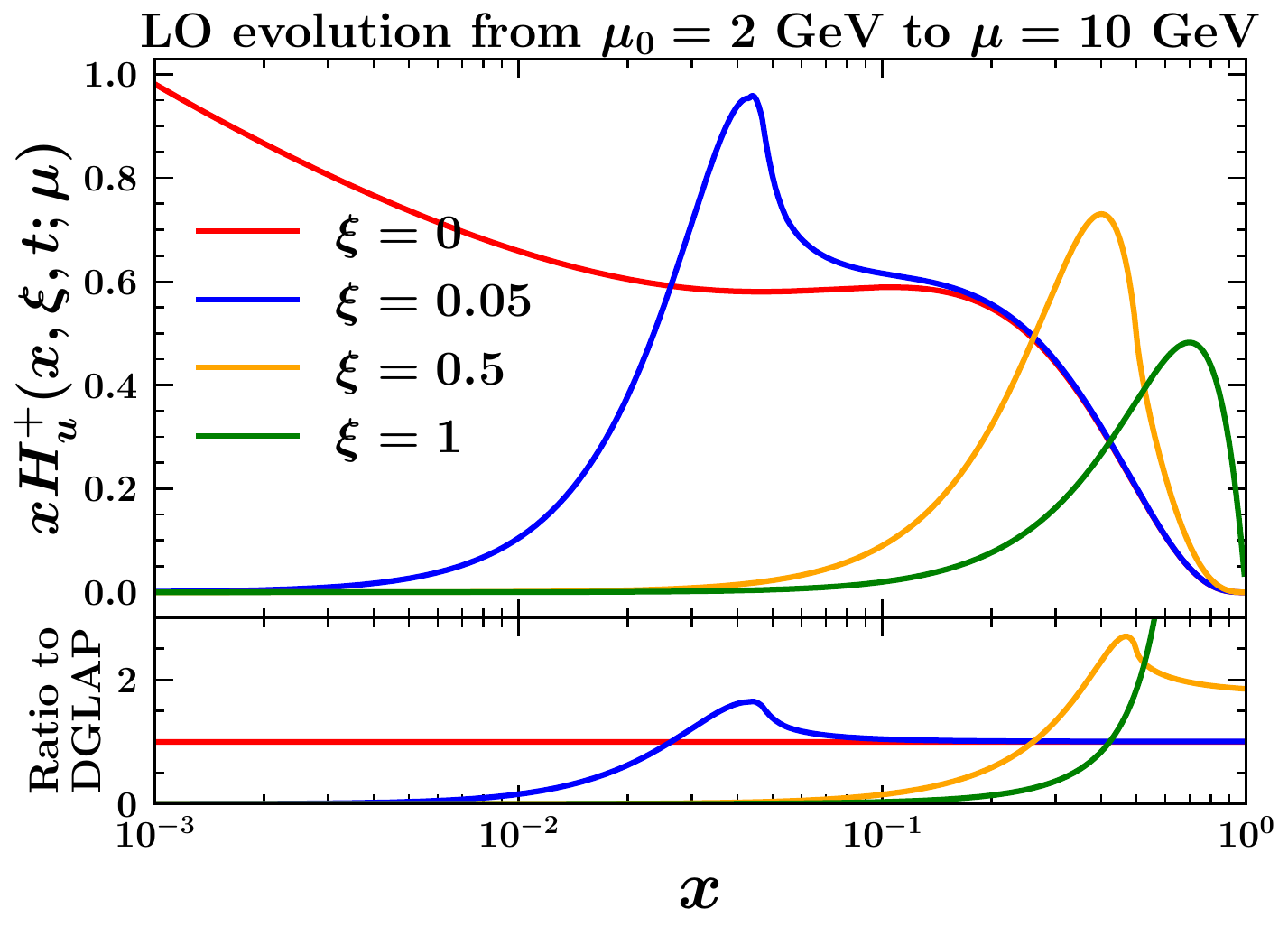}\label{GPDEvolutionUnpSinglet}}
\subfloat[c][]{\includegraphics[width = 0.33\textwidth]{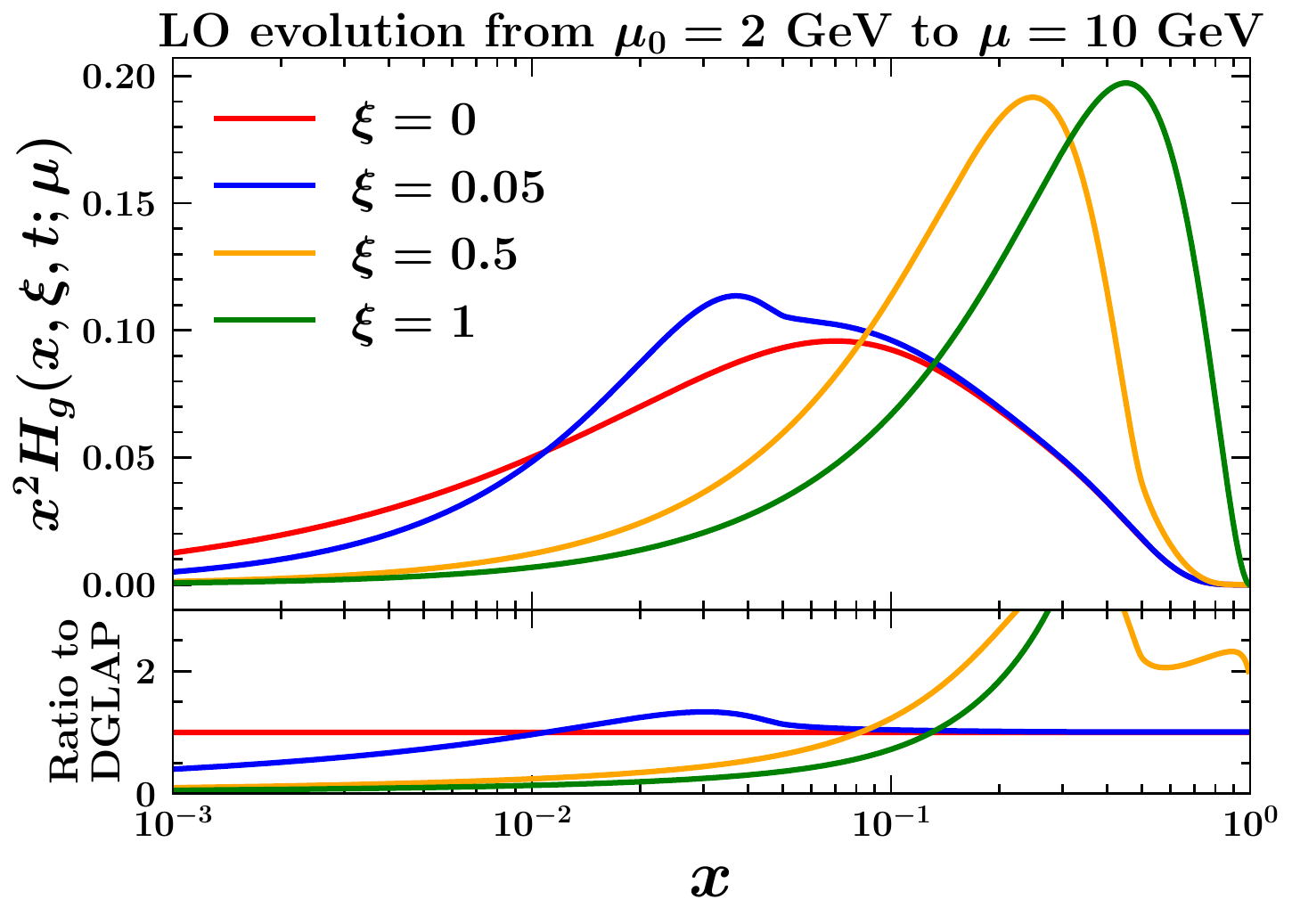}\label{GPDEvolutionUnpGluon}}
\caption{Evolution from $\mu_0=2$~GeV to $\mu = 10$~GeV for the
  unpolarised GPDs $H$. Panel~(\ref{GPDEvolutionUnpNonSinglet}) shows
  the non-singlet up-quark GPD; panel~(\ref{GPDEvolutionUnpSinglet})
  shows the singlet up-quark GPD, and~(\ref{GPDEvolutionUnpGluon})
  shows the gluon GPD.}
\label{GPDEvolutionUnp}
\end{figure}
\begin{figure}[h]
\centering
\subfloat[a][]{\includegraphics[width = 0.33\textwidth]{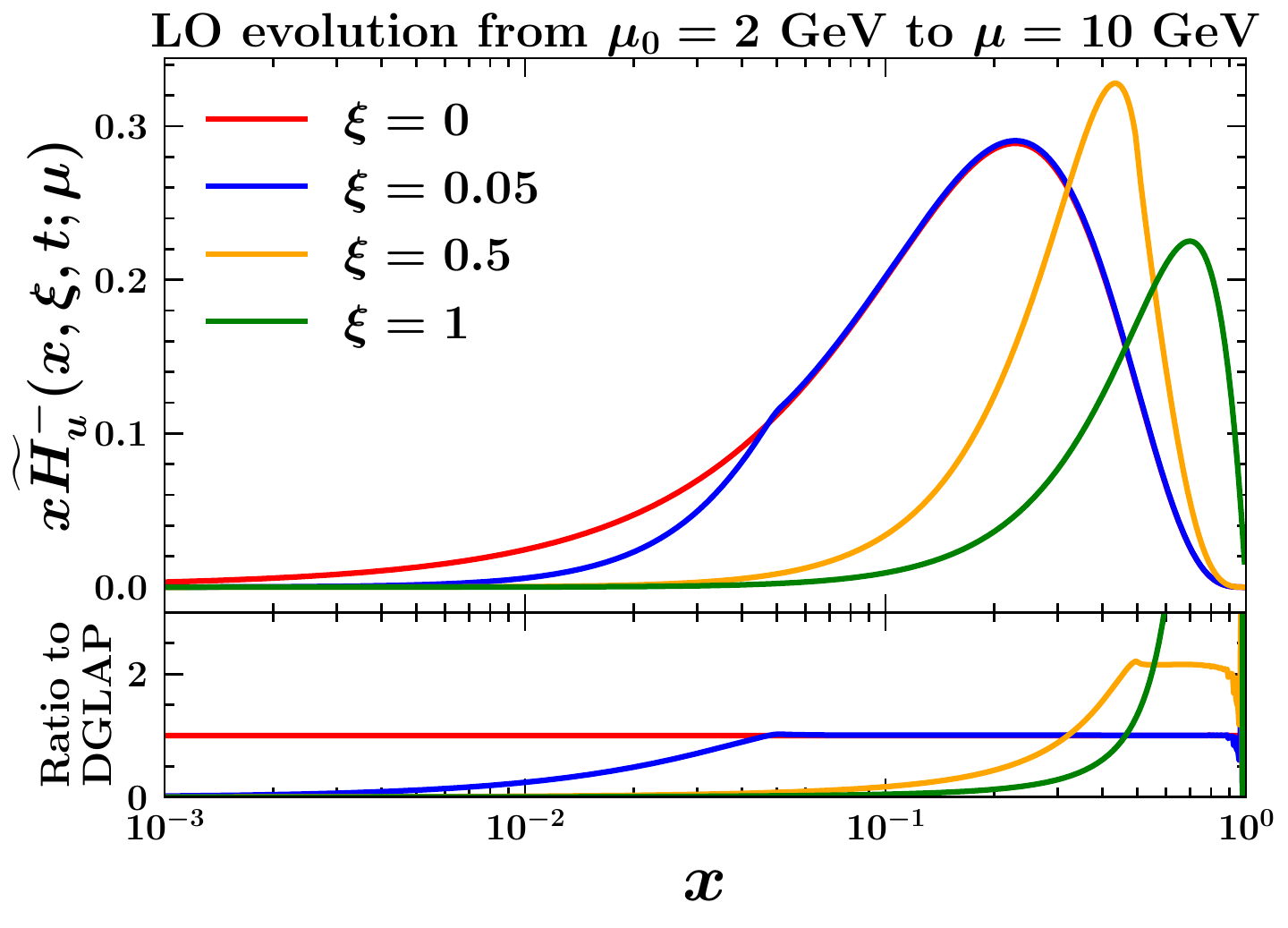}\label{GPDEvolutionPolNonSinglet}}
\subfloat[b][]{\includegraphics[width = 0.33\textwidth]{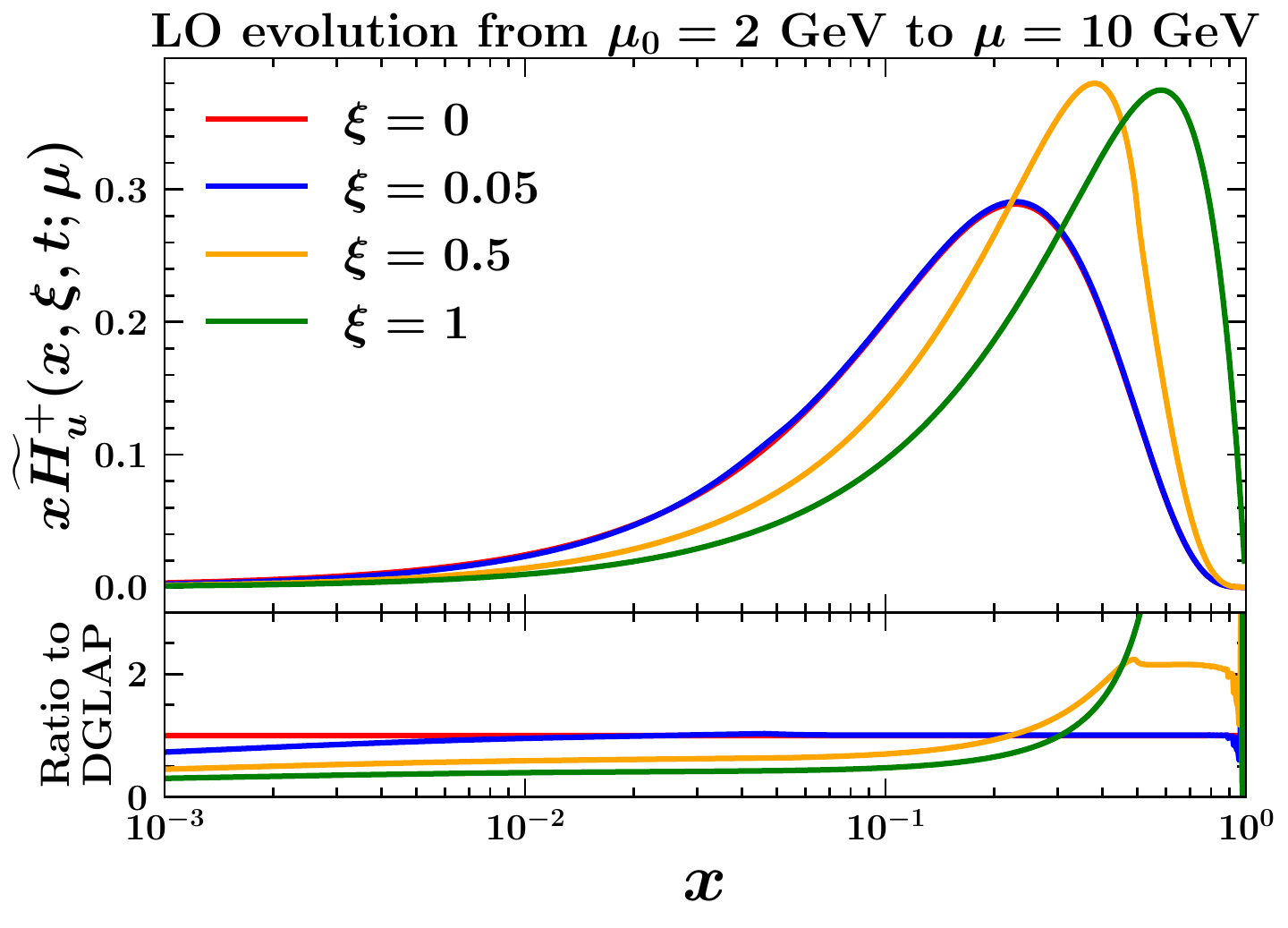}\label{GPDEvolutionPolSinglet}}
\subfloat[c][]{\includegraphics[width = 0.33\textwidth]{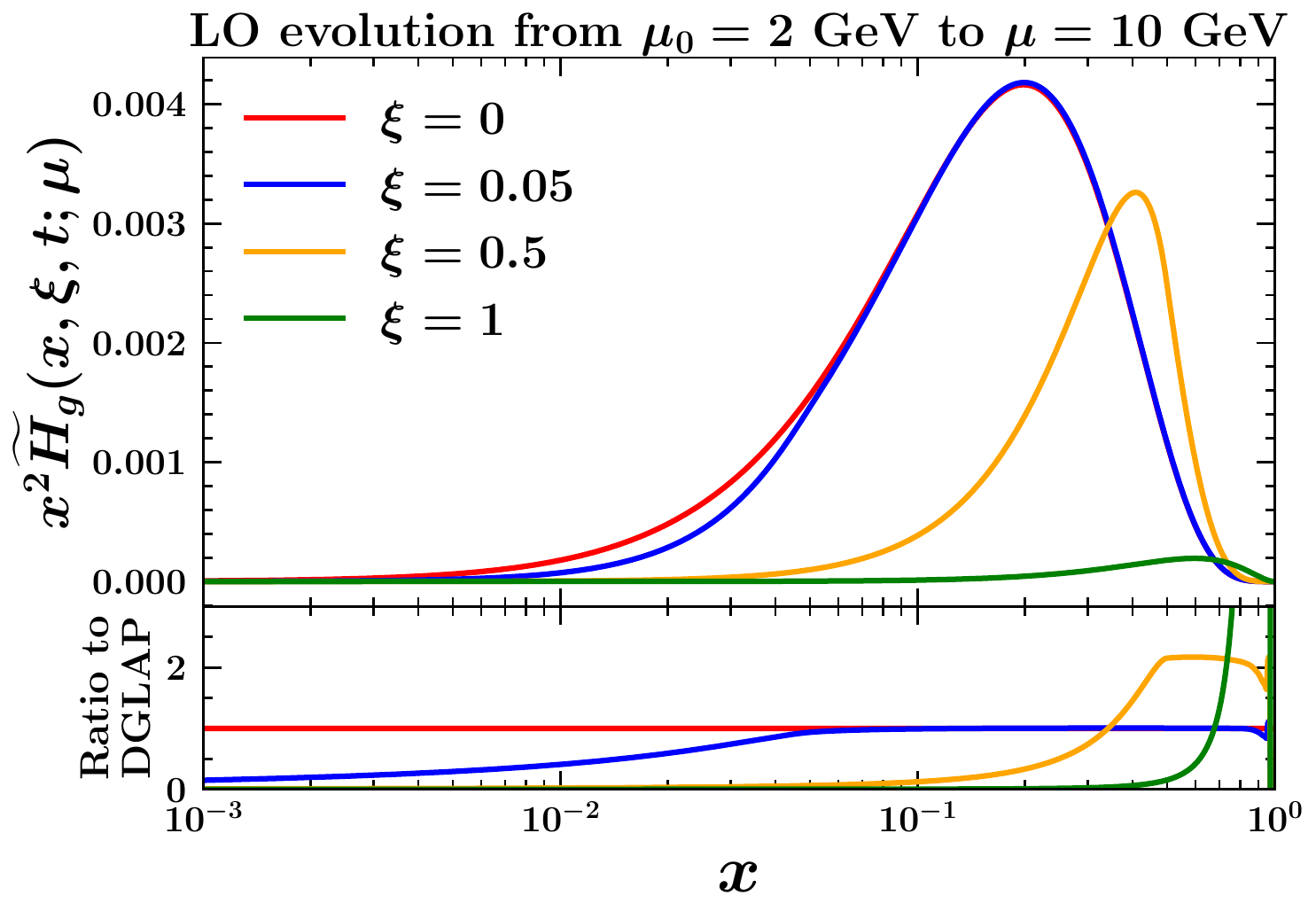}\label{GPDEvolutionPolGluon}}
\caption{Same as Fig.~\ref{GPDEvolutionUnp} but for longitudinally
  polarised GPDs $\widetilde{H}$.}
\label{GPDEvolutionPol}
\end{figure}
\begin{figure}[h]
\centering
\subfloat[a][]{\includegraphics[width = 0.33\textwidth]{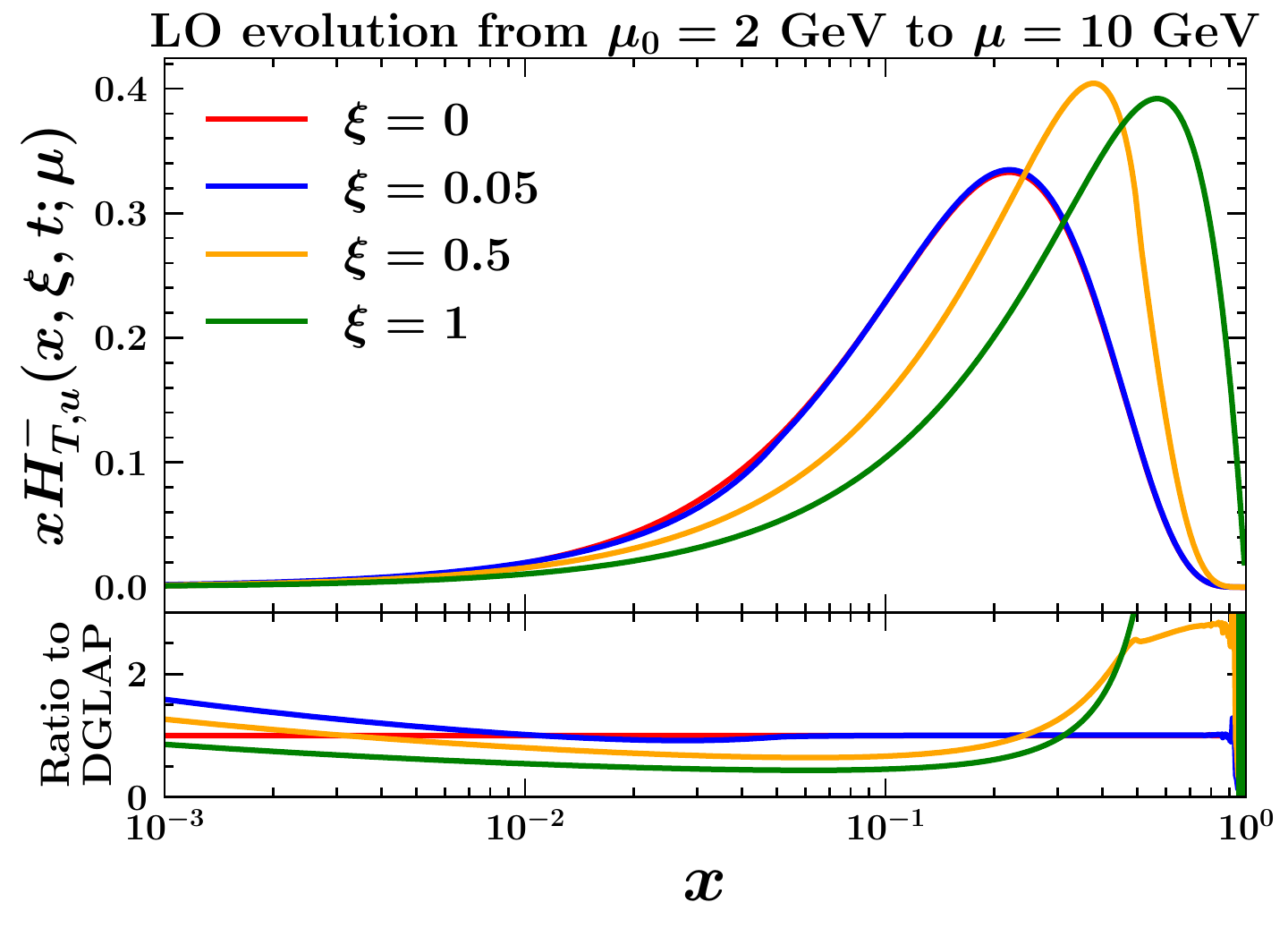}\label{GPDEvolutionTransNonSinglet}}
\subfloat[b][]{\includegraphics[width = 0.33\textwidth]{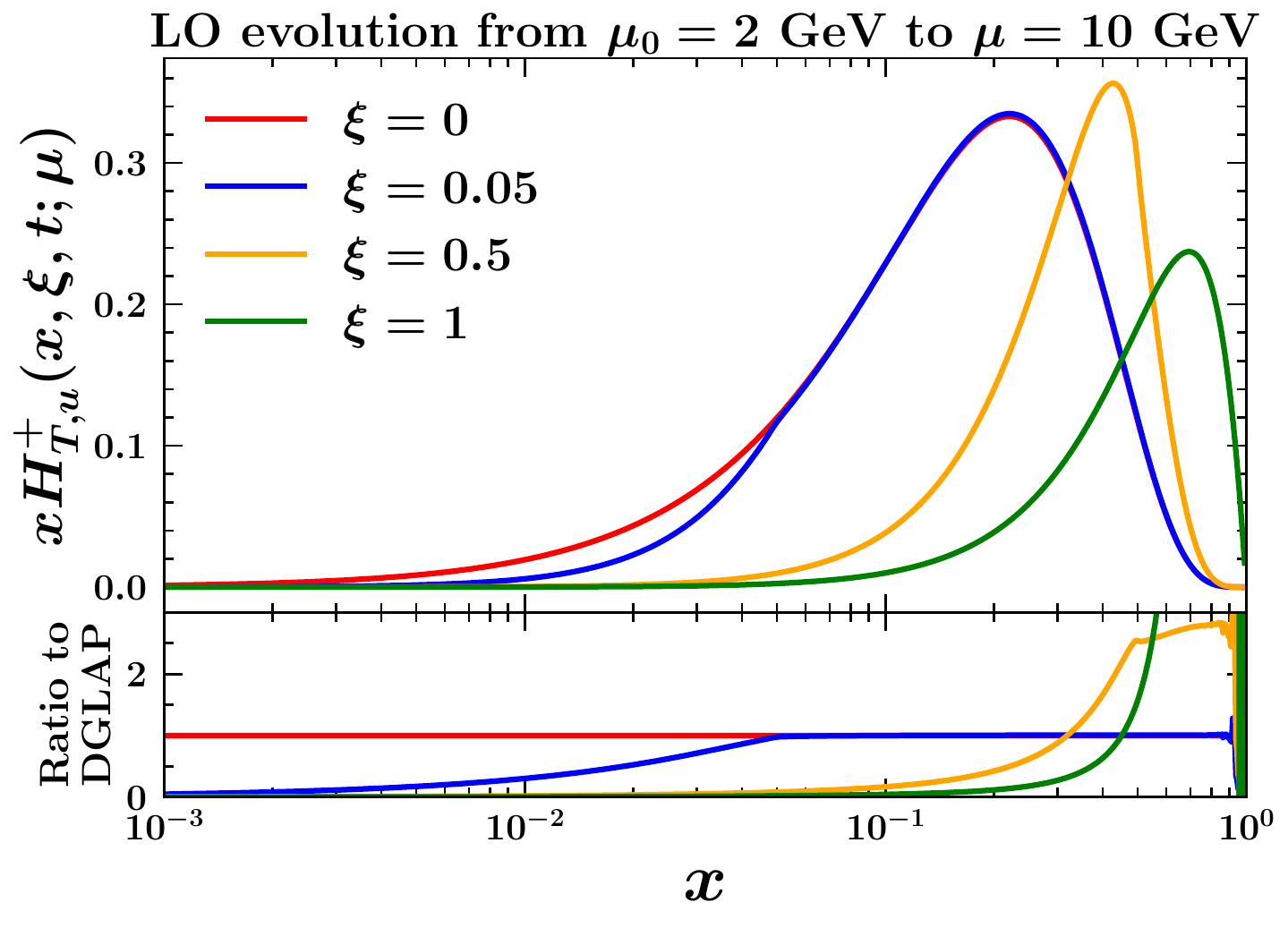}\label{GPDEvolutionTransSinglet}}
\subfloat[c][]{\includegraphics[width = 0.33\textwidth]{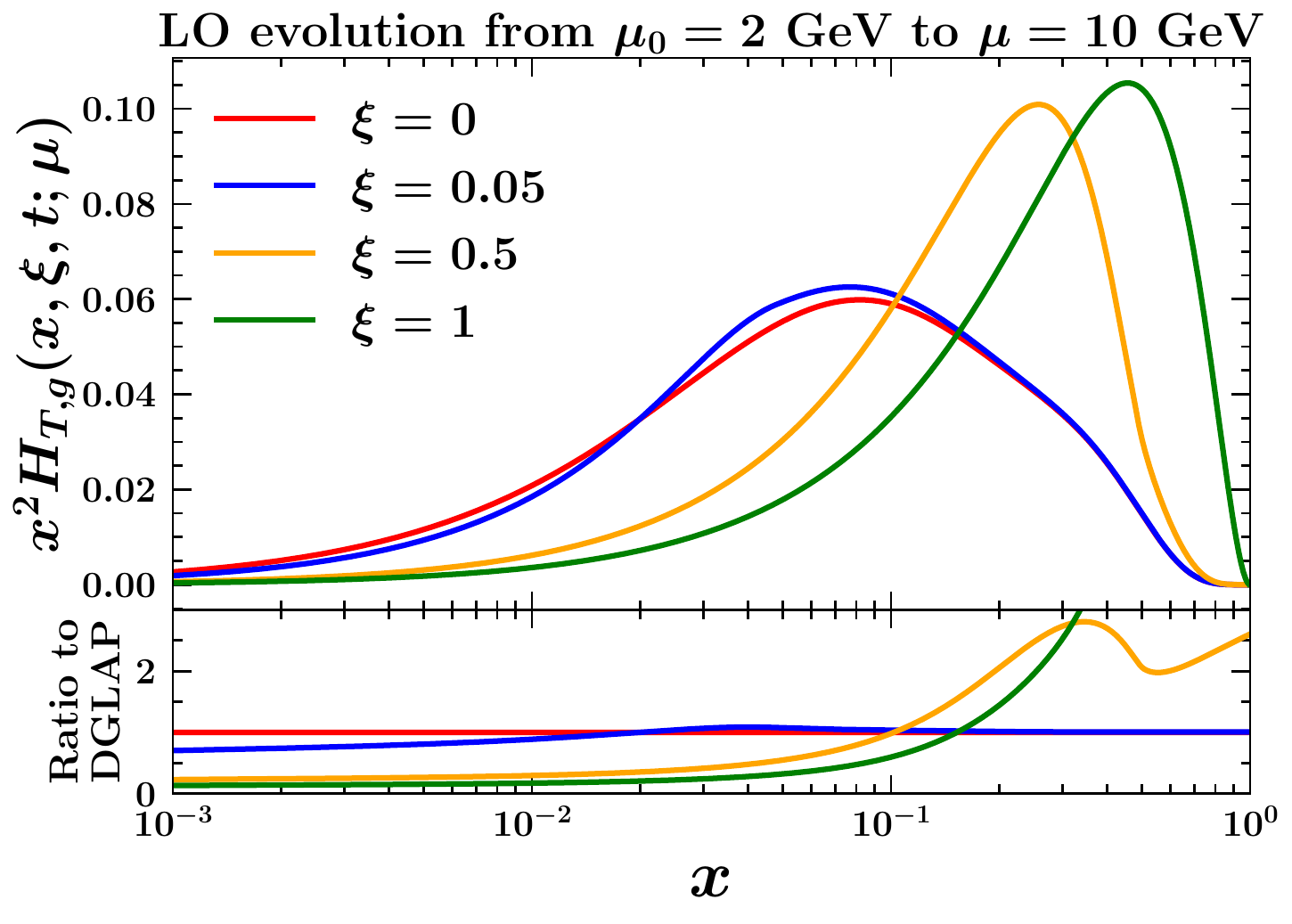}\label{GPDEvolutionTransGluon}}
\caption{Same as Fig.~\ref{GPDEvolutionUnp} but for
  transversely/circularly polarised GPDs $H_T$.}
\label{GPDEvolutionTrans}
\end{figure}
Figs.~\ref{GPDEvolutionUnp},~\ref{GPDEvolutionPol},
and~\ref{GPDEvolutionTrans} show the effect on GPDs of unpolarised,
longitudinally polarised, and transversely/circularly polarised
evolutions, respectively. GPDs are displayed as functions of $x$ for
four different values of $\xi$, including the DGLAP ($\xi=0$) and ERBL
($\xi=1$) limits. The upper panels display the absolute distributions
at the final scale $\mu=10$~GeV multiplied by a factor of $x$ for the
quark GPDs and $x^2$ for the gluon GPDs for a better
visualisation. The lower panels display their ratio to the
corresponding distributions evolved using the DGLAP equations.

We first note that for all three cases, setting $\xi=0$ exactly
reproduces the DGLAP evolution, as expected. For increasing values of
$\xi$, the evolution gradually deviates from DGLAP for all considered
distributions. The deviations are particularly pronounced for
$x \lesssim \xi$ where GPD evolution causes a strong slowdown of the
evolution as compared to DGLAP. We also observe that GPDs at the
crossover point $x=\xi$ are continuous, as expected from the
discussion in Sect.~\ref{subsec:continuity}. However, they develop a
cusp (discontinuity of the derivative in $x$) that is a consequence of
the fact that the evolution kernels are continuous but not smooth at
$x=\xi$.

\begin{figure}[h]
\centering
\subfloat[][]{\includegraphics[width = 0.45\textwidth]{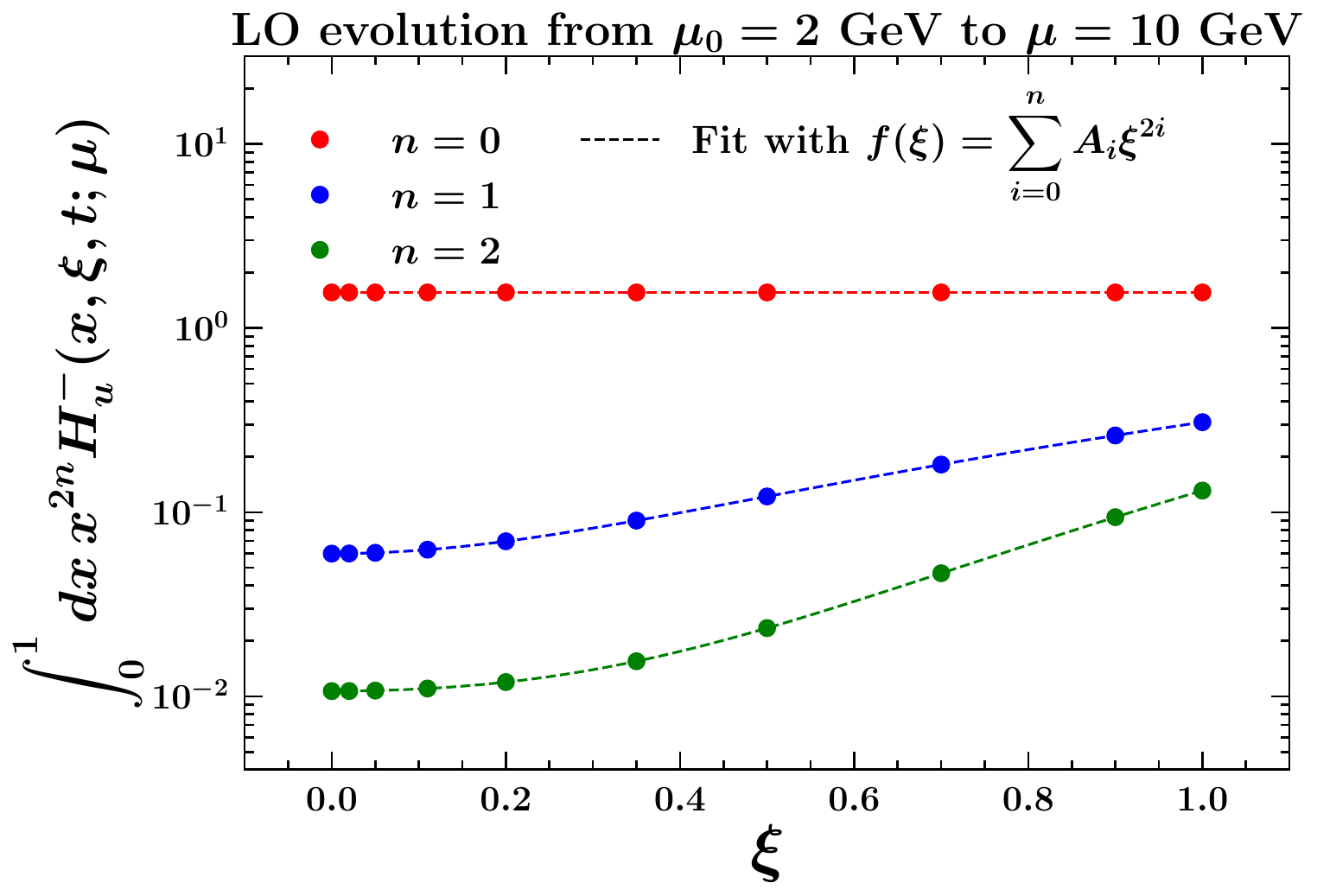}\label{GPDMomentsUnpMinus}}
\subfloat[][]{\includegraphics[width = 0.45\textwidth]{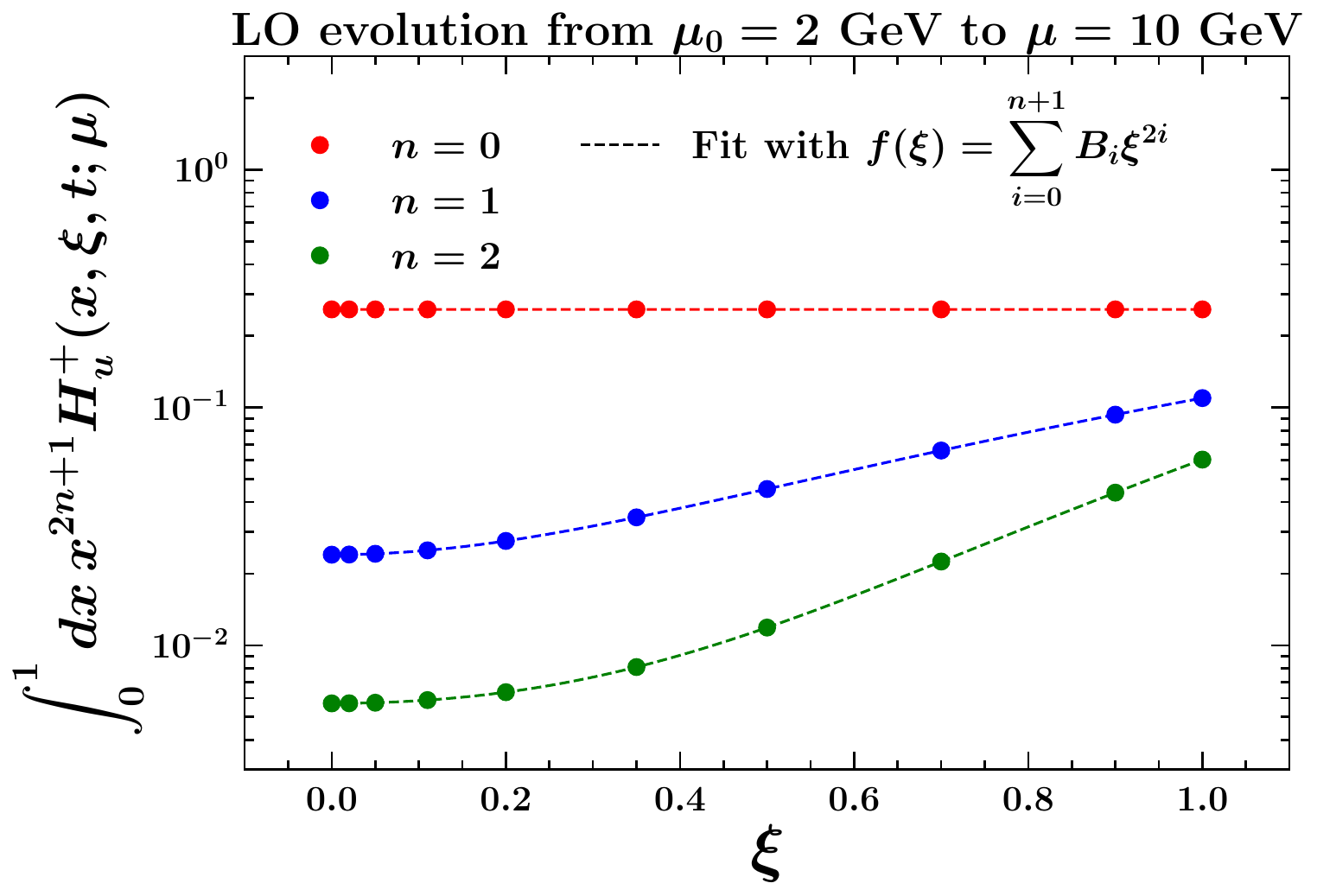}\label{GPDMomentsUnpPlus}}
\caption{Effect of the evolution on the Mellin moments of the
  unpolarised up-quark GPD $H$. The bullets display the value of the
  moments computed numerically as integrals of the distributions,
  whereas the dashed lines show the fits to the bullets using the
  expected polynomial law. The left panel~(\ref{GPDMomentsUnpMinus})
  displays the first three even moments related to the non-singlet
  combination, while the right panel~(\ref{GPDMomentsUnpPlus})
  displays the first three odd moments that are instead related to the
  singlet combination.}
\label{GPDMomentsUnp}
\end{figure}

\begin{figure}[h]
\centering
\subfloat[][]{\includegraphics[width = 0.45\textwidth]{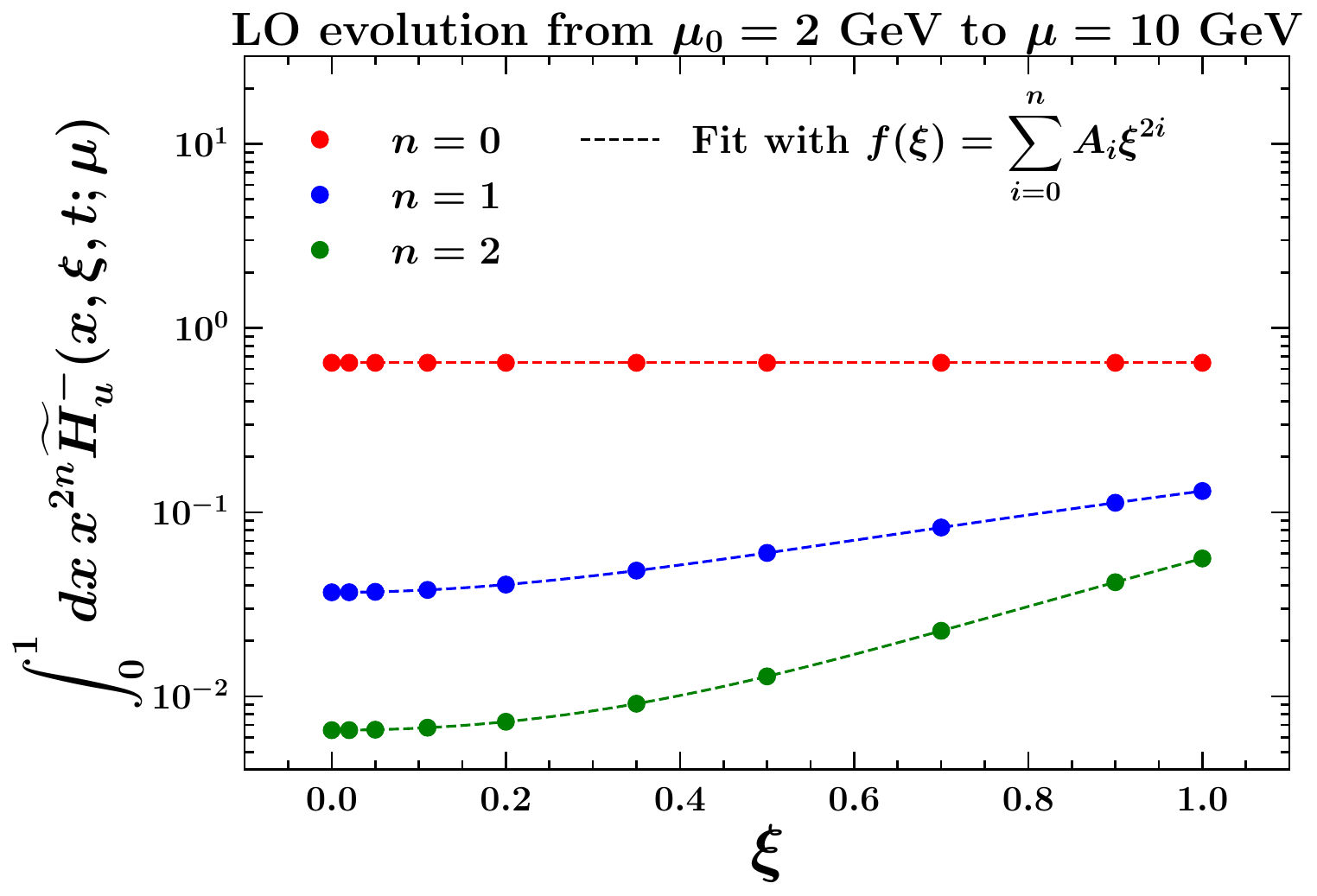}\label{GPDMomentsPolMinus}}
\subfloat[][]{\includegraphics[width = 0.45\textwidth]{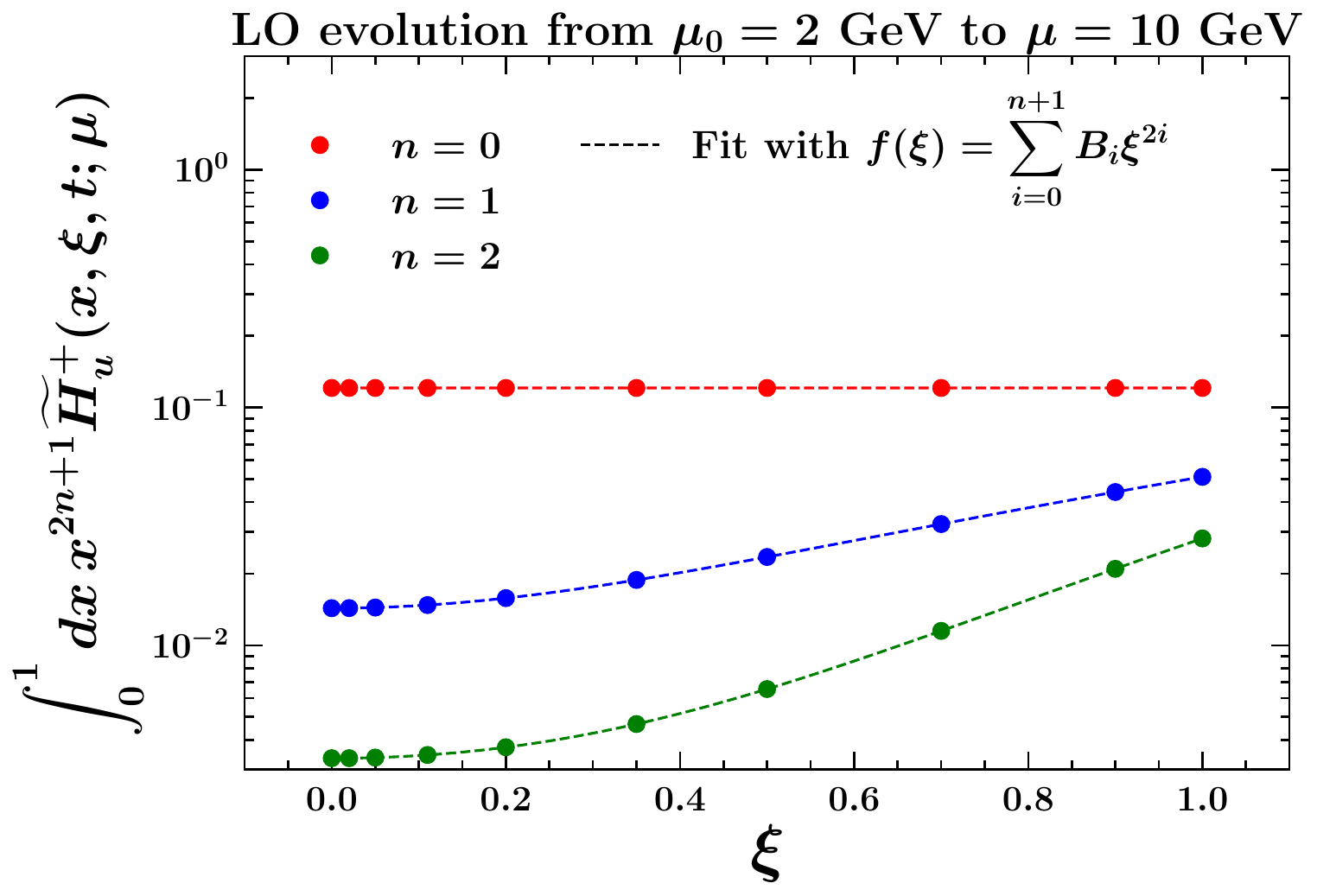}\label{GPDMomentsPolPlus}}
\caption{Same as Fig.~\ref{GPDMomentsUnp} but for longitudinally
  polarised GPDs.}
\label{GPDMomentsLong}
\end{figure}

\begin{figure}[h]
\centering
\subfloat[][]{\includegraphics[width = 0.45\textwidth]{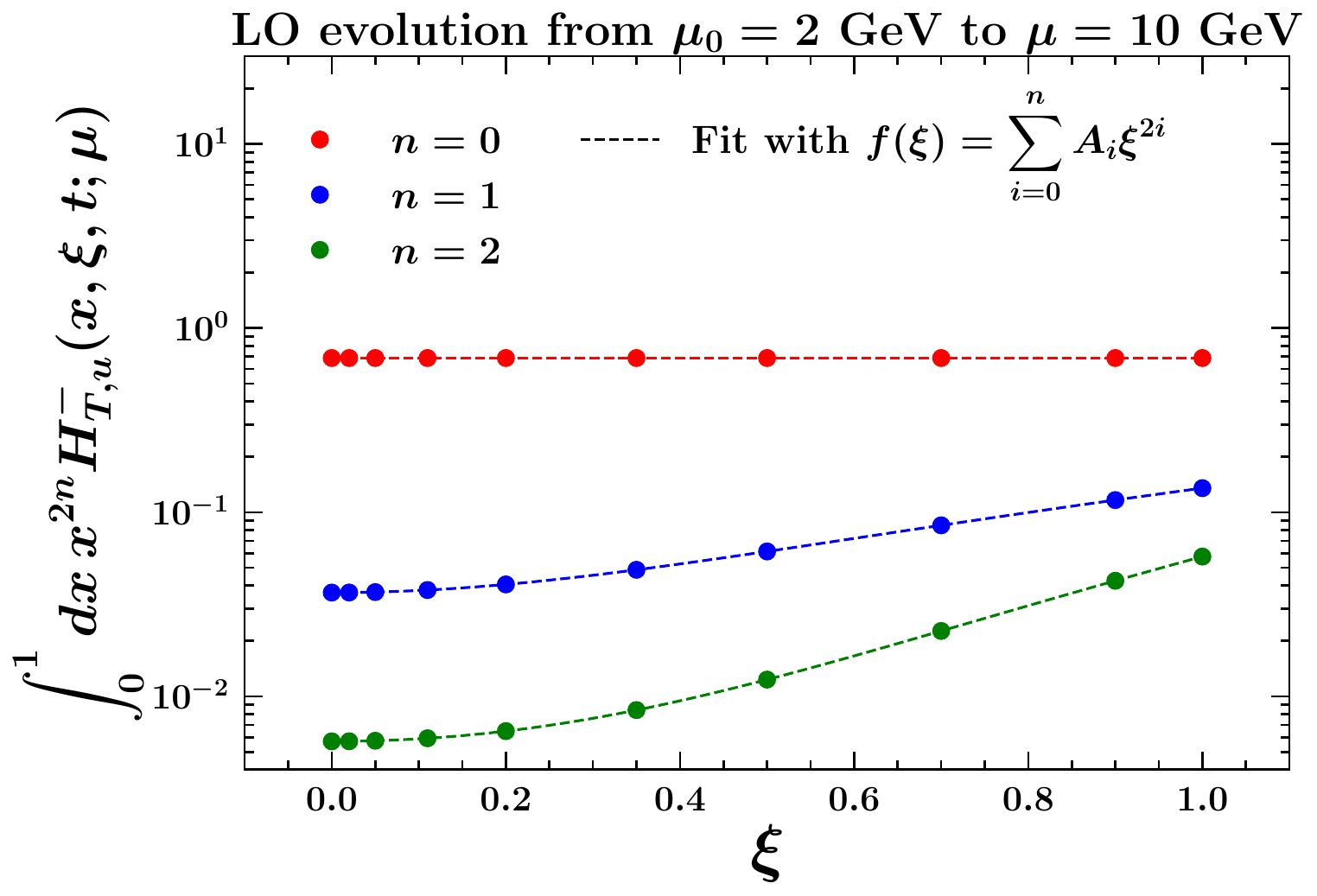}\label{GPDMomentsTransMinus}}
\subfloat[][]{\includegraphics[width = 0.45\textwidth]{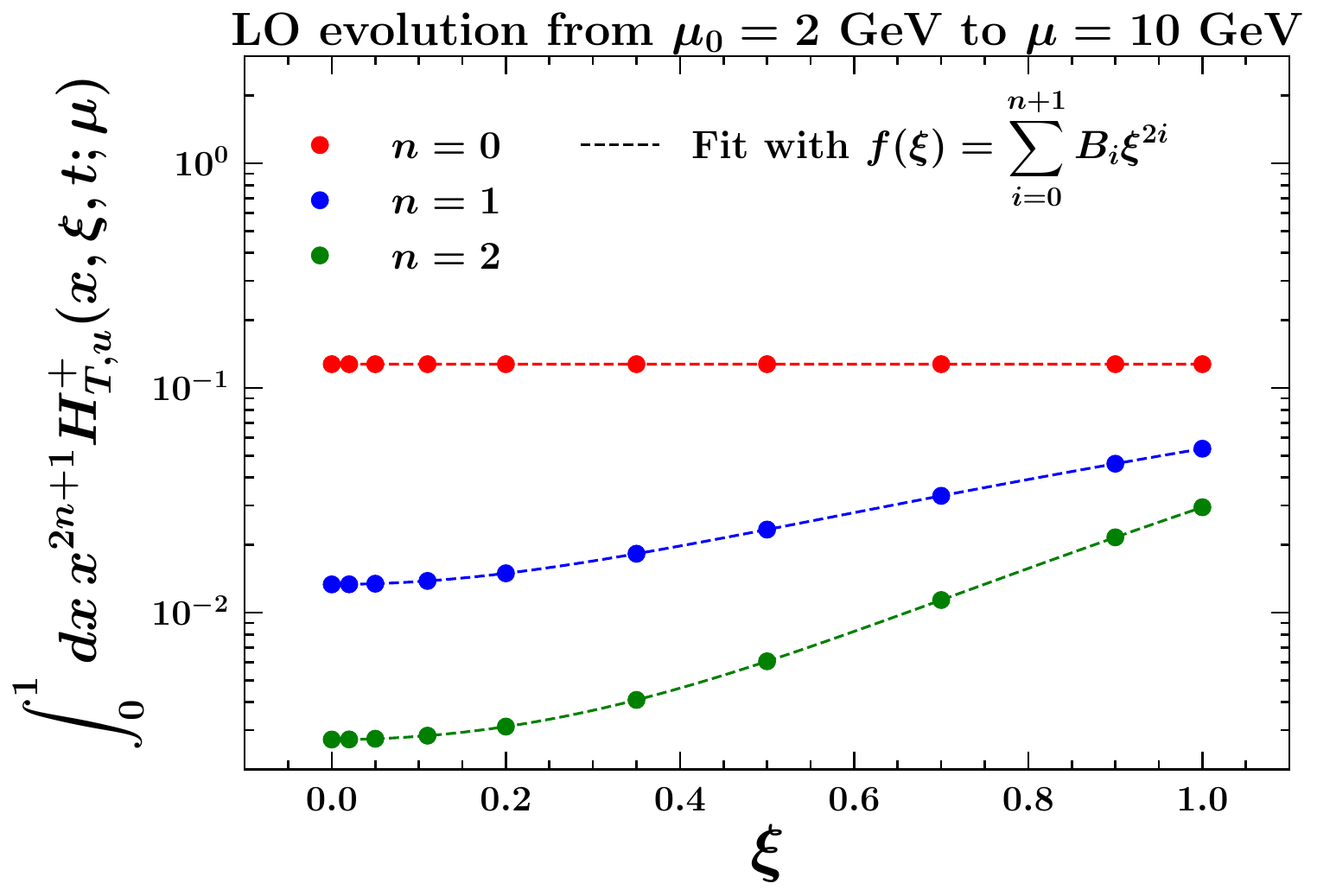}\label{GPDMomentsTransPlus}}
\caption{Same as Fig.~\ref{GPDMomentsUnp} but for
  transversely/circularly polarised GPDs.}
\label{GPDMomentsTrans}
\end{figure}
As discussed in Sect.~\ref{subsec:PolynomCons}, a crucial property of
GPDs, that must be preserved by the evolution, is
polynomiality. Figs.~\ref{GPDMomentsUnp},~\ref{GPDMomentsLong},
and~\ref{GPDMomentsTrans} show the behaviour as functions of $\xi$ of
the first three even (left plots) and odd (right plots) moments of the
up-quark distributions $H$, $\widetilde{H}$, and $H_T$,
respectively. The bullets correspond to the values obtained by
integrating numerically the evolved GPDs for different values of
$\xi$, while the dashed lines show the fits using the expected
polynomial laws in $\xi$. It is clear that in all cases, the expected
behaviour is accurately reproduced. It is also interesting to observe
that both even and odd first moments ($n=0$) for all polarisations are
constant in $\xi$. In fact, this is the expected behaviour in all
cases, except for the unpolarised even moment that would in principle
admit a quadratic term in $\xi$. However, this contribution, often
referred to as $D$-term, evolves independently from the rest of the
GPD. Since the GK model does not include any $D$-term, the evolution
does not generate it and it is thus absent at all scales, finally
producing a constant first even moment also for the unpolarised GPD
$H$.

\begin{figure}
\centering
\subfloat[][]{\includegraphics[width=0.33 \textwidth]{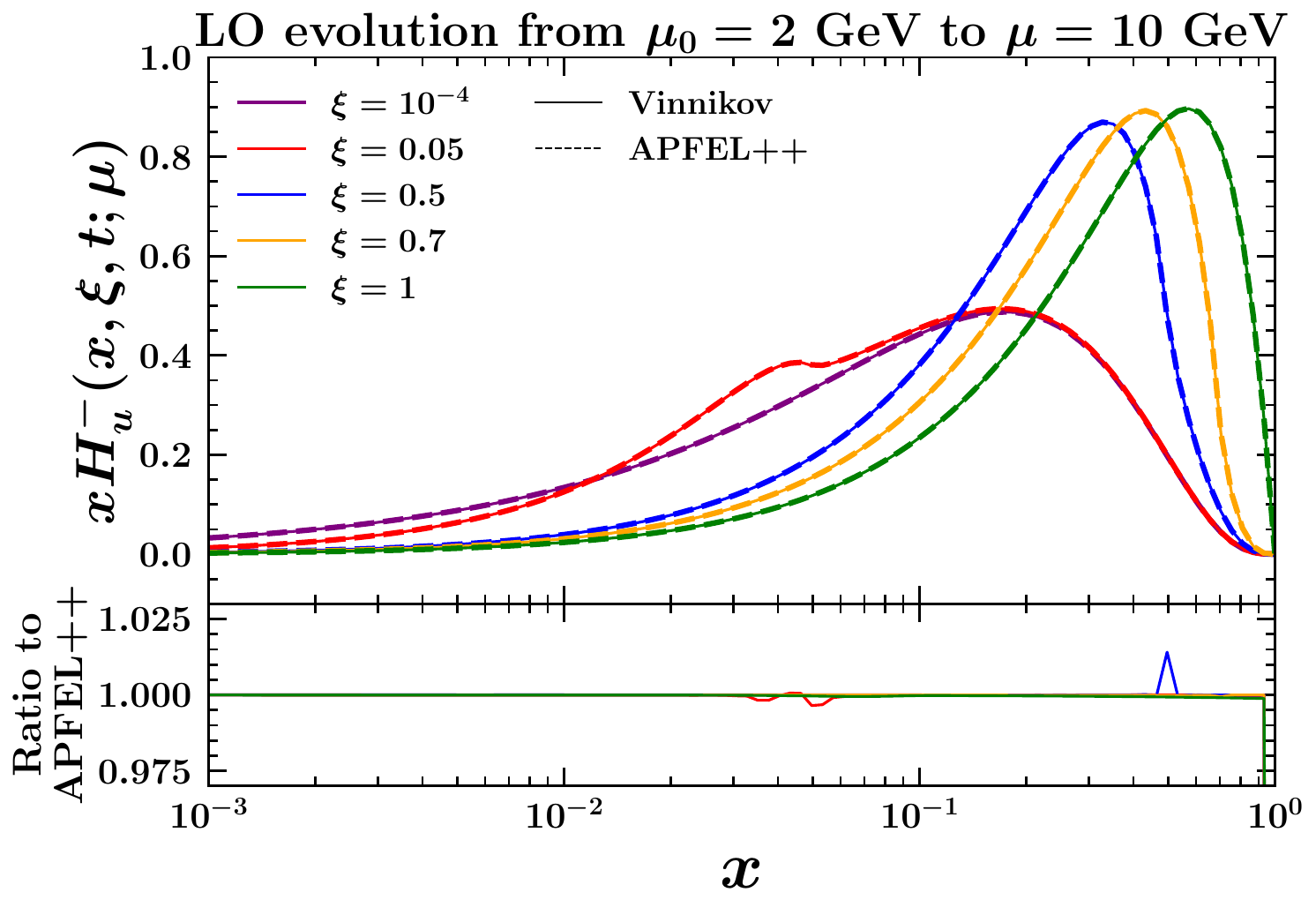}\label{ApfelVsVinnikovUnpMinus}}
\subfloat[][]{\includegraphics[width=0.33 \textwidth]{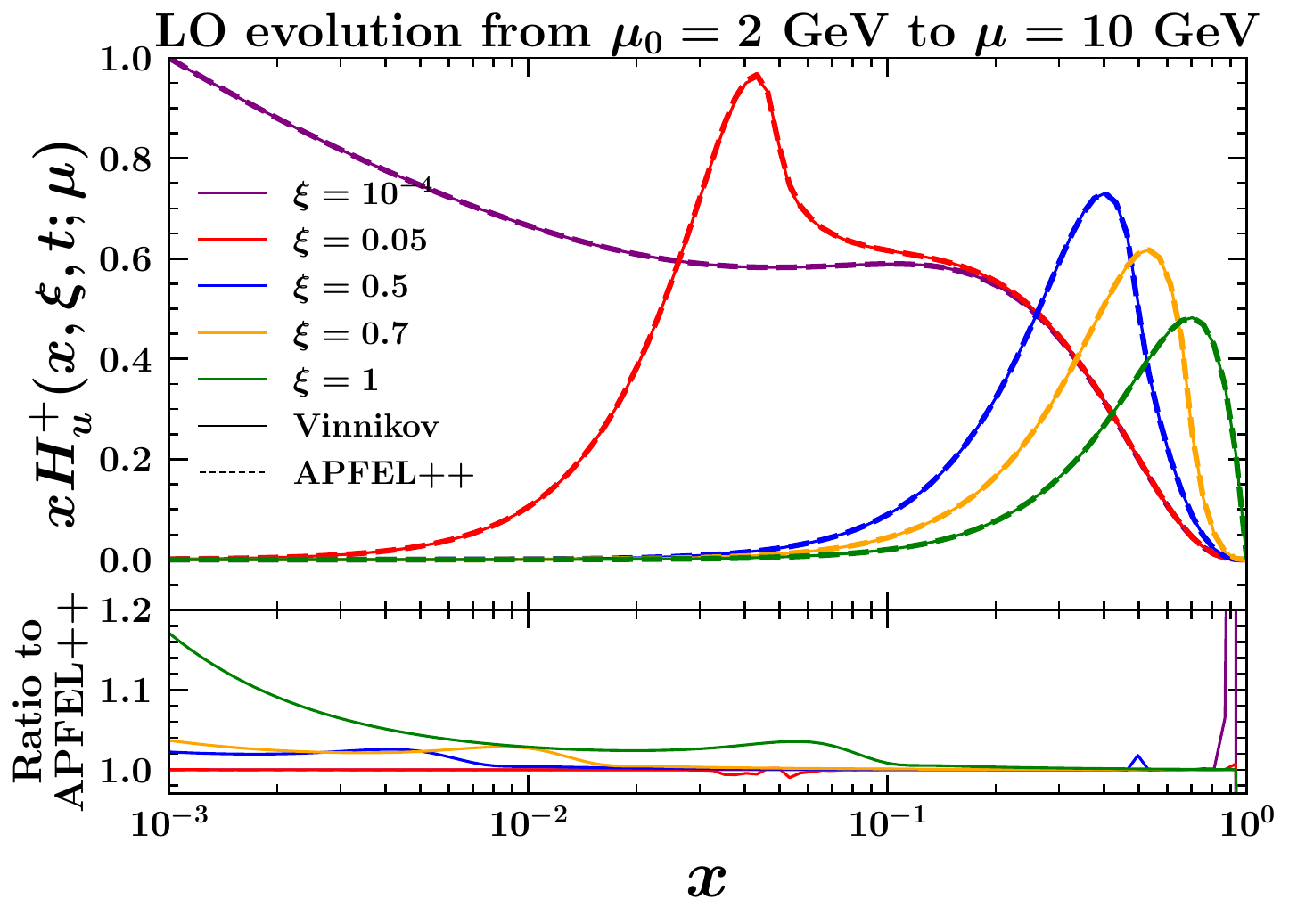}\label{ApfelVsVinnikovUnpPlus}}
\subfloat[][]{\includegraphics[width=0.33 \textwidth]{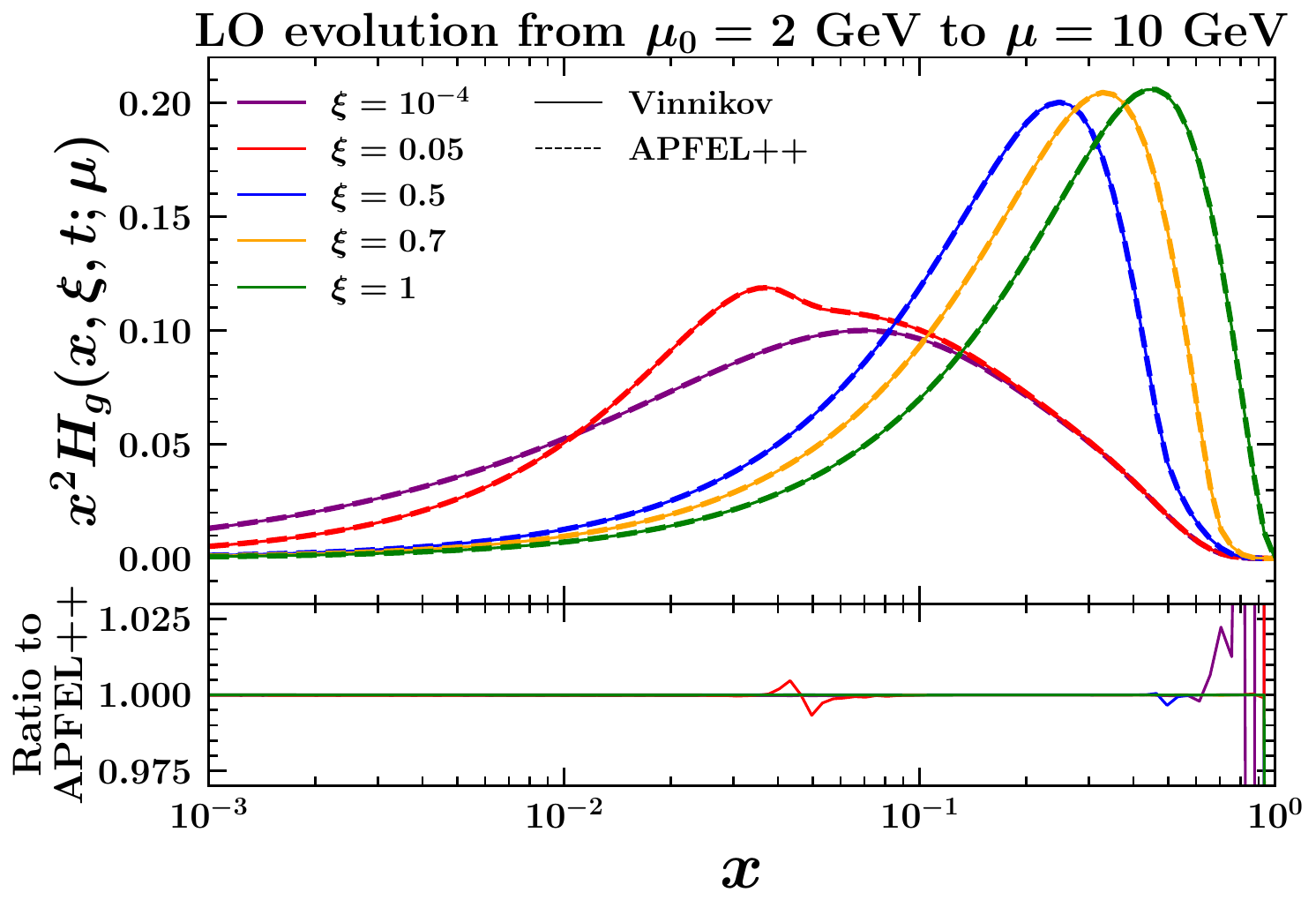}\label{ApfelVsVinnikovUnpGluon}
}
\caption{Comparison of the unpolarised evolution performed using the
  code of Ref.~\cite{Vinnikov:2006xw} (Vinnikov) and {\tt
    APFEL++}. The panels (\ref{ApfelVsVinnikovUnpMinus}),
  (\ref{ApfelVsVinnikovUnpPlus}), and (\ref{ApfelVsVinnikovUnpGluon})
  correspond to the non-singlet up-quark, singlet up-quark, and gluon
  GPDs, respectively.}
\label{ApfelVsVinnikovUnp}
\end{figure}

\begin{figure}
\centering
\subfloat[][]{\includegraphics[width=0.33 \textwidth]{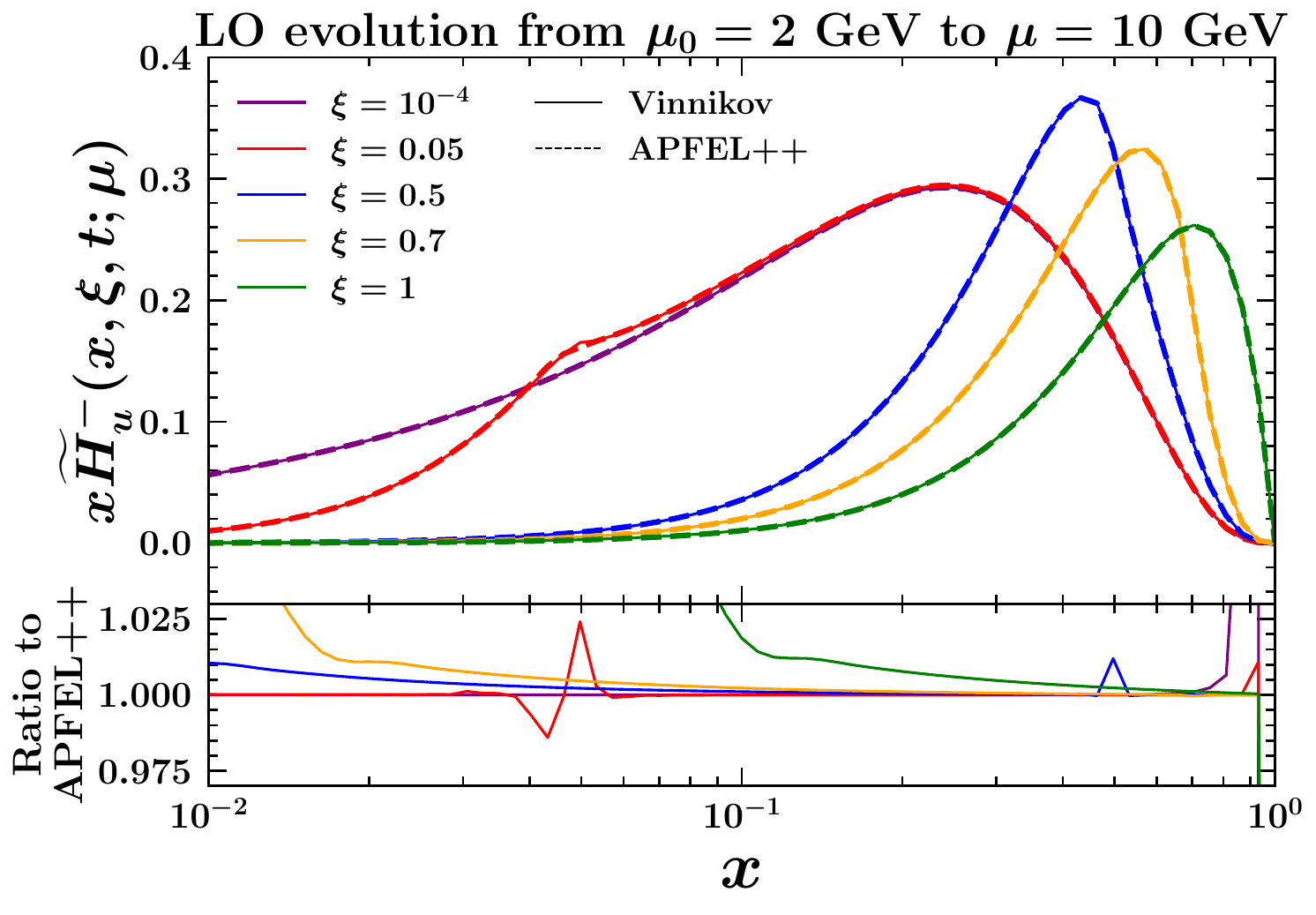}\label{ApfelVsVinnikovPolMinus}}
\subfloat[][]{\includegraphics[width=0.33 \textwidth]{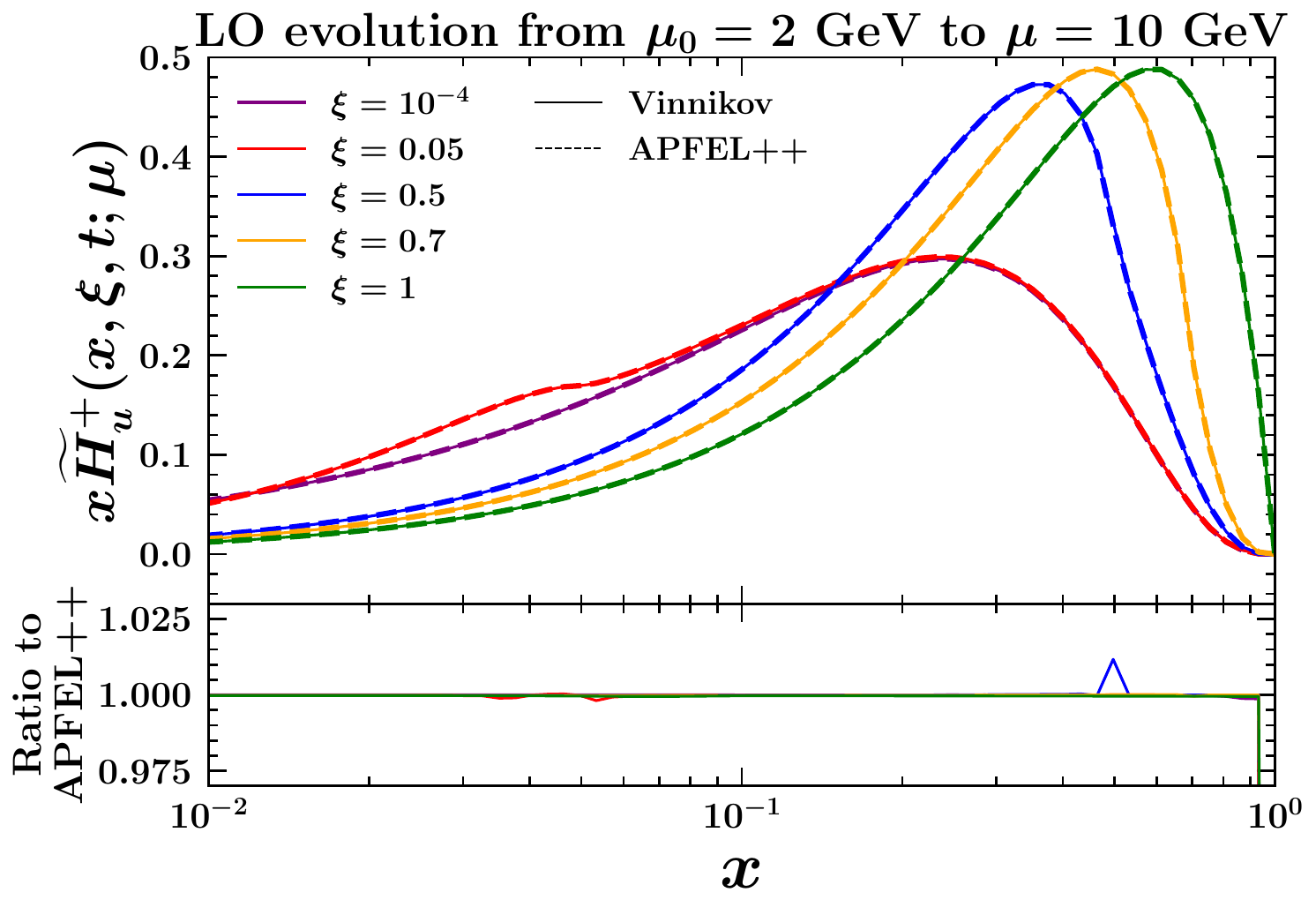}\label{ApfelVsVinnikovPolPlus}}
\subfloat[][]{\includegraphics[width=0.33 \textwidth]{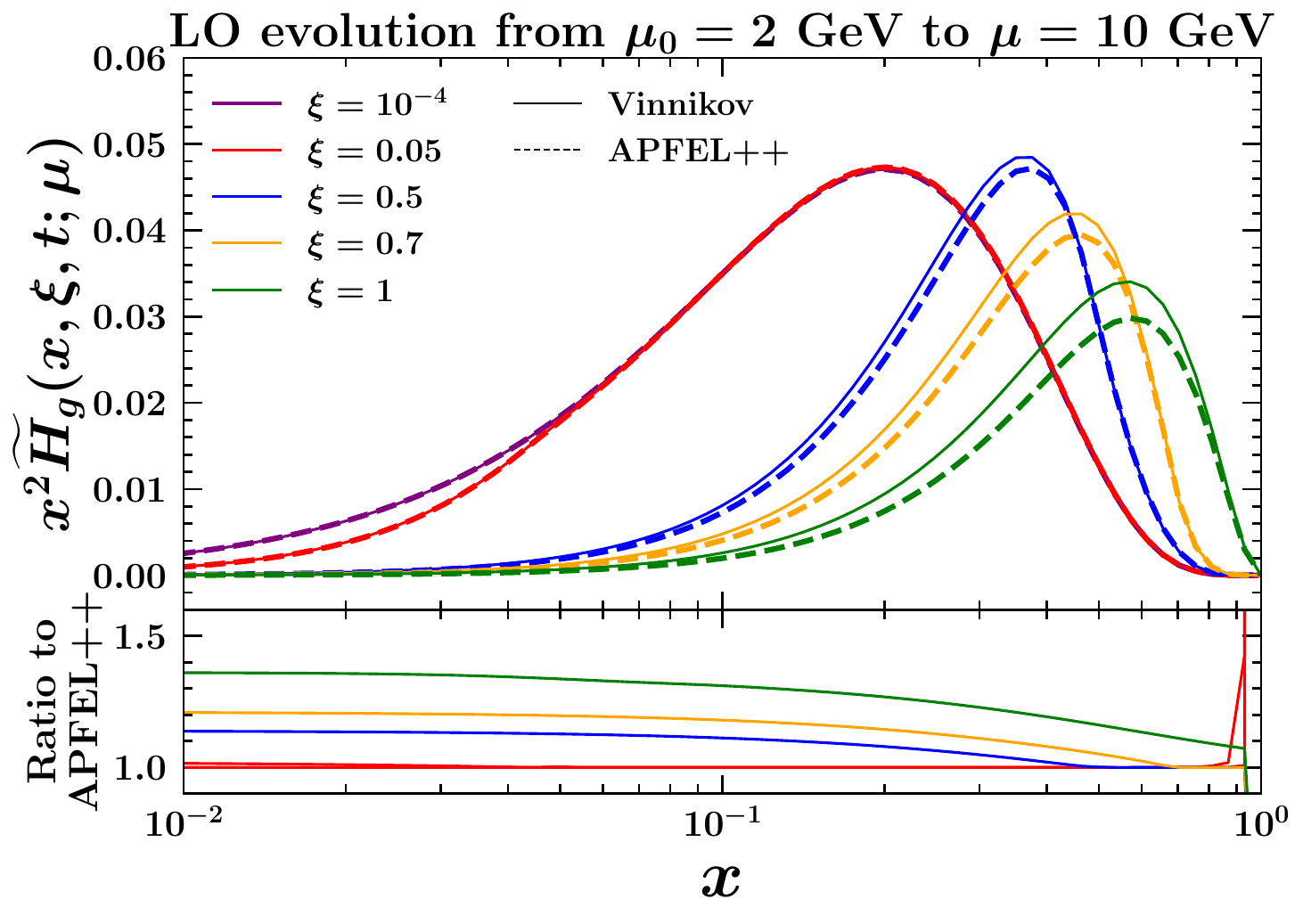}\label{ApfelVsVinnikovPolGluon}
}
\caption{Same as Fig.~\ref{ApfelVsVinnikovUnp} but for
  longitudinally polarised GPDs.}
\label{ApfelVsVinnikovPol}
\end{figure}

Finally, in Figs.~\ref{ApfelVsVinnikovUnp}
and~\ref{ApfelVsVinnikovPol}, we present a comparative analysis of the
evolution of unpolarised and longitudinally polarised distributions
between the code developed in Ref.~\cite{Vinnikov:2006xw}, which we
refer to as Vinnikov's code, and our implementation in {\tt
  APFEL++}.\footnote{Vinnikov's code does not include the
  implementation of transversely/circularly polarised evolution.} To
match the capabilities of Vinnikov's code, the comparison is performed
without heavy-flavour threshold crossing. All other settings are
consistent with those applied in the numerical results presented
above.

This comparison was already presented in Ref.~\cite{Bertone:2022frx}
in the unpolarised case, where it was observed that, for small enough
values of $\xi$ ($\xi\lesssim 0.6$), a generally good agreement
between the two codes was achieved. However, for larger values of
$\xi$ ($\xi\gtrsim 0.6$) a significant deterioration in the agreement
was noted. Subsequently, we conducted a deeper investigation of this
issue\footnote{Prompted by the referee's encouragement, we express our
  gratitude for this valuable suggestion.} that revealed that
Vinnikov’s code was indeed affected by a bug in the region $\xi>
2/3$. This issue is due to the way the $x$-space interpolation grid is
constructed (see Eq.~(6) of
Ref.~\cite{Vinnikov:2006xw}). Specifically, the definition of the grid
parameter $\gamma$ in Eq.~(6) of Ref.~\cite{Vinnikov:2006xw}
guarantees that the constraints in Eqs.~(7) and~(9) are fulfilled only
when $\xi\leq2/3$, while they are broken for $\xi> 2/3$. Following the
resolution of this issue,\footnote{The fix has been implemented in the
  current master branch of {\tt PARTONS}.} as illustrated in
Figs.~\ref{ApfelVsVinnikovUnp} and \ref{ApfelVsVinnikovPol}, a better
agreement between {\tt APFEL++} and Vinnikov's code is found across a
broad range of $\xi$, extending beyond $\xi\simeq 0.6$, for both
unpolarised and longitudinally polarised evolutions.

In the unpolarised case (Fig.~\ref{ApfelVsVinnikovUnp}), the agreement
generally remains at or below the percent level, with the only
exception of the up-quark singlet distribution $H_u^+$ at $\xi=1$,
which displays a larger departure of the order of 10-20\% for small
values of $x$. As far as the longitudinally polarised evolution is
concerned (Fig.~\ref{ApfelVsVinnikovPol}), the agreement between the
two codes is excellent for $\widetilde{H}_u^+$, while it tends to
deteriorate at small $x$ and large $\xi$ for $\widetilde{H}_u^-$. The
disagreement is more pronounced for $\widetilde{H}_g$ where
differences tend to increase with growing $\xi$, reaching the 40\%
level at $\xi=1$ and small $x$. Unfortunately, we could not identify
the origin of this residual discrepancy.

\section{Conclusions}\label{sec:Conclusions}

In this paper, we have revisited the evolution of all twist-2 GPDs,
\textit{i.e.} unpolarised, longitudinally polarised, and
transversely/circularly polarised, at one-loop accuracy.

Our re-derivation of the evolution kernels closely follows that of
Ref.~\cite{Bertone:2022frx}, where the unpolarised evolution kernels
were obtained, and extends it to the two remaining twist-2
polarisations. One of our main purposes is to obtain an efficient
numerical implementation of the evolution of GPDs. This is achieved by
recasting the GPD evolution equations in a form that resembles the
DGLAP equations, which allowed us to exploit well-established
numerical techniques to obtain a solid implementation in the numerical
code {\tt APFEL++}~\cite{Bertone:2013vaa, Bertone:2017gds}. The
computation is performed following a Feynman-graph approach in
momentum space using the light-cone gauge (see
Appendix~\ref{sec:Appendix}), paving the way for the extraction of the
two-loop evolution kernels along the same lines of
Ref.~\cite{Curci:1980uw}. This will eventually allow us to achieve GPD
evolution at next-to-leading order accuracy that, with the advent of
the Electron-Ion Collider~\cite{AbdulKhalek:2021gbh}, will soon become
a necessary ingredient for accurate phenomenology.

We have performed a number of analytic checks of the expressions
obtained for the evolution kernels. Specifically, we have verified the
correctness of the DGLAP ($\xi\rightarrow 0$) and ERBL
($\xi\rightarrow 1$) limits, ensured that the evolution does not cause
any discontinuity of the GPDs at $x=\xi$, and proven that the kernels
preserve polynomiality. Finally, we have thoroughly checked our
numerical implementation of the evolution using as a set of
initial-scale GPDs the realistic GK
model~\cite{Goloskokov:2005sd,Goloskokov:2007nt,Goloskokov:2009ia} as
provided by {\tt PARTONS}~\cite{Berthou:2015oaw}. We showed that the
DGLAP limit is accurately recovered for all polarisations and that
polynomiality is conserved upon evolution. Our implementation was also
compared against an independent code (Vinnikov's
code~\cite{Vinnikov:2006xw}), revealing a generally good agreement.

\section*{Acknowledgements}

We thank P. Sznajder for helping us spot a bug in the code of
Ref.~\cite{Vinnikov:2006xw}. This project is partially supported in
part by the State Agency for Research of the Spanish Ministry of
Science and Innovation through the grants PID2019-106080GB-C21,
PCI2022-132984, PID2022-136510NB-C31, PID2022-136510NB-C33 and
CNS2022-135186, by the Basque Government through the grant IT1628-22,
as well as by the European Union Horizon 2020 research and innovation
program under grant agreement Num. 824093 (STRONG-2020).  OdR is
supported by the MIU (Ministerio de Universidades, Spain) fellowship
FPU20/03110.  SR is supported by the Deutsche Forschungsgemeinschaft
(DFG, German Research Foundation) – grant number 409651613 (Research
Unit FOR 2926), subproject 430915355.

\appendix

\section{GPD evolution kernel integrals}\label{sec:Appendix}

In this appendix, we give a pedagogical review of the calculation of
the one-loop evolution kernels for all twist-2
polarisations. Specifically, we show how the set of functions
$p_{i/k}^{\Gamma}$ in Eq.~(\ref{KernelGPDs}) are obtained.

\subsection{Quark-in-quark GPD}

\begin{figure}[h]
\centering
\includegraphics[width=0.4\textwidth]{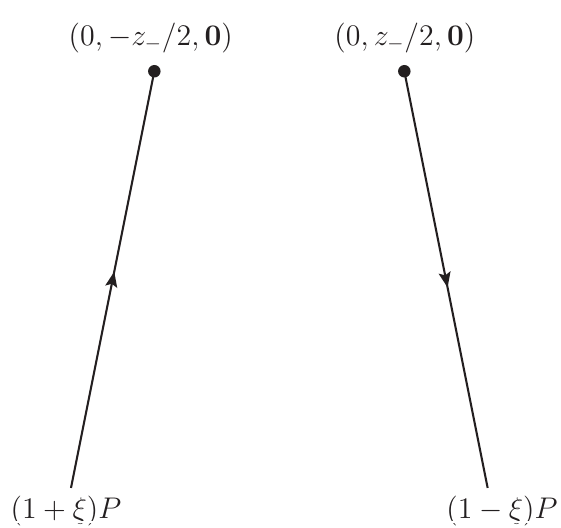}
\caption{Leading-order diagram contributing to the quark-in-quark
  GPDs.}
\label{fig:QQTreeLevel}
\end{figure}

Since we are working in light-cone gauge, we can avoid considering
gluons attaching to the Wilson line. Also, as the evolution kernels do
not depend on $\Delta_T^\mu$, we can substitute in
Eq.~(\ref{GPD_definition})
$\left| P-\Delta/2\right\rangle\,\longrightarrow\,\left|
  (1+\xi)P\right\rangle_i$ and
$\left\langle P+\Delta/2\right|\,\longrightarrow\, _i\left\langle
  (1-\xi)P\right|$, with $i=q,g$. Hence, we rewrite the bare
quark-in-quark GPD with interacting fields as follows:
\begin{equation}\label{eq:GPDQQ}
\begin{split}
  &\hat{F}^{[\Gamma]}_{q\leftarrow
    q}(x,\xi,\varepsilon)=\frac{1}{2N_c}\int\frac{dz}{2\pi}e^{-ixP_+z}\leftindex_q{\left\langle(1-\xi)P\left|\bar{q}^j\left(\frac{z}2\right)
      \Gamma_{ji}q^i\left(-\frac{z}2\right)
    \right|(1+\xi)P\right\rangle}_q\,.
\end{split}
\end{equation}
Before moving to the computation of the one-loop graphs, it is
instructive to compute the tree-level graph first
(Fig.~\ref{fig:QQTreeLevel}).

In the absence of interactions, the quark fields can directly act on
the momentum and spin partonic eigenstates, so that:
\begin{equation}
\begin{split}
    F^{[\Gamma],[0]}_{q\leftarrow q}(x,\xi)&= \frac{1}{4N_c}\int\frac{dz}{2\pi}e^{i(1-x)P_+z}\delta_{aa}\bar{u}_{s',j}\left((1-\xi)P\right)\Gamma_{ji}u_{s,i}((1+\xi)P)\Lambda_{ss'}^{[\Gamma]}
    \\
    &=\frac{1}{4}\int\frac{dz}{2\pi}e^{i(1-x)P_+z}\text{Tr}\left[\Gamma u_{s}((1+\xi)P)\Lambda_{ss'}^{[\Gamma]}\bar{u}_{s'}((1-\xi)P)\right]\\
    &=\frac{1}{4}\int\frac{dz}{2\pi}e^{i(1-x)P_+z}\sqrt{1-\xi^2}\,\text{Tr}\left[\Gamma\Lambda^{[\Gamma]}\right]\,.
\end{split}
\label{eq:QQGPDLO}
\end{equation}
where $\Lambda_{ss'}^{[\Gamma]}$ is a projector introduced for
convenience to select the desired polarisation of the external quark
state. The three relevant polarisations (unpolarised, longitudinally
polarised, and transversely polarised) produce the following outcome
when summed over the spin states $s$ and $s'$:
$\Lambda^{[\Gamma]}=u_s(P)\,\Lambda_{ss'}^{[\Gamma]}\,\bar{u}_{s'}(P)\in
\{\slashed{P},\slashed{P}\gamma_5,i\sigma^{\sigma\rho}P_\rho\gamma_5\}$,
respectively. The trace in Eq.~(\ref{eq:QQGPDLO}) is easily computed
yielding:
\begin{equation}
    F^{[U],[0]}_{q\leftarrow q}(x,\xi)=F^{[L],[0]}_{q\leftarrow q}(x,\xi)=F^{[T],[0]}_{q\leftarrow q}(x,\xi)=\sqrt{1-\xi^2}\,\delta(1-x)\,.
\end{equation}

As expected, in the limit $\xi\rightarrow 0$, the leading-order
quark-in-quark PDFs are recovered. However, it is also interesting to
observe that in the limit $\xi\rightarrow 1$ the leading-order GPD
vanishes. This is indeed the correct behavior because the limit
$\xi\rightarrow 1$ corresponds to a distribution amplitude that
encodes the creation of meson bound state out of a pair of incoming
quark-antiquark. Of course, this cannot happen without any interaction
between the quark and the anti-quark. Therefore, the leading-order graph has to vanish.

\begin{figure}[H]
    \centering
    \includegraphics[width=0.5\textwidth]{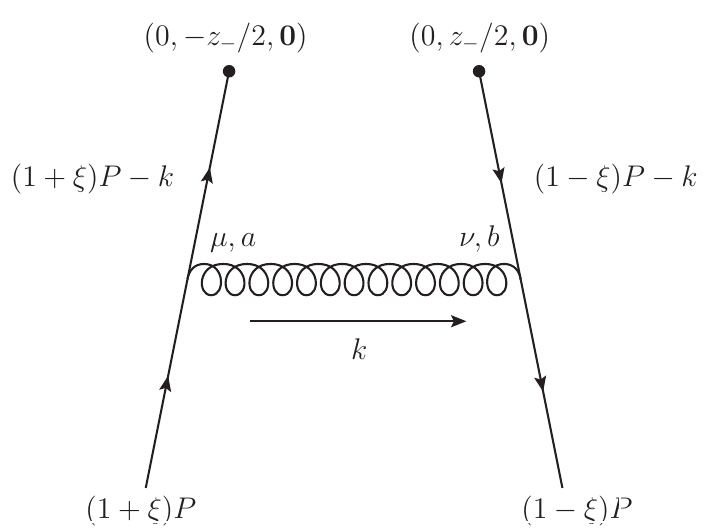}
    \caption{One-loop diagram contributing to the quark-in-quark GPD.}
    \label{fig:QQOneLoop}
\end{figure}

Now, we can use the same procedure to compute the one-loop correction
to the quark-in-quark GPDs whose relevant diagram is displayed in
Fig.~\ref{fig:QQOneLoop}.
The integral to compute is:
\begin{equation}\label{eq:Fqqoneloop}
a_s\hat{F}^{[\Gamma],[1]}_{q\leftarrow q}(x,\xi,\varepsilon) =
\sqrt{1-\xi^2}\int_{-\infty}^{\infty}\frac{dz}{2\pi} e^{i(1-x)P_+z}\mbox{Tr}\left[R_{qq}^{[\Gamma]}(z,\xi,\varepsilon)\Lambda^{[\Gamma]}\right]\,,
\end{equation}
with
\begin{equation}
R^{[\Gamma]}_{qq}(z,\xi,\varepsilon) =ia_sC_F  \int\frac{d^{4-2\epsilon}k}{(2\pi)^{2-2\epsilon}}e^{-ik_+z}\frac{\gamma^\mu[(1+\xi)\slashed{P}-\slashed{k}]\,\Gamma\,[(1-\xi)\slashed{P}-\slashed{k}]\gamma^\nu\mathcal{D}_{\mu\nu}(k)}{[((1+\xi)P-k)^2+i\varepsilon][((1-\xi)P-k)^2+i\varepsilon]}\,.
\end{equation}
This integral will give different results depending on the relative
position of $x$ and $\xi$. In particular, the region $x<\xi$
corresponds the ERBL region, while $x>\xi$ corresponds to the DGLAP
region. The functions $p_{q/q}^{\Gamma}$ can then be obtained by
extracting the $\overline{\mbox{MS}}$ UV pole part of this integral in
these regions as shown in the Appendix of Ref.~\cite{Bertone:2022frx}.

\subsection{Gluon-in-gluon GPD}

We now consider the gluon-in-gluon GPD whose operator definition in
light-cone gauge reduces to:
\begin{equation}
\hat{F}^{[\Gamma]}_{g\leftarrow g}(x,\xi,\varepsilon)=-\frac{P_+(x^2-\xi^2)}{2(N_c^2-1)x}\int\frac{dy}{2\pi}e^{-ixp_+y}\leftindex_g{\left\langle
      (1-\xi)P\left|A_a^{\mu}\left(\frac{z}2\right)\Gamma_{\mu\nu}
        A_a^{\nu}\left(-\frac{z}2\right)\right|(1+\xi)P\right\rangle}_g\,.
\end{equation}
The leading-order contribution is obtained using free fields and
plugging in the Lorentz structures $\Gamma_{\mu\nu}$ given in
Eq.~(\ref{eq:gluonprojectors}). The result is:
\begin{equation}
  F^{[U],[0]}_{g\leftarrow g}(x,\xi)=F^{[L],[0]}_{g\leftarrow g}(x,\xi)=F^{[T],[0]}_{g\leftarrow g}(x,\xi)=(1-\xi^2)\,\delta(1-x)\,.
\end{equation}

\begin{figure}[H]
    \centering
    \includegraphics[width=\textwidth]{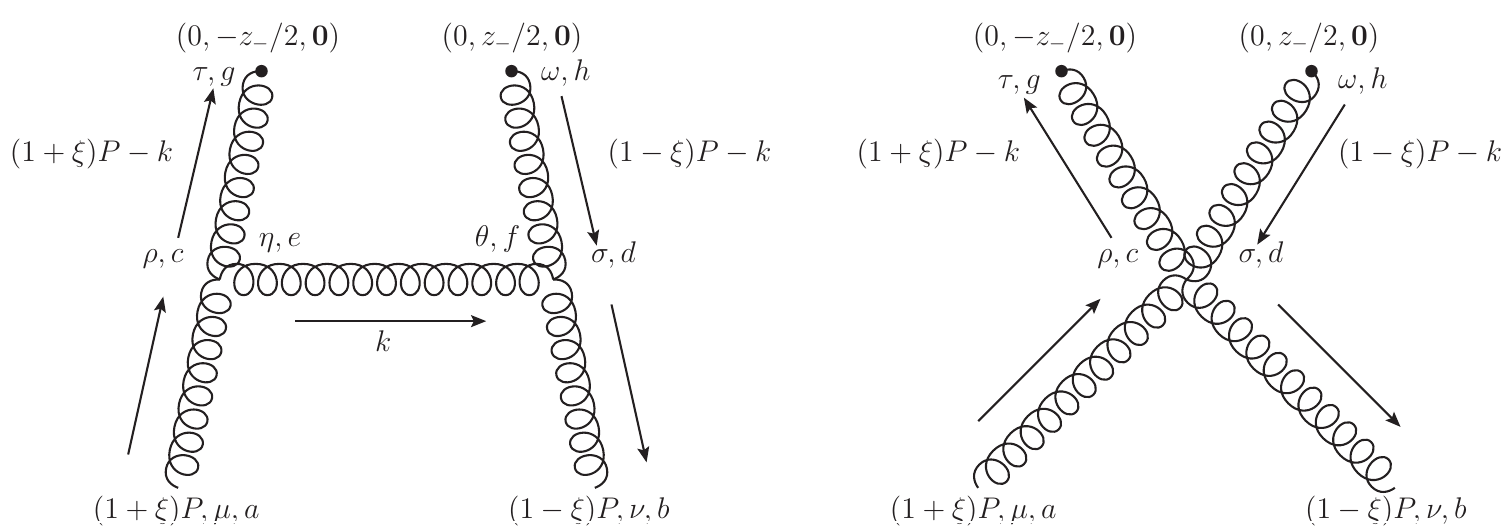}
    \caption{One-loop diagrams contributing to the gluon-in-gluon
      GPD.}
    \label{fig:GGOneLoop}
\end{figure}

The next-to-leading order correction is obtained considering the
diagrams in Fig.~\ref{fig:GGOneLoop}.

The integral to compute for the left diagram is:
\begin{equation}\label{eq:GPDggR}
  a_s\hat{F}_{g\leftarrow g,3g}^{[\Gamma],[1]}(x,\xi,\varepsilon) =
  -\frac{P_+(x^2-\xi^2)}{2(N_c^2-1)x}\int_{-\infty}^{\infty}\frac{dz}{2\pi}e^{i(1-x)P_+z}R_{3g}^{[\Gamma]}(z,\xi,\varepsilon)\,,
\end{equation}
with:
\begin{equation}
\begin{split}
  &R_{3g}^{[\Gamma]}(z,\xi)=-4ia_s f ^{ace} f ^{eca}
    \int\frac{d^{4-2\epsilon}k}{(2\pi)^{2-2\epsilon}}e^{-ik_+z} V_{\mu\rho\eta}((1+\xi)P,-(1+\xi)P+k,-k)\\
\\
&\times\frac{\Lambda^{[\Gamma]\mu\nu}d^{\rho\tau}((1+\xi)P-k)\,d^{\eta\theta}(k)\,d^{\omega\sigma}((1-\xi)P-k)\,\Gamma_{\tau\omega}}{[((1+\xi)P-k)^2+i\varepsilon][k^2+i\varepsilon][((1-\xi)P-k)^2+i\varepsilon]}\,V_{\theta\sigma\nu}(k,(1-\xi)P-k,-(1-\xi)P)\,,
\end{split}
\end{equation}
where:
\begin{equation}
    V^{\mu\nu\rho}(q,l,r)=-\left[g^{\mu\nu}(q-l)^\rho+g^{\nu\rho}(l-r)^\mu+g^{\rho\mu}(r-q)^\nu\right]\,.
\end{equation}
Similarly to the quark-in-quark distribution, the projector
$\Lambda^{[\Gamma]\mu\nu}$ selects the relevant gluon polarisation
according to:
\begin{equation}
    \Lambda^{[\Gamma]\mu\nu}=e^{*\,\mu}_{s'}(P)\,\Lambda_{ss'}^{[\Gamma]}\,e^\nu_{s}(P)\,,
\end{equation}
where the tensor $\Lambda^{[\Gamma]\mu\nu}$ is to be selected amongst
the following Lorentz structures:
\begin{equation}
  -d^{\mu\nu}(P), \ -i\e_T^{\mu\nu}(P)\equiv -i\frac{\e^{\alpha\beta\mu\nu}P_\alpha n_\beta}{(nP)},\, \ -R^\mu R^\nu-L^\mu L^\nu\,,
\end{equation}
for unpolarised, longitudinally polarised, and circularly polarised
gluons, respectively.

The integral corresponding to the four-gluon-vertex diagram on the
r.h.s. of Fig.~\ref{fig:GGOneLoop} is:
\begin{equation}\label{eq:GPDggR4g}
  a_s\hat{F}_{g\leftarrow g,4g}^{[\Gamma],[1]}(x,\xi,\varepsilon) =
  -\frac{P_+(x^2-\xi^2)}{2(N_c^2-1)x}\int_{-\infty}^{\infty}\frac{dz}{2\pi}e^{i(1-x)P_+z}
  R_{4g}^{[\Gamma]}(z,\xi)\,,
\end{equation}
with:
\begin{equation}
 R_{4g}^{[\Gamma]}(z,\xi)=4ia_s\int\frac{d^{4-2\epsilon}k}{(2\pi)^{2-2\epsilon}}e^{-ik_+z}
    \frac{d^{\rho\tau}((1+\xi)P-k)}{((1+\xi)P-k)^2+i\varepsilon}\frac{d^{\omega\sigma}((1-\xi)P-k)}{((1-\xi)P-k)^2+i\varepsilon}\,\Lambda^{[\Gamma]\mu\nu}W_{\mu\rho\sigma\nu}^{bggb}\Gamma_{\tau\omega}\,,
\end{equation}
where $W_{\mu\rho\sigma\nu}^{bggb}$ is the four-gluon-vertex Feynman
rule given by:
\begin{equation}
    W_{\mu\nu\rho\sigma}^{abcd}=f^{eab}f^{ecd}(g_{\mu\rho}g_{\nu\sigma}-g_{\mu\sigma}g_{\nu\rho})+f^{eac}f^{ebd}(g_{\mu\nu}g_{\rho\sigma}-g_{\mu\sigma}g_{\nu\rho})+f^{ead}f^{ebc}(g_{\mu\nu}g_{\rho\sigma}-g_{\mu\rho}g_{\nu\sigma})\,.
\end{equation}
Extracting the UV pole part of these diagrams for the different
polarisations leads to the functions $p_{g/g}^{\Gamma}$ given in
Eqs.~(\ref{KernelFunction_ggU}), (\ref{KernelFunction_ggL}),
and~(\ref{KernelFunction_ggT}).

\subsection{Quark-in-gluon and gluon-in-quark GPDs}

\begin{figure}[h]
    \centering
    \includegraphics[width=\textwidth]{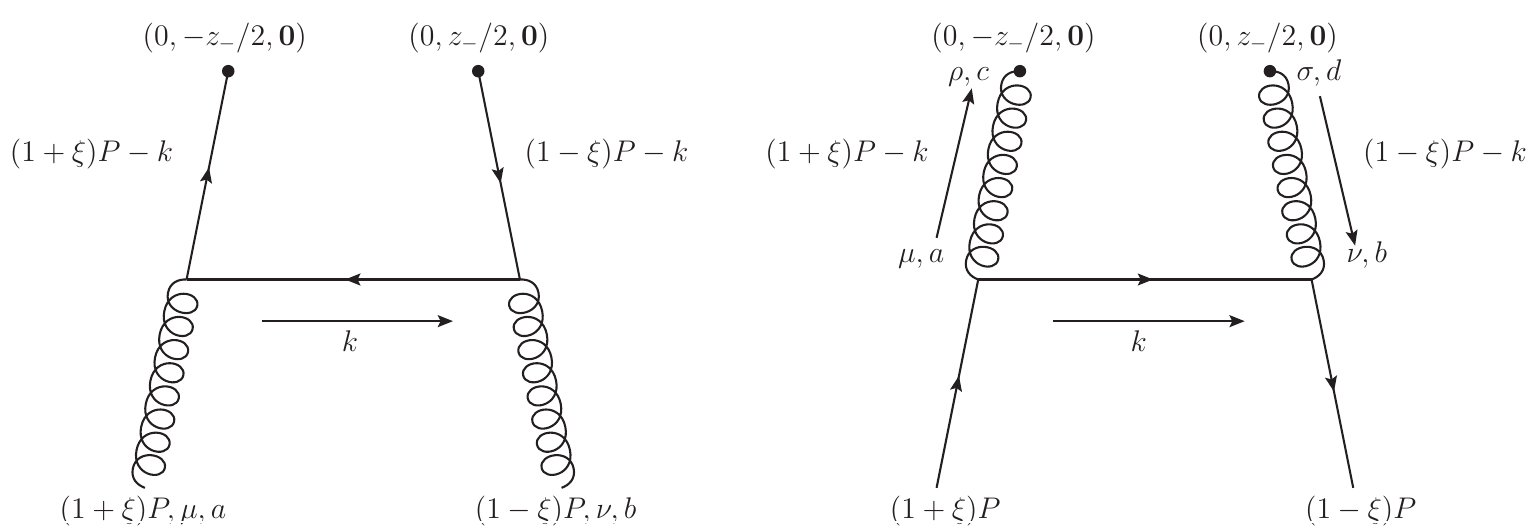}
    \caption{One-loop diagrams contributing to the quark-in-gluon
      (left) and the gluon-in-quark (right) GPDs.}
    \label{fig:OneLoopQGandGQ}
\end{figure}

We now consider the off-diagonal quark-in-gluon and gluon-in-quark GPDs whose operator definitions are: 
\begin{equation}
\begin{split}
     &\hat{F}^{[\Gamma]}_{g\leftarrow
       q}(x,\xi,\varepsilon)=\frac{1}{4(N_c^2-1)}\int\frac{dz}{2\pi}e^{-ixP_+z}\leftindex_g{\left\langle(1-\xi)P\left|\bar{q}^j\left(\frac{z}2\right)\Gamma_{ji}q^i\left(-\frac{z}2\right)
       \right|(1+\xi)P\right\rangle}_g\,,\\
  &\hat{F}^{[\Gamma]}_{q\leftarrow g}(x,\xi,\varepsilon)=-\frac{p_+(x^2-\xi^2)}{2N_cx}\int\frac{dy}{2\pi}e^{-ixp_+y}\leftindex_q{\left\langle
    (1-\xi)P\left|A_a^{\mu}\left(\frac{z}2\right)\Gamma_{\mu\nu}
      A_a^{\nu}\left(-\frac{z}2\right)\right|(1+\xi)P\right\rangle}_q\,.
\end{split}
\end{equation}
These GPDs have no tree-level contribution and the first non-vanishing
contribution appears at $\mathcal{O}(\alpha_s)$. The corresponding
diagrams are shown in Fig.~\ref{fig:OneLoopQGandGQ}.

The integral to be computed for the quark-in-gluon diagram is:
\begin{equation}\label{eq:FqgR}
a_s\hat{F}^{[\Gamma],[1]}_{g\leftarrow q}(x,\xi,\varepsilon)=
\int_{-\infty}^{\infty}\frac{dz}{2\pi} e^{i(1-x)P_+z}\text{Tr}\left[R_{qg}^{[\Gamma]}(z,\xi,\varepsilon)\right]\,,
\end{equation}
where:
\begin{equation}
\begin{split}
R_{qg}^{[\Gamma]}(z,\xi,\varepsilon)=&ia_sT_R\int\frac{d^{2-2\epsilon}\mathbf{k}_T}{(2\pi)^{2-2\epsilon}}dk_+dk_-e^{-ik_+z}\Lambda^{[\Gamma]\mu\nu}\frac{\slashed{k}}{k^2+i\varepsilon}\gamma_\mu\\
&\times\frac{((1+\xi)\slashed{P}-\slashed{k})}{((1+\xi)P-k)^2+i\varepsilon}\,\Gamma\,\frac{(1-\xi)\slashed{P}-\slashed{k}}{((1-\xi)P-k)^2+i\varepsilon}\gamma_\nu\,.
\end{split}
\end{equation}
While the one-loop gluon-in-quark diagram is computed as:
\begin{equation}\label{eq:FqgR}
a_s\hat{F}^{[\Gamma],[1]}_{q\leftarrow g}(x,\xi,\varepsilon) =
-\frac{P_+(x^2-\xi^2)\sqrt{1-\xi^2}}{2\,x}\int_{-\infty}^{\infty}\frac{dz}{2\pi}e^{i(1-x)P_+z}{\rm
Tr} \left[R_{gq}^{[\Gamma]}(z,\xi,\varepsilon)\Lambda^{[\Gamma]  }\right]\,,
\end{equation}
where:
\begin{equation}
\begin{split}
    R_{gq}^{[\Gamma]}(z,\xi,\varepsilon) =&
    4ia_sC_F\int\frac{d^{2-2\epsilon}\mathbf{k}_T}{(2\pi)^{2-2\epsilon}}dk_+dk_-e^{-ik_+z
    }\gamma_\mu\frac{\slashed{k}}{k^2+i\varepsilon}\\
    &\times\frac{
    d^{\mu\rho}((1+\xi)P-k)}{((1+\xi)P-k)^2+i\varepsilon}\,\Gamma_{\rho\sigma}\,\frac{d^{\sigma\nu}((1-\xi)P-k)}{((1-\xi)P-k)^2+i\varepsilon}\gamma_\nu\,.
\end{split}
\end{equation}
Once again, the UV pole part of these integrals allows us to obtain
the functions $p_{q/g}^{\Gamma}$ and $p_{g/q}^{\Gamma}$.

\bibliographystyle{utphys}

\begin{thebibliography}{10}

\bibitem{Muller:1994ses}
D.~M\"uller, D.~Robaschik, B.~Geyer, F.~M. Dittes, and J.~Ho\v{r}ej\v{s}i,
  ``{Wave functions, evolution equations and evolution kernels from light ray
  operators of QCD},'' \href{http://dx.doi.org/10.1002/prop.2190420202}{{\em
  Fortsch. Phys.} {\bfseries 42} (1994) 101--141},
  \href{http://arxiv.org/abs/hep-ph/9812448}{{\ttfamily arXiv:hep-ph/9812448}}.

\bibitem{Ji:1996ek}
X.-D. Ji, ``{Gauge-Invariant Decomposition of Nucleon Spin},''
  \href{http://dx.doi.org/10.1103/PhysRevLett.78.610}{{\em Phys. Rev. Lett.}
  {\bfseries 78} (1997) 610--613},
  \href{http://arxiv.org/abs/hep-ph/9603249}{{\ttfamily arXiv:hep-ph/9603249}}.

\bibitem{Radyushkin:1996nd}
A.~V. Radyushkin, ``{Scaling limit of deeply virtual Compton scattering},''
  \href{http://dx.doi.org/10.1016/0370-2693(96)00528-X}{{\em Phys. Lett. B}
  {\bfseries 380} (1996) 417--425},
  \href{http://arxiv.org/abs/hep-ph/9604317}{{\ttfamily arXiv:hep-ph/9604317}}.

\bibitem{Radyushkin:1996ru}
A.~V. Radyushkin, ``{Asymmetric gluon distributions and hard diffractive
  electroproduction},''
  \href{http://dx.doi.org/10.1016/0370-2693(96)00844-1}{{\em Phys. Lett. B}
  {\bfseries 385} (1996) 333--342},
  \href{http://arxiv.org/abs/hep-ph/9605431}{{\ttfamily arXiv:hep-ph/9605431}}.

\bibitem{Ji:1996nm}
X.-D. Ji, ``{Deeply virtual Compton scattering},''
  \href{http://dx.doi.org/10.1103/PhysRevD.55.7114}{{\em Phys. Rev. D}
  {\bfseries 55} (1997) 7114--7125},
  \href{http://arxiv.org/abs/hep-ph/9609381}{{\ttfamily arXiv:hep-ph/9609381}}.

\bibitem{Radyushkin:1997ki}
A.~V. Radyushkin, ``{Nonforward parton distributions},''
  \href{http://dx.doi.org/10.1103/PhysRevD.56.5524}{{\em Phys. Rev. D}
  {\bfseries 56} (1997) 5524--5557},
  \href{http://arxiv.org/abs/hep-ph/9704207}{{\ttfamily arXiv:hep-ph/9704207}}.

\bibitem{Ji:1998pc}
X.-D. Ji, ``{Off forward parton distributions},''
  \href{http://dx.doi.org/10.1088/0954-3899/24/7/002}{{\em J. Phys. G}
  {\bfseries 24} (1998) 1181--1205},
  \href{http://arxiv.org/abs/hep-ph/9807358}{{\ttfamily arXiv:hep-ph/9807358}}.

\bibitem{Diehl:2003ny}
M.~Diehl, ``{Generalized parton distributions},''
  \href{http://dx.doi.org/10.1016/j.physrep.2003.08.002}{{\em Phys. Rept.}
  {\bfseries 388} (2003) 41--277},
  \href{http://arxiv.org/abs/hep-ph/0307382}{{\ttfamily arXiv:hep-ph/0307382}}.

\bibitem{Belitsky:2005qn}
A.~V. Belitsky and A.~V. Radyushkin, ``{Unraveling hadron structure with
  generalized parton distributions},''
  \href{http://dx.doi.org/10.1016/j.physrep.2005.06.002}{{\em Phys. Rept.}
  {\bfseries 418} (2005) 1--387},
  \href{http://arxiv.org/abs/hep-ph/0504030}{{\ttfamily arXiv:hep-ph/0504030}}.

\bibitem{Boffi:2007yc}
S.~Boffi and B.~Pasquini, ``{Generalized parton distributions and the structure
  of the nucleon},'' \href{http://dx.doi.org/10.1393/ncr/i2007-10025-7}{{\em
  Riv. Nuovo Cim.} {\bfseries 30} no.~9, (2007) 387--448},
  \href{http://arxiv.org/abs/0711.2625}{{\ttfamily arXiv:0711.2625 [hep-ph]}}.

\bibitem{Goeke:2001tz}
K.~Goeke, M.~V. Polyakov, and M.~Vanderhaeghen, ``{Hard exclusive reactions and
  the structure of hadrons},''
  \href{http://dx.doi.org/10.1016/S0146-6410(01)00158-2}{{\em Prog. Part. Nucl.
  Phys.} {\bfseries 47} (2001) 401--515},
  \href{http://arxiv.org/abs/hep-ph/0106012}{{\ttfamily arXiv:hep-ph/0106012}}.

\bibitem{Kumericki:2016ehc}
K.~Kumericki, S.~Liuti, and H.~Moutarde, ``{GPD phenomenology and DVCS
  fitting}: {Entering the high-precision era},''
  \href{http://dx.doi.org/10.1140/epja/i2016-16157-3}{{\em Eur. Phys. J. A}
  {\bfseries 52} no.~6, (2016) 157},
  \href{http://arxiv.org/abs/1602.02763}{{\ttfamily arXiv:1602.02763
  [hep-ph]}}.

\bibitem{Braun:2020yib}
V.~M. Braun, A.~N. Manashov, S.~Moch, and J.~Schoenleber, ``{Two-loop
  coefficient function for DVCS: vector contributions},''
  \href{http://dx.doi.org/10.1007/JHEP09(2020)117}{{\em JHEP} {\bfseries 09}
  (2020) 117}, \href{http://arxiv.org/abs/2007.06348}{{\ttfamily
  arXiv:2007.06348 [hep-ph]}}. [Erratum: JHEP 02, 115 (2022)].

\bibitem{Braun:2022bpn}
V.~M. Braun, Y.~Ji, and J.~Schoenleber, ``{Deeply Virtual Compton Scattering at
  Next-to-Next-to-Leading Order},''
  \href{http://dx.doi.org/10.1103/PhysRevLett.129.172001}{{\em Phys. Rev.
  Lett.} {\bfseries 129} no.~17, (2022) 172001},
  \href{http://arxiv.org/abs/2207.06818}{{\ttfamily arXiv:2207.06818
  [hep-ph]}}.

\bibitem{Braun:2022ezk}
V.~Braun, A.~Manashov, S.-O. Moch, and J.~Schoenleber, ``{Vector and
  axial-vector coefficient functions for DVCS at NNLO},''
  \href{http://dx.doi.org/10.22323/1.416.0074}{{\em PoS} {\bfseries LL2022}
  (2022) 074}.

\bibitem{Schoenleber:2022myb}
J.~Schoenleber, ``{Resummation of threshold logarithms in deeply-virtual
  Compton scattering},'' \href{http://dx.doi.org/10.1007/JHEP02(2023)207}{{\em
  JHEP} {\bfseries 02} (2023) 207},
  \href{http://arxiv.org/abs/2209.09015}{{\ttfamily arXiv:2209.09015
  [hep-ph]}}.

\bibitem{Ji:2023xzk}
Y.~Ji and J.~Schoenleber, ``{Two-loop coefficient functions in deeply virtual
  Compton scattering: flavor-singlet axial-vector and transversity case},''
  \href{http://dx.doi.org/10.1007/JHEP01(2024)053}{{\em JHEP} {\bfseries 01}
  (2024) 053}, \href{http://arxiv.org/abs/2310.05724}{{\ttfamily
  arXiv:2310.05724 [hep-ph]}}.

\bibitem{Burkardt:2002hr}
M.~Burkardt, ``{Impact parameter space interpretation for generalized parton
  distributions},'' \href{http://dx.doi.org/10.1142/S0217751X03012370}{{\em
  Int. J. Mod. Phys. A} {\bfseries 18} (2003) 173--208},
  \href{http://arxiv.org/abs/hep-ph/0207047}{{\ttfamily arXiv:hep-ph/0207047}}.

\bibitem{Polyakov:2018zvc}
M.~V. Polyakov and P.~Schweitzer, ``{Forces inside hadrons: pressure, surface
  tension, mechanical radius, and all that},''
  \href{http://dx.doi.org/10.1142/S0217751X18300259}{{\em Int. J. Mod. Phys. A}
  {\bfseries 33} no.~26, (2018) 1830025},
  \href{http://arxiv.org/abs/1805.06596}{{\ttfamily arXiv:1805.06596
  [hep-ph]}}.

\bibitem{Dutrieux:2021nlz}
H.~Dutrieux, C.~Lorc\'e, H.~Moutarde, P.~Sznajder, A.~Trawi\'nski, and
  J.~Wagner, ``{Phenomenological assessment of proton mechanical properties
  from deeply virtual Compton scattering},''
  \href{http://dx.doi.org/10.1140/epjc/s10052-021-09069-w}{{\em Eur. Phys. J.
  C} {\bfseries 81} no.~4, (2021) 300},
  \href{http://arxiv.org/abs/2101.03855}{{\ttfamily arXiv:2101.03855
  [hep-ph]}}.

\bibitem{Freese:2022jlu}
A.~Freese, A.~Metz, B.~Pasquini, and S.~Rodini, ``{The gravitational form
  factors of the electron in quantum electrodynamics},''
  \href{http://dx.doi.org/10.1016/j.physletb.2023.137768}{{\em Phys. Lett. B}
  {\bfseries 839} (2023) 137768},
  \href{http://arxiv.org/abs/2212.12197}{{\ttfamily arXiv:2212.12197
  [hep-ph]}}.

\bibitem{Lorce:2021xku}
C.~Lorc\'e, A.~Metz, B.~Pasquini, and S.~Rodini, ``{Energy-momentum tensor in
  QCD: nucleon mass decomposition and mechanical equilibrium},''
  \href{http://dx.doi.org/10.1007/JHEP11(2021)121}{{\em JHEP} {\bfseries 11}
  (2021) 121}, \href{http://arxiv.org/abs/2109.11785}{{\ttfamily
  arXiv:2109.11785 [hep-ph]}}.

\bibitem{Duran:2022xag}
B.~Duran {\em et~al.}, ``{Determining the gluonic gravitational form factors of
  the proton},'' \href{http://dx.doi.org/10.1038/s41586-023-05730-4}{{\em
  Nature} {\bfseries 615} no.~7954, (2023) 813--816},
  \href{http://arxiv.org/abs/2207.05212}{{\ttfamily arXiv:2207.05212
  [nucl-ex]}}.

\bibitem{AbdulKhalek:2021gbh}
R.~Abdul~Khalek {\em et~al.}, ``{Science Requirements and Detector Concepts for
  the Electron-Ion Collider}: {EIC Yellow Report},''
  \href{http://dx.doi.org/10.1016/j.nuclphysa.2022.122447}{{\em Nucl. Phys. A}
  {\bfseries 1026} (2022) 122447},
  \href{http://arxiv.org/abs/2103.05419}{{\ttfamily arXiv:2103.05419
  [physics.ins-det]}}.

\bibitem{Accardi:2023chb}
A.~Accardi {\em et~al.}, ``{Strong Interaction Physics at the Luminosity
  Frontier with 22 GeV Electrons at Jefferson Lab},''
  \href{http://arxiv.org/abs/2306.09360}{{\ttfamily arXiv:2306.09360
  [nucl-ex]}}.

\bibitem{Bertone:2021yyz}
V.~Bertone, H.~Dutrieux, C.~Mezrag, H.~Moutarde, and P.~Sznajder,
  ``{Deconvolution problem of deeply virtual Compton scattering},''
  \href{http://dx.doi.org/10.1103/PhysRevD.103.114019}{{\em Phys. Rev. D}
  {\bfseries 103} no.~11, (2021) 114019},
  \href{http://arxiv.org/abs/2104.03836}{{\ttfamily arXiv:2104.03836
  [hep-ph]}}.

\bibitem{Moffat:2023svr}
E.~Moffat, A.~Freese, I.~Clo\"et, T.~Donohoe, L.~Gamberg, W.~Melnitchouk,
  A.~Metz, A.~Prokudin, and N.~Sato, ``{Shedding light on shadow generalized
  parton distributions},''
  \href{http://dx.doi.org/10.1103/PhysRevD.108.036027}{{\em Phys. Rev. D}
  {\bfseries 108} no.~3, (2023) 036027},
  \href{http://arxiv.org/abs/2303.12006}{{\ttfamily arXiv:2303.12006
  [hep-ph]}}.

\bibitem{Mueller:2005ed}
D.~Mueller and A.~Schafer, ``{Complex conformal spin partial wave expansion of
  generalized parton distributions and distribution amplitudes},''
  \href{http://dx.doi.org/10.1016/j.nuclphysb.2006.01.019}{{\em Nucl. Phys. B}
  {\bfseries 739} (2006) 1--59},
  \href{http://arxiv.org/abs/hep-ph/0509204}{{\ttfamily arXiv:hep-ph/0509204}}.

\bibitem{Braun:2017cih}
V.~M. Braun, A.~N. Manashov, S.~Moch, and M.~Strohmaier, ``{Three-loop
  evolution equation for flavor-nonsinglet operators in off-forward
  kinematics},'' \href{http://dx.doi.org/10.1007/JHEP06(2017)037}{{\em JHEP}
  {\bfseries 06} (2017) 037}, \href{http://arxiv.org/abs/1703.09532}{{\ttfamily
  arXiv:1703.09532 [hep-ph]}}.

\bibitem{Braun:2012bg}
V.~M. Braun, A.~N. Manashov, and B.~Pirnay, ``{Finite-t and target mass
  corrections to DVCS on a scalar target},''
  \href{http://dx.doi.org/10.1103/PhysRevD.86.014003}{{\em Phys. Rev. D}
  {\bfseries 86} (2012) 014003},
  \href{http://arxiv.org/abs/1205.3332}{{\ttfamily arXiv:1205.3332 [hep-ph]}}.

\bibitem{Braun:2012hq}
V.~M. Braun, A.~N. Manashov, and B.~Pirnay, ``{Finite-t and target mass
  corrections to deeply virtual Compton scattering},''
  \href{http://dx.doi.org/10.1103/PhysRevLett.109.242001}{{\em Phys. Rev.
  Lett.} {\bfseries 109} (2012) 242001},
  \href{http://arxiv.org/abs/1209.2559}{{\ttfamily arXiv:1209.2559 [hep-ph]}}.

\bibitem{Braun:2014sta}
V.~M. Braun, A.~N. Manashov, D.~M\"uller, and B.~M. Pirnay, ``{Deeply Virtual
  Compton Scattering to the twist-four accuracy: Impact of finite-$t$ and
  target mass corrections},''
  \href{http://dx.doi.org/10.1103/PhysRevD.89.074022}{{\em Phys. Rev. D}
  {\bfseries 89} no.~7, (2014) 074022},
  \href{http://arxiv.org/abs/1401.7621}{{\ttfamily arXiv:1401.7621 [hep-ph]}}.

\bibitem{Braun:2019qtp}
V.~M. Braun, A.~N. Manashov, S.~Moch, and M.~Strohmaier, ``{Two-loop evolution
  equations for flavor-singlet light-ray operators},''
  \href{http://dx.doi.org/10.1007/JHEP02(2019)191}{{\em JHEP} {\bfseries 02}
  (2019) 191}, \href{http://arxiv.org/abs/1901.06172}{{\ttfamily
  arXiv:1901.06172 [hep-ph]}}.

\bibitem{Braun:2021tzi}
V.~M. Braun, A.~N. Manashov, S.~Moch, and M.~Strohmaier, ``{Three-loop
  off-forward evolution kernel for axial-vector operators in
  Larin\textquoteright{}s scheme},''
  \href{http://dx.doi.org/10.1103/PhysRevD.103.094018}{{\em Phys. Rev. D}
  {\bfseries 103} no.~9, (2021) 094018},
  \href{http://arxiv.org/abs/2101.01471}{{\ttfamily arXiv:2101.01471
  [hep-ph]}}.

\bibitem{Braun:2022byg}
V.~M. Braun, K.~G. Chetyrkin, and A.~N. Manashov, ``{NNLO anomalous dimension
  matrix for twist-two flavor-singlet operators},''
  \href{http://dx.doi.org/10.1016/j.physletb.2022.137409}{{\em Phys. Lett. B}
  {\bfseries 834} (2022) 137409},
  \href{http://arxiv.org/abs/2205.08228}{{\ttfamily arXiv:2205.08228
  [hep-ph]}}.

\bibitem{Braun:2022qly}
V.~M. Braun, Y.~Ji, and A.~N. Manashov, ``{Next-to-leading-power kinematic
  corrections to DVCS: a scalar target},''
  \href{http://dx.doi.org/10.1007/JHEP01(2023)078}{{\em JHEP} {\bfseries 01}
  (2023) 078}, \href{http://arxiv.org/abs/2211.04902}{{\ttfamily
  arXiv:2211.04902 [hep-ph]}}.

\bibitem{Moch:2021cdq}
S.~Moch and S.~Van~Thurenhout, ``{Renormalization of non-singlet quark operator
  matrix elements for off-forward hard scattering},''
  \href{http://dx.doi.org/10.1016/j.nuclphysb.2021.115536}{{\em Nucl. Phys. B}
  {\bfseries 971} (2021) 115536},
  \href{http://arxiv.org/abs/2107.02470}{{\ttfamily arXiv:2107.02470
  [hep-ph]}}.

\bibitem{VanThurenhout:2022nmx}
S.~Van~Thurenhout, ``{Off-forward anomalous dimensions of non-singlet
  transversity operators},''
  \href{http://dx.doi.org/10.1016/j.nuclphysb.2022.115835}{{\em Nucl. Phys. B}
  {\bfseries 980} (2022) 115835},
  \href{http://arxiv.org/abs/2204.02140}{{\ttfamily arXiv:2204.02140
  [hep-ph]}}.

\bibitem{Vinnikov:2006xw}
A.~V. Vinnikov, ``{Code for prompt numerical computation of the leading order
  GPD evolution},'' \href{http://arxiv.org/abs/hep-ph/0604248}{{\ttfamily
  arXiv:hep-ph/0604248}}.

\bibitem{Berthou:2015oaw}
B.~Berthou {\em et~al.}, ``{PARTONS: PARtonic Tomography Of Nucleon Software}:
  {A computing framework for the phenomenology of Generalized Parton
  Distributions},''
  \href{http://dx.doi.org/10.1140/epjc/s10052-018-5948-0}{{\em Eur. Phys. J. C}
  {\bfseries 78} no.~6, (2018) 478},
  \href{http://arxiv.org/abs/1512.06174}{{\ttfamily arXiv:1512.06174
  [hep-ph]}}.

\bibitem{Freund:2001bf}
A.~Freund and M.~F. McDermott, ``{Next-to-leading order evolution of
  generalized parton distributions for DESY HERA and HERMES},''
  \href{http://dx.doi.org/10.1103/PhysRevD.66.079903}{{\em Phys. Rev. D}
  {\bfseries 65} (2002) 056012},
  \href{http://arxiv.org/abs/hep-ph/0106115}{{\ttfamily arXiv:hep-ph/0106115}}.
  [Erratum: Phys.Rev.D 66, 079903 (2002)].

\bibitem{Kumericki:2007sa}
K.~Kumericki, D.~Mueller, and K.~Passek-Kumericki, ``{Towards a fitting
  procedure for deeply virtual Compton scattering at next-to-leading order and
  beyond},'' \href{http://dx.doi.org/10.1016/j.nuclphysb.2007.10.029}{{\em
  Nucl. Phys. B} {\bfseries 794} (2008) 244--323},
  \href{http://arxiv.org/abs/hep-ph/0703179}{{\ttfamily arXiv:hep-ph/0703179}}.

\bibitem{gepard}
K.~Kumeri\v{c}ki, ``Gepard: Tool for studying the 3d quark and gluon
  distributions in the nucleon,''. \url{https://gepard.phy.hr/credits.html}.
  \url{https://gepard.phy.hr/credits.html}.

\bibitem{Bertone:2022frx}
V.~Bertone, H.~Dutrieux, C.~Mezrag, J.~M. Morgado, and H.~Moutarde,
  ``{Revisiting evolution equations for generalised parton distributions},''
  \href{http://dx.doi.org/10.1140/epjc/s10052-022-10793-0}{{\em Eur. Phys. J.
  C} {\bfseries 82} no.~10, (2022) 888},
  \href{http://arxiv.org/abs/2206.01412}{{\ttfamily arXiv:2206.01412
  [hep-ph]}}.

\bibitem{Belitsky:2000yn}
A.~V. Belitsky, A.~Freund, and D.~Mueller, ``{NLO evolution kernels for skewed
  transversity distributions},''
  \href{http://dx.doi.org/10.1016/S0370-2693(00)01129-1}{{\em Phys. Lett. B}
  {\bfseries 493} (2000) 341--349},
  \href{http://arxiv.org/abs/hep-ph/0008005}{{\ttfamily arXiv:hep-ph/0008005}}.

\bibitem{Bertone:2013vaa}
V.~Bertone, S.~Carrazza, and J.~Rojo, ``{APFEL: A PDF Evolution Library with
  QED corrections},'' \href{http://dx.doi.org/10.1016/j.cpc.2014.03.007}{{\em
  Comput. Phys. Commun.} {\bfseries 185} (2014) 1647--1668},
  \href{http://arxiv.org/abs/1310.1394}{{\ttfamily arXiv:1310.1394 [hep-ph]}}.

\bibitem{Bertone:2017gds}
V.~Bertone, ``{APFEL++: A new PDF evolution library in C++},''
  \href{http://dx.doi.org/10.22323/1.297.0201}{{\em PoS} {\bfseries DIS2017}
  (2018) 201}, \href{http://arxiv.org/abs/1708.00911}{{\ttfamily
  arXiv:1708.00911 [hep-ph]}}.

\bibitem{Braun:2009mi}
V.~M. Braun, A.~N. Manashov, and B.~Pirnay, ``{Scale dependence of twist-three
  contributions to single spin asymmetries},''
  \href{http://dx.doi.org/10.1103/PhysRevD.80.114002}{{\em Phys. Rev. D}
  {\bfseries 80} (2009) 114002},
  \href{http://arxiv.org/abs/0909.3410}{{\ttfamily arXiv:0909.3410 [hep-ph]}}.
  [Erratum: Phys.Rev.D 86, 119902 (2012)].

\bibitem{Kroll:2022roq}
P.~Kroll and K.~Passek-Kumeri\v{c}ki, ``{Transition GPDs and exclusive
  electroproduction of \ensuremath{\pi}-\ensuremath{\Delta}(1232) final
  states},'' \href{http://dx.doi.org/10.1103/PhysRevD.107.054009}{{\em Phys.
  Rev. D} {\bfseries 107} no.~5, (2023) 054009},
  \href{http://arxiv.org/abs/2211.09474}{{\ttfamily arXiv:2211.09474
  [hep-ph]}}.

\bibitem{Belitsky:2002sm}
A.~V. Belitsky, X.~Ji, and F.~Yuan, ``{Final state interactions and gauge
  invariant parton distributions},''
  \href{http://dx.doi.org/10.1016/S0550-3213(03)00121-4}{{\em Nucl. Phys. B}
  {\bfseries 656} (2003) 165--198},
  \href{http://arxiv.org/abs/hep-ph/0208038}{{\ttfamily arXiv:hep-ph/0208038}}.

\bibitem{Altarelli:1977zs}
G.~Altarelli and G.~Parisi, ``{Asymptotic Freedom in Parton Language},''
  \href{http://dx.doi.org/10.1016/0550-3213(77)90384-4}{{\em Nucl. Phys. B}
  {\bfseries 126} (1977) 298--318}.

\bibitem{Delduc:1980ef}
F.~Delduc, M.~Gourdin, and E.~G. Oudrhiri-Safiani, ``{Photon Structure
  Functions in Quantum Chromodynamics. 2. Feynman Diagram Method},''
  \href{http://dx.doi.org/10.1016/0550-3213(80)90195-9}{{\em Nucl. Phys. B}
  {\bfseries 174} (1980) 157--165}.

\bibitem{Artru:1989zv}
X.~Artru and M.~Mekhfi, ``{Transversely Polarized Parton Densities, their
  Evolution and their Measurement},''
  \href{http://dx.doi.org/10.1007/BF01556280}{{\em Z. Phys. C} {\bfseries 45}
  (1990) 669}.

\bibitem{Vogelsang:1998yd}
W.~Vogelsang, ``{Q**2 evolution of spin dependent parton densities},'' {\em
  Acta Phys. Polon. B} {\bfseries 29} (1998) 1189--1200,
  \href{http://arxiv.org/abs/hep-ph/9805295}{{\ttfamily arXiv:hep-ph/9805295}}.

\bibitem{Lepage:1980fj}
G.~P. Lepage and S.~J. Brodsky, ``{Exclusive Processes in Perturbative Quantum
  Chromodynamics},'' \href{http://dx.doi.org/10.1103/PhysRevD.22.2157}{{\em
  Phys. Rev. D} {\bfseries 22} (1980) 2157}.

\bibitem{Leader:2013jra}
E.~Leader and C.~Lorc\'e, ``{The angular momentum controversy:
  What\textquoteright{}s it all about and does it matter?},''
  \href{http://dx.doi.org/10.1016/j.physrep.2014.02.010}{{\em Phys. Rept.}
  {\bfseries 541} no.~3, (2014) 163--248},
  \href{http://arxiv.org/abs/1309.4235}{{\ttfamily arXiv:1309.4235 [hep-ph]}}.

\bibitem{Lorce:2017wkb}
C.~Lorc\'e, L.~Mantovani, and B.~Pasquini, ``{Spatial distribution of angular
  momentum inside the nucleon},''
  \href{http://dx.doi.org/10.1016/j.physletb.2017.11.018}{{\em Phys. Lett. B}
  {\bfseries 776} (2018) 38--47},
  \href{http://arxiv.org/abs/1704.08557}{{\ttfamily arXiv:1704.08557
  [hep-ph]}}.

\bibitem{Goloskokov:2005sd}
S.~V. Goloskokov and P.~Kroll, ``{Vector meson electroproduction at small
  Bjorken-x and generalized parton distributions},''
  \href{http://dx.doi.org/10.1140/epjc/s2005-02298-5}{{\em Eur. Phys. J. C}
  {\bfseries 42} (2005) 281--301},
  \href{http://arxiv.org/abs/hep-ph/0501242}{{\ttfamily arXiv:hep-ph/0501242}}.

\bibitem{Goloskokov:2007nt}
S.~V. Goloskokov and P.~Kroll, ``{The Role of the quark and gluon GPDs in hard
  vector-meson electroproduction},''
  \href{http://dx.doi.org/10.1140/epjc/s10052-007-0466-5}{{\em Eur. Phys. J. C}
  {\bfseries 53} (2008) 367--384},
  \href{http://arxiv.org/abs/0708.3569}{{\ttfamily arXiv:0708.3569 [hep-ph]}}.

\bibitem{Goloskokov:2009ia}
S.~V. Goloskokov and P.~Kroll, ``{An Attempt to understand exclusive pi+
  electroproduction},''
  \href{http://dx.doi.org/10.1140/epjc/s10052-009-1178-9}{{\em Eur. Phys. J. C}
  {\bfseries 65} (2010) 137--151},
  \href{http://arxiv.org/abs/0906.0460}{{\ttfamily arXiv:0906.0460 [hep-ph]}}.

\bibitem{Curci:1980uw}
G.~Curci, W.~Furmanski, and R.~Petronzio, ``{Evolution of Parton Densities
  Beyond Leading Order: The Nonsinglet Case},''
  \href{http://dx.doi.org/10.1016/0550-3213(80)90003-6}{{\em Nucl. Phys. B}
  {\bfseries 175} (1980) 27--92}.

\end{thebibliography}
\providecommand{\href}[2]{#2}\begingroup\raggedright\endgroup

\end{document}